\documentclass[onecolumn, 12pt]{article}

\usepackage{amsfonts}
\usepackage{epsfig}

\usepackage{graphicx}
\usepackage{float} 
\usepackage{subfigure}
\usepackage{indentfirst}
\usepackage{setspace}
\usepackage[justification=centering]{caption}
\usepackage{latexsym}
\usepackage{multirow}
\usepackage{amsmath}
\usepackage{epstopdf}
\usepackage{cases}
\usepackage{bbding}

\usepackage{bigdelim}
\usepackage{amssymb}
\usepackage{mathrsfs}
\usepackage{caption}
\usepackage{amsthm}
\usepackage{pstricks}
\usepackage[bookmarks,colorlinks,linkcolor=blue,anchorcolor=blue,citecolor=blue]{hyperref}
\usepackage[left=1.25in, right=1.25in, top=1.25in, bottom=1.25in]{geometry}
\usepackage{amsthm}
\usepackage[title]{appendix}
\usepackage{caption}
\usepackage{longtable}
\usepackage{epstopdf}
\usepackage{lineno}

\newtheorem{definition}{Definition}
\newtheorem{lemma}{Lemma}

\newtheorem{theorem}{Theorem}
\newtheorem{corollary}{Corollary}
\newtheorem{example}{Example}
\allowdisplaybreaks[4]

\begin{document}

\title{\Large A Characterization of Sequential Equilibrium through $\varepsilon$-Perfect $\gamma$-Sequential Equilibrium with Local Sequential Rationality and Its Computation\footnote{This work was partially supported by GRF: CityU 11215123 of Hong Kong SAR Government.}}

\author{Yiyin Cao\textsuperscript{\ref{fnote1}} and Chuangyin Dang\textsuperscript{\ref{fnote2}}\footnote{Corresponding Author. Email: mecdang@cityu.edu.hk}}

\date{}
\maketitle             

\footnotetext[1]{School of Management, Xi'an Jiaotong University, Xi'an, China, yiyincao2-c@my.cityu.edu.hk\label{fnote1}}
\footnotetext[2]{Department of Systems Engineering, City University of Hong Kong, Kowloon, Hong Kong  \label{fnote2}}

\begin{abstract}

Sequential equilibrium requires a consistent assessment and sequential rationality, where the consistent assessment emerges from a convergent sequence of totally mixed behavioral strategies and associated beliefs. However, the original definition lacks explicit guidance on constructing such convergent sequences. To overcome this difficulty, this paper presents a characterization of sequential equilibrium by introducing $\varepsilon$-perfect $\gamma$-sequential equilibrium with local sequential rationality. For any $\gamma>0$, we establish a perfect $\gamma$-sequential equilibrium as a limit point of a sequence of $\varepsilon_k$-perfect $\gamma$-sequential equilibrium with $\varepsilon_k\to 0$. A sequential equilibrium is then derived from a limit point of a sequence of perfect $\gamma_q$-sequential equilibrium with $\gamma_q\to 0$. This characterization systematizes the construction of convergent sequences and enables the analytical determination of sequential equilibria and the development of a polynomial system serving as a necessary and sufficient condition for $\varepsilon$-perfect $\gamma$-sequential equilibrium. Exploiting the characterization, we develop a differentiable path-following method to compute a sequential equilibrium. 

\end{abstract}


\maketitle

\section{Introduction}

 Nash equilibrium prescribes a notion of rational behavior in extensive-form games. When players behave according to a Nash equilibrium, the sequence of their actions generates an equilibrium path. Nevertheless, a Nash equilibrium may induce irrational behavior off the equilibrium path. To overcome this deficiency, Selten~\cite{Selten (1975)} introduced the concept of perfect equilibrium through strategy perturbations. However, it remains a difficult task to verify whether a Nash equilibrium is perfect or not. To ease such a task as one of the purposes, Kreps and Wilson~\cite{Kreps and Wilson (1982)} developed the concept of sequential equilibrium by demanding global sequential rationality  and consistency off the equilibrium path. A sequential equilibrium consists of two types of probability assessments: behavioral strategy profiles and beliefs for each player. The beliefs of a player at an information set concern a probability distribution on the histories in the information set, indicating which history has been reached. The global sequential rationality condition imposes that taking the beliefs and other players' strategies as fixed, no player would prefer to change their strategies at any information set. Furthermore, the consistency requirement necessitates that the assessment should be a limit point of a sequence of totally mixed strategy profiles and associated beliefs, which are derived from the strategy profiles using Bayes' rule. Due to the nonconstructive feature of the definition of sequential equilibrium, Kreps and Wilson~\cite{Kreps and Wilson (1982)} could only derive the existence of a sequential equilibrium from the existence of a perfect equilibrium. Kohlberg and Reny~\cite{Kohlberg and Reny (1997)} gave a natural interpretation of the consistent assessment.
 

To facilitate a better understanding of the practical applications of sequential equilibrium, we provide a selection of scenarios where this concept is commonly employed. Camerer and Weigelt~\cite{Camerer and Weigelt (1988)} showed through the experiments that the concept of sequential equilibrium predicts the behavior of players in a lending game. Geanakoplos et al.~\cite{Geanakoplos et al. (1989)} verified the existence of a sequential equilibrium in psychological games. Battigalli and Dufwenberg~\cite{Battigalli and Dufwenberg (2009)} extended the definition of consistency of sequential equilibrium by adding a requirement concerning the higher-order beliefs that need to be specified in psychological games. Besanko and Spulber~\cite{Besanko and Spulber (1990)} constituted a sequential equilibrium model of private antitrust enforcement. Manelli~\cite{Manelli (1996)} proved the existence of sequential equilibria in a cheap-talk extension of signaling games. Adao and Temzelides~\cite{Adao and Temzelides (1998)} analytically found sequential equilibria in a Diamond-Dybvig model with multiple banks. Huang and Werner~\cite{Huang and Werner (2000)} studied the nature of asset price bubbles in sequential markets through sequential equilibrium. Choi et al.~\cite{Choi et al. (2008)} analyzed the set of sequential equilibria in voluntary contribution games. The wide range of applications mentioned above demonstrates the valuable role of sequential equilibrium in understanding and analyzing complex decision-making scenarios.

In advancing the practical application of sequential equilibrium, the analytical or computational determination of such equilibria is pivotal.  As stated in Kreps and Wilson~\cite{Kreps and Wilson (1982)}, an assessment of strategies and beliefs is a sequential equilibrium if it satisfies the criteria of global sequential rationality and consistency. To check whether a consistent assessment is a sequential equilibrium, the existing approaches tend to link the backward induction procedure or dynamic programming to the sequential rationality since the procedure ensures that players' strategies characterize optimal behavior at every information set of the game. To simplify the verification procedure of consistency in the original definition, Kreps and Wilson~\cite{Kreps and Wilson (1982)} proposed a labeling procedure. Nevertheless, the consistency verification of an assessment remains a computationally challenging problem within the paradigm of sequential equilibrium. While Kreps and Wilson's definition offers valuable insights into sequential equilibrium, it falls short in providing explicit guidance on constructing a convergent sequence to verify whether a Nash equilibrium in the agent normal-form or associated normal-form representation of an extensive-form game qualifies as a sequential equilibrium. Concerning this point, Kreps and Wilson~\cite{Kreps and Wilson (1982)} themselves convey dissatisfaction with their definition of sequential equilibrium: ``What ought to be the definition of a consistent assessment that, with sequential rationality, will give the proper definition of a sequential equilibrium." Kreps~\cite{Kreps (1990)} believed that a lot of bodies are buried in the definition of consistency. Osborne and Rubinstein~\cite{Osborne and Rubinstein (1994)} regarded the consistency requirement as a rather opaque technical assumption. In a similar vein, Watson~\cite{Watson (2023)} highlighted the intricate challenge of constructing a sequential equilibrium for complex games, emphasizing the need to delineate a sequence of totally mixed behavioral strategies that determines beliefs across all information sets and converges towards the equilibrium strategy profile. These concerns inspire the development of a mathematical characterization of sequential equilibrium within our paper, which offers an effective scheme on constructing the necessary convergent sequences to meet the consistency requirement.

To the best of our knowledge, there are very few methods available for directly computing a sequential equilibrium. As a numerical implementation of the logistic agent quantal response equilibrium in Mckelvey and Palfey~\cite{Mckelvey and Palfey (1998)}, a differentiable path-following method was proposed in Turocy~\cite{Turocy (2010)} for computing sequential equilibria. However, the method may have difficulties in numerically finding such an equilibrium with the desired accuracy. Some relevant studies on computing Nash equilibria, normal-form perfect equilibria, and quasi-perfect equilibria in $n$-person or two-person extensive-form games can be found in Wilson~\cite{Wilson (1972)},  von Stengel et al.~\cite{von Stengel et al. (2002)}, and Miltersen and Sorensen~\cite{Miltersen and Sorensen (2010)}. Recently, Dang et al.~\cite{Dang et al. (2022)} and Eibelsh$\ddot{a}$user et al.~\cite{Eibelshauser et al. (2023)} introduced approaches for computing stationary equilibria in stochastic games. Various differentiable path-following methods have been proposed in the literature to compute Nash equilibria, perfect equilibria, and proper equilibria for $n$-person normal-form games, including Herings and Peeters~\cite{Herings and Peeters (2001)}, Chen and Dang~\cite{Chen and Dang (2021)}, and Cao et al.~\cite{Cao et al. (2024)}. 
Although the existing methods have contributed a lot to equilibrium computation in extensive-form games, most of these methods primarily focus on computing equilibria other than sequential equilibrium. The methods designed for computing equilibria weaker than sequential equilibrium limit their effectiveness in finding sequential equilibria, while the methods targeting a refinement of sequential equilibrium can be computationally more demanding. 

To enhance the analytical and computational efficacy in identifying sequential equilibria, we present a characterization of sequential equilibrium through $\varepsilon$-perfect $\gamma$-sequential equilibrium with local sequential rationality. For any given $\gamma>0$, a perfect $\gamma$-sequential equilibrium is generated as a limit point of a sequence of $\varepsilon_k$-perfect $\gamma$-sequential equilibrium with $\varepsilon_k\to 0$. A sequential equilibrium is then obtained from a limit point of a sequence of perfect $\gamma_q$-sequential equilibrium with $\gamma_q\to 0$. This characterization enables the construction of a required convergent sequence and the analytical determination of all sequential equilibria for small extensive-form games without the need to verify whether an assessment fulfills the criteria of consistency and global sequential rationality. Furthermore, the implementation of local sequential rationality as mandated in our characterization results in a polynomial system serving as a necessary and sufficient condition for sequential equilibrium. 
Exploiting the characterization, we develop an entropy-barrier differentiable path-following method to compute sequential equilibria. The method involves formulating an entropy-barrier extensive-form game, where each player at each of their information sets solves a strictly convex optimization problem. For practical applications, our method has been coded in MATLAB. Comprehensive numerical experiments have been carried out to demonstrate the efficiency of the method.

The rest of the paper is organized as follows. Section 2 introduces the necessary notation and reviews Kreps and Wilson's concept of sequential equilibrium. Section 3 presents our characterization of sequential equilibrium. Section 4 provides two illustrative examples demonstrating the analytical determination of all sequential equilibria in small extensive-form games. Section 5 develops an entropy-barrier differentiable path-following method. Numerical experiments are reported in Section 6. The concluding remarks are made in Section 7. 

\section{Preliminaries}

This paper is concerned with a finite extensive-form game with perfect recall.  To describe such a game in accordance with Osborne and Rubinstein~\cite{Osborne and Rubinstein (1994)}, we need to introduce some essential notations, which are presented in Table~\ref{Table} for ease of reference. Figure~\ref{Notation} illustrates a selection of these notations to further clarify their meanings.
\begin{table}[H]\setlength{\abovedisplayskip}{1.2pt}
\setlength{\belowdisplayskip}{1.2pt}
\linespread{1.2} 
\footnotesize
\centering
\caption{Notation for Extensive-Form Games}
\label{Table}
\begin{tabular*}{\hsize}{@{}@{\extracolsep{\fill}}l|l@{}}
\hline
Notation & Terminology\\
\hline
$N=\{1,2,\ldots,n\}$ & Set of players without the chance player\\ 
$h=\langle a_1,a_2,\ldots,a_L\rangle$ &  A history, which is a sequence of actions taken by players\\ 
$H$, $\emptyset\in H$ &  Set of histories, $\langle a_1,\ldots,a_L\rangle\in H$ if $\langle a_1,\ldots,a_K\rangle\in H$ and $L<K$\\ 
$Z$ & Set of terminal histories\\ 
$A(h)=\{a:\langle h,a\rangle\in H\}$ & Set of actions after a nonterminal history $h\in H$\\ 
$P(h)$ & Player who takes an action after a history $h\in H$\\ 
${\cal I}_i$ & Collection of information partitions of player $i$\\ 
$I_i^j\in {\cal I}_i$, $j\in M_i=\{1,\ldots,m_i\}$ & $j$th information set of player $i$, $A(h)=A(h')$ whenever $h,h'\in I_i^j$\\ 
$A(I^j_i)=A(h)$ for any $h\in I^j_i$ & Set of actions of player $i$ at information set $I^j_i$\\ 
$\beta=(\beta^i_{I_i^j}(a):i\in N,I_i^j\in{\cal I}_i,a\in A(I_i^j))$ & Profile of behavioral strategies\\ 
$\beta^i=(\beta^i_{I_i^j}:j\in M_i)$ & Behavioral strategy of player $i$\\  
$\beta^{-i}=(\beta^p_{I^p_q}:p\in N\backslash\{i\},q\in M_p)$ & Profile of behavioral strategies without $\beta^i$\\ 
$\beta^i_{I_i^j}=(\beta^i_{I_i^j}(a):a\in A(I_i^j))^\top$ & Probability measure over $A(I_i^j)$ and $\beta^i_h=\beta^i_{I^j_i}$ for any $h\in I^j_i$\\
$\beta^{-I^j_i}=(\beta^p_{I^p_q}:p\in N,q\in M_p,I^q_p\ne I^j_i)$ & Profile of behavioral strategies without $\beta^i_{I^j_i}$\\ 
$f_c(\cdot|h)=(f_c(a|h):a\in A(h))^\top$ & Probability measure of the chance player $c$ over $A(h)$ \\ 
$\mu=(\mu^i_{I_i^j}(h):i\in N,I_i^j\in{\cal I}_i,h\in I_i^j)$ & Belief system, $\sum_{h\in I_i^j}\mu^i_{I_i^j}(h)=1$ and $\mu^i_{I_i^j}(h)\ge 0$ for all $h\in I_i^j$\\ 
$h\cap A(I^j_i)$; $a\in h$ & $\{a_1,\ldots,a_L\}\cap A(I^j_i)$; $a\in \{a_1,\ldots,a_L\}$ for $h=\langle a_1,\ldots,a_L\rangle$\\
$u^i:Z\rightarrow\mathbb{R}$ & Payoff function of player $i$\\ \hline
\end{tabular*}
\end{table}
\begin{figure}[ht]\setlength{\abovedisplayskip}{1.2pt}
\setlength{\belowdisplayskip}{1.2pt}
    \centering
    \begin{minipage}{0.4\textwidth}
        \centering
        \includegraphics[width=0.9\textwidth, height=0.25\textheight]{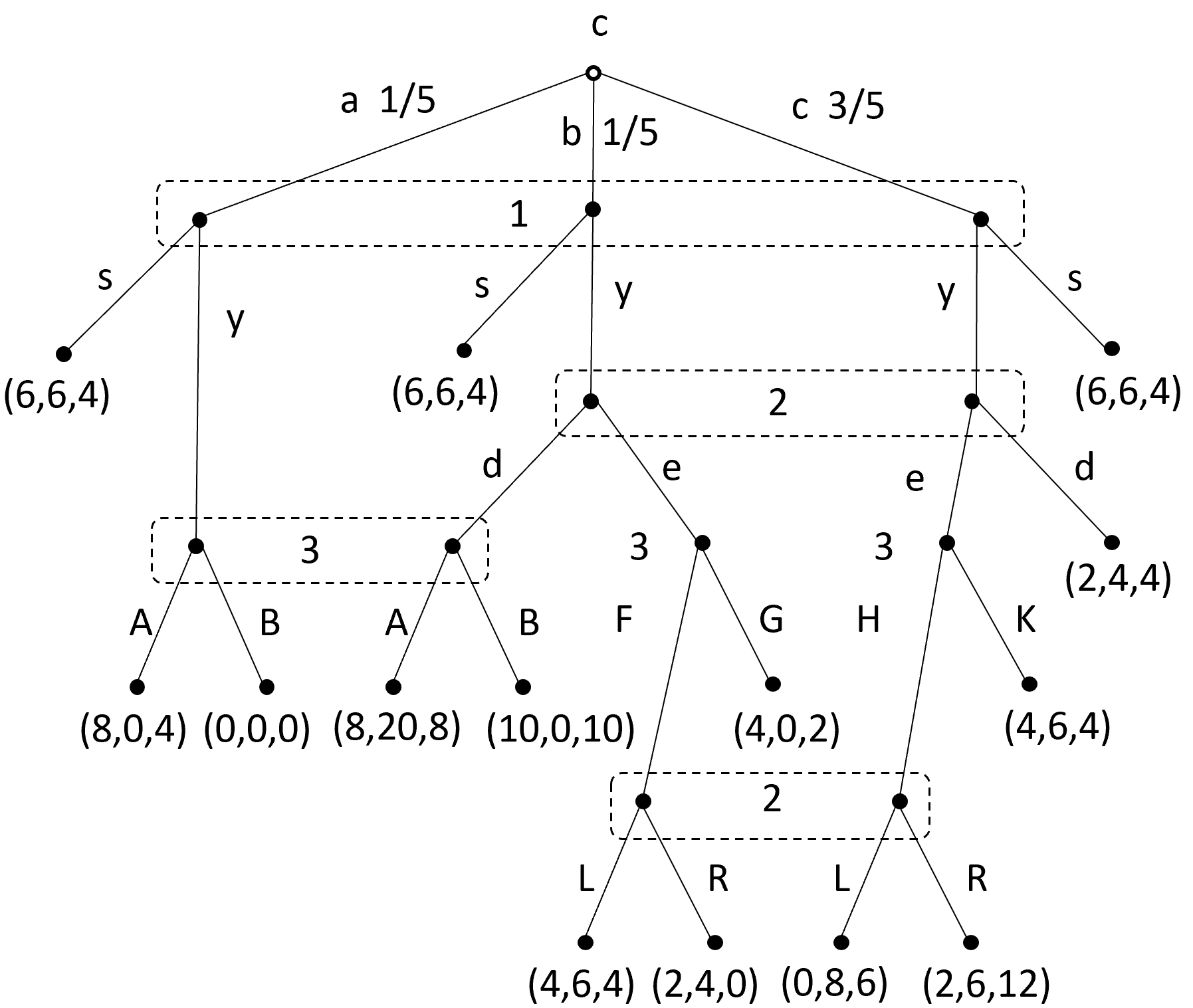}
\end{minipage}\hfill
    \begin{minipage}{0.6\textwidth}
       {\scriptsize
       Information sets: ${\cal I}_1=\{I^1_1\}$, ${\cal I}_2=\{I^1_2,I^2_2\}$,  ${\cal I}_3=\{I^1_3,I^2_3,I^3_3\}$, $I^1_1=\{\langle a\rangle,\langle b\rangle,\langle c\rangle\}$, $I^1_2=\{\langle b, y\rangle, \langle c, y\rangle\}$, $I^2_2=\{\langle b, y, e, F\rangle, \langle c, y, e, H\rangle\}$, $I^1_3=\{\langle a,  y \rangle, \langle b,  y, d \rangle\}$, $I^2_3=\{\langle b,  y, e \rangle\}$, $I^3_3=\{\langle c,  y, e \rangle\}$.\newline Player who takes an action after a history: $P(\emptyset)=c$, $P(\langle b\rangle)=1$, $P(\langle b, y\rangle)= 2$, $P(\langle c, y, e, H\rangle)=2$, $P(\langle b, y, d\rangle)=3$, $P(\langle b, y, e\rangle)=3$.\newline
       Action sets: $A(\emptyset)=\{a,b,c\}$, $A(I^1_1)=\{s, y\}$,  $A(I^1_2)=\{d, e\}$, $A(I^3_3)=\{H, K\}$. \newline
       $h\cap A(I^j_i)$: $h\cap A(I^1_2)=\{d\}$ for $h=\langle b, y, d, A\rangle$, $h\cap A(I^1_1)=\{y\}$ for $h=\langle c, y, e\rangle$.\newline
       Behavioral strategies: $f_c(\cdot|\emptyset)=(f_c(a|\emptyset), f_c(b|\emptyset),f_c(c|\emptyset))^\top=(\frac{1}{5},\frac{1}{5},\frac{3}{5})^\top$,
       $\beta^1_{I^1_1}=(\beta^1_{I^1_1}(s),\beta^1_{I^1_1}(y))^\top$, $\beta^2_{I^2_2}=(\beta^2_{I^2_2}(L),\beta^2_{I^2_2}(R))^\top$,  $\beta^3_{I^3_3}=(\beta^3_{I^3_3}(H),\beta^3_{I^3_3}(K))^\top$.\newline
       Beliefs: $\mu^2_{I^1_2}=(\mu^2_{I^1_2}(\langle b, y\rangle), \mu^2_{I^1_2}(\langle c, y\rangle))^\top$, $\mu^3_{I^1_3}=(\mu^3_{I^1_3}(\langle a, y\rangle), \mu^3_{I^1_3}(\langle b, y, d\rangle))^\top$.}
        \end{minipage}
\caption{\label{Notation} Illustrations of Some Notations}
\end{figure}
Given these notations, an extensive-form game is represented as $\Gamma=\langle N, H, P, f_c, \{{\cal I}_i\}_{i\in N}\rangle$. When $P(\langle a_1,\ldots,a_k\rangle)=c$, $\beta^c_{\langle a_1,\ldots,a_k\rangle}(a_{k+1})=f_c(a_{k+1}|\langle a_1,\ldots,a_k\rangle)$. For $h=\langle a_1,a_2,\ldots,a_k\rangle$, we have $h=\emptyset$ when $k=0$. A finite extensive-form game means an extensive-form game with a finite number of histories. Let $X_i(h)$ be the record of player $i$'s experience along the history $h$. Then, $X_i(h)$ is the sequence consisting of the information sets that player $i$  encounters in the history $h$ and the actions he takes at them in the order that these events occur. An extensive-form game has perfect recall if, for each player $i$, we have $X_i(h')=X_i(h'')$ whenever the histories $h'$ and $h''$ are in the same information set of player $i$.

We denote by $\text{int}(C)$ and $|C|$ the interior of a set $C$ and the cardinality of a finite set $C$, respectively. Let
$\triangle=\mathop{\times}\limits_{i\in N,\;j\in M_i}\triangle^i_{I^j_i}$ and $\triangle^i=\mathop{\times}\limits_{j\in M_i}\triangle^i_{I^j_i}$, where $\triangle^i_{I^j_i}=\{\beta^i_{I^j_i}\in\mathbb{R}_+^{|A(I^j_i)|}|\sum\limits_{a\in A(I^j_i)}\beta^i_{I^j_i}(a)=1\}$. Let $\Xi=\mathop{\times}\limits_{i\in N,\;j\in M_i}\Xi^i_{I^j_i}$, where $\Xi^i_{I^j_i}=\{\mu^i_{I^j_i}=(\mu^i_{I^j_i}(h):h\in I^j_i)^{\top}|\sum\limits_{h\in I^j_i}\mu^i_{I^j_i}(h)=1,\;0\le\mu^i_{I^j_i}(h)\}$. 
An {\bf assessment} in an extensive-form game is a pair $(\beta,\mu)\in\triangle\times\Xi$, where $\beta$ is a behavioral strategy profile and $\mu$ is a function that assigns to every information set a probability measure on the set of histories in the information set. 
We describe $\mu$ as a {\bf belief system}. A player's belief at an information set $I^j_i$ along a specific history $h \in I^j_i$, denoted as $\mu^i_{I^j_i}(h)$, represents the probability that the history $h$ has occurred when it is player $i$'s turn to choose an action at $I^j_i$. A behavioral strategy profile $\beta=(\beta^i_{I^j_i}(a):i\in N,j\in M_i,a\in A(I^j_i))$ is said to be totally mixed if $\beta^i_{I^j_i}(a)>0$ for all $i\in N$, $j\in M_i$, and $a\in A(I^j_i)$. Moreover, a belief system $\mu$ is said to be totally mixed if $\mu^i_{I^j_i}(h)>0$ for all $i\in N,j\in M_i,h\in I^j_i$. In this paper, $\beta>0$ and $\mu>0$ indicate that $\beta$ and $\mu$ are totally mixed, respectively. 

For a behavioral strategy profile $\beta$ and a history $h=\langle a_1,\ldots,a_L\rangle\in H$, the realization probability of $h$ at $\beta$ equals \begin{equation}\label{Aseeq1}\setlength{\abovedisplayskip}{1.2pt}
\setlength{\belowdisplayskip}{1.2pt}\omega(h|\beta)=\prod\limits_{k=0}^{L-1}\beta^{P(\langle a_1,\ldots,a_k\rangle)}_{\langle a_1,\ldots,a_k\rangle}(a_{k+1}),\end{equation} where $\omega(h|\beta)$ represents the probability that the moves along $h$ are played when $\beta$ is taken by players. For $i\in N$, $j\in M_i$, and $h=\langle a_1,\ldots,a_L\rangle\in H$, when action $a\in A(I^j_i)$ is taken, the realization probability of $h$ at $(a,\beta^{-I^j_i})$ equals \begin{equation}\label{Aseeq2}\setlength{\abovedisplayskip}{1.2pt}
\setlength{\belowdisplayskip}{1.2pt}\omega(h|a,\beta^{-I^j_i})=\mathop{\prod\limits_{k=0}^{L-1}}\limits_{\langle a_1,\ldots,a_k\rangle\notin I^j_i}\beta^{P(\langle a_1,\ldots,a_k\rangle)}_{\langle a_1,\ldots,a_k\rangle}(a_{k+1}).\end{equation}  For the game in Fig.~\ref{Notation},  $\omega(\langle b, y\rangle|\beta)=f_c(b|\emptyset)\beta^1_{\langle b\rangle}(y)=\frac{1}{5}\beta^1_{I^1_1}(y)$ and $\omega(\langle b, y, d\rangle|y, \beta^{-I^1_1})=f_c(b|\emptyset)\beta^2_{\langle b, y\rangle}(d)=\frac{1}{5}\beta^2_{I^1_2}(d)$. The expected payoff of player $i$ at a behavioral strategy profile $\beta$ is given by
\begin{equation}\label{Aseeq3}\setlength{\abovedisplayskip}{1.2pt}
\setlength{\belowdisplayskip}{1.2pt}
u^i(\beta)=\sum\limits_{h\in Z}u^i(h)\omega(h|\beta).\end{equation}
To represent conditional expected payoffs on information sets with a belief system $\mu$,
for $i\in N$, $j\in M_i$, $a\in A(I^j_i)$, and $h=\langle a_1,\ldots,a_K\rangle\in Z$, let {\small
\begin{equation}\label{seedeqA}\setlength{\abovedisplayskip}{1.2pt}
\setlength{\belowdisplayskip}{1.2pt}
\nu^i_{I^j_i}(h|\beta,\mu)=\left\{\begin{array}{ll} 
\mu^i_{I^j_i}(\hat h)\prod\limits_{k=L}^{K-1}\beta^{P(\langle a_1,\ldots,a_k\rangle)}_{\langle a_1,\ldots,a_k\rangle}(a_{k+1}) & \text{if $\hat h=\langle a_1,\ldots,a_L\rangle\in I^j_i$,}\\
0 & \text{if there is no subhistory of $h$ in $I^j_i$,}
\end{array}\right.\end{equation}}
and {\small
\begin{equation}\label{seedeqB}\setlength{\abovedisplayskip}{1.2pt}
\setlength{\belowdisplayskip}{1.2pt}
\nu^i_{I^j_i}(h|a,\beta^{-I^j_i},\mu)=\left\{\begin{array}{ll}
\mu^i_{I^j_i}(\hat h)\prod\limits_{k=L+1}^{K-1}\beta^{P(\langle a_1,\ldots,a_k\rangle)}_{\langle a_1,\ldots,a_k\rangle}(a_{k+1}) & \text{if $\hat h=\langle a_1,\ldots,a_L\rangle\in I^j_i$,}\\
0 & \text{if there is no subhistory of $h$ in $I^j_i$,}
\end{array}\right.\end{equation}}where $\nu^i_{I^j_i}(h|\beta,\mu)$ represents the probability that the moves along $h$ are played if $\beta$ is taken by players given that  $I^j_i$ has been reached and the belief system is $\mu$. Within $\nu^i_{I^j_i}(h|\beta,\mu)$, $\mu^i_{I^j_i}(\hat{h})$ represents  the probability that a subhistory $\hat{h}\in I^j_i$ along $h$ has occurred conditioned on information set $I^j_i$ has been reached. For the game in Fig.~\ref{Notation},  $\nu^2_{I^1_2}(\langle b, y, d, A\rangle|\beta,\mu)=\mu^2_{I^1_2}(\langle b, y\rangle)\beta^2_{I^1_2}(d)\beta^3_{I^1_3}(A)$ and $\nu^2_{I^1_2}(\langle b, y, d, A\rangle|d, \beta^{-I^1_2},\mu)=\mu^2_{I^1_2}(\langle b, y\rangle)\beta^3_{I^1_3}(A)$.

With these notations, the conditional expected payoff of player $i$ on $I^j_i$ at $(\beta,\mu)$ can be denoted as
\begin{equation}\label{cpeq1}\setlength{\abovedisplayskip}{1.2pt}\setlength{\belowdisplayskip}{1.2pt}
u^i(\beta,\mu|I^j_i)=\sum\limits^
{h\cap A(I^j_i)\ne\emptyset}_{h\in Z}u^i(h)\nu^i_{I^j_i}(h|\beta,\mu),\end{equation}
and when player $i$ takes a pure action $a\in A(I^j_i)$ at $I^j_i$, the conditional expected payoff of player $i$ on $I^j_i$ at $(a,\beta^{-I^j_i},\mu)$ can be denoted as
\begin{equation}\label{cpeq2}\setlength{\abovedisplayskip}{1.2pt}\setlength{\belowdisplayskip}{1.2pt}
u^i(a,\beta^{-I^j_i},\mu|I^j_i)=\sum\limits_{a\in h\in Z}u^i(h)\nu^i_{I^j_i}(h|a,\beta^{-I^j_i},\mu).\end{equation}
For the game in Fig.~\ref{Notation},  $u^2(\beta,\mu| I^2_2)=\mu^2_{I^2_2}(\langle b, y, e, F\rangle)(6\beta^2_{I^2_2}(L)+4\beta^2_{I^2_2}(R))+ \mu^2_{I^2_2}(\langle c, y, e, H\rangle)$ $(8\beta^2_{I^2_2}(L)+6\beta^2_{I^2_2}(R))$ and $u^2(R, \beta^{-I^2_2},\mu| I^2_2)=4\mu^2_{I^2_2}(\langle b, y, e, F\rangle)+ 6\mu^2_{I^2_2}(\langle c, y, e, H\rangle)$. 

For $i\in N$ and $j\in M_i$, the realization probability of information set $I^j_i$ is given by
\begin{equation}\label{Bseeq1}\setlength{\abovedisplayskip}{1.2pt}
\setlength{\belowdisplayskip}{1.2pt}\omega(I^j_i|\beta)=\sum\limits_{h\in I^j_i}\omega(h|\beta),\end{equation} where $\omega(I^j_i|\beta)$ represents the probability that $I^j_i$ is reached when $\beta$ is played. For the game in Fig.~\ref{Notation},  $\omega(I^1_2|\beta)=\omega(\langle b, y\rangle|\beta)+\omega(\langle c, y\rangle|\beta)=(f_c(b|\emptyset)+f_c(c|\emptyset))\beta^1_{I^1_1}(y)=\frac{4}{5}\beta^1_{I^1_1}(y)$.
As $\beta>0$, we define $\mu(\beta)=(\mu^i_{I^j_i}(h|\beta):i\in N,j\in M_i,h\in I^j_i)$ with  $\mu^i_{I^j_i}(h|\beta)=\frac{\omega(h|\beta)}{\omega(I^j_i|\beta)}$ for $h\in I^j_i$. Clearly, $\mu^i_{I^j_i}(h|\beta)$ is a continuous function on $\text{int}(\triangle)$. 
For $i\in N$, $j\in M_i$ and $a\in A(I^j_i)$, we denote \begin{equation}\label{seedeqC}\setlength{\abovedisplayskip}{1.2pt}
\setlength{\belowdisplayskip}{1.2pt} u^i(\beta\land I^j_i)=\sum\limits^{h\cap A(I^j_i)\ne\emptyset}_{h\in Z}u^i(h)\omega(h|\beta)\text{\; and \;}u^i((a,\beta^{-I^j_i})\land I^j_i)=\sum\limits_{a\in h\in Z}u^i(h)\omega(h|a,\beta^{-I^j_i}),\end{equation} where $u^i(\beta\land I^j_i)$ and $u^i((a,\beta^{-I^j_i})\land I^j_i)$ indicate the expected payoffs of player $i$ along terminal histories that intersect with the action set of player $i$ at information set $I^j_i$ when $\beta$ and $(a,\beta^{-I^j_i})$ are played, respectively. For the game in Fig.~\ref{Notation}, $u^3(\beta\land I^1_3)=\frac{1}{5}\beta^1_{I^1_1}(y)(4\beta^3_{I^1_3}(A)+ \beta^2_{I^1_2}(d)(8\beta^3_{I^1_3}(A)+10\beta^3_{I^1_3}(B)))$ and $u^3((B, \beta^{-I^1_3})\land I^1_3)= 2\beta^1_{I^1_1}(y)\beta^2_{I^1_2}(d)$.
When $\beta>0$ and $\mu=\mu(\beta)$, it follows from Eqs.~(\ref{seedeqA})-(\ref{seedeqC}) that \begin{equation}\label{seedeqD}\setlength{\abovedisplayskip}{1.2pt}
\setlength{\belowdisplayskip}{1.2pt} u^i(\beta,\mu|I^j_i)=\frac{u^i(\beta\land I^j_i)}{\omega(I^j_i|\beta) }\text{ and }u^i(a,\beta^{-I^j_i},\mu|I^j_i)=\frac{u^i((a,\beta^{-I^j_i})\land I^j_i)}{\omega(I^j_i|\beta)}.\end{equation} For $i\in N$ and $j\in M_i$, when $\omega(I^j_i|\beta)>0$,
let \begin{equation}\label{seedeqE}\setlength{\abovedisplayskip}{1.2pt}
\setlength{\belowdisplayskip}{1.2pt}  u^i(\beta| I^j_i)=
\frac{u^i(\beta\land I^j_i)}{\omega(I^j_i|\beta) }\text{ and }u^i(a,\beta^{-I^j_i}|I^j_i)=\frac{u^i((a,\beta^{-I^j_i})\land I^j_i)}{\omega(I^j_i|\beta)}.\end{equation}

A Nash equilibrium in behavioral strategies of an extensive-form game is a behavioral strategy profile $\beta^*$ with the property that, for every player $i\in N$, we have \(u^i(\beta^*)\ge u^i(\beta^i, \beta^{*-i})\) for every behavioral strategy $\beta^i$ of player $i$. A subgame perfect equilibrium is a Nash equilibrium whose restriction on every subgame remains to be a Nash equilibrium of the subgame. 
As a strict refinement of subgame perfect equilibrium, the concept of sequential equilibrium was proposed by Kreps and Wilson~\cite{Kreps and Wilson (1982)} for a finite extensive-form game with perfect recall. 
Given the above notations, the concept of sequential equilibrium in Kreps and Wilson~\cite{Kreps and Wilson (1982)} can be stated as follows.

\begin{definition}[{\bf Sequential Equilibrium}, Kreps and Wilson~\cite{Kreps and Wilson (1982)}]\label{sed1} {\em An assessment $(\beta^*,\mu^*)$ is {\bf sequentially rational} if, for every player $i$ and every information set $I^j_i\in {\cal I}_i$, we have 
\(u^i(\beta^*,\mu^*|I^j_i)\ge u^i(\beta^i,\beta^{*-i},\mu^*|I^j_i)\)
for every $\beta^i$ of player $i$. An assessment $(\beta^*,\mu^*)$ is {\bf consistent} if there is a sequence of assessments, $\{(\beta^\ell,\mu^\ell), \ell=1,2,\ldots\}$, that converges to $(\beta^*,\mu^*)$ in Euclidean space and has the properties that each behavioral strategy profile $\beta^\ell$ is totally mixed and that each belief system $\mu^\ell$ is derived from $\beta^\ell$ with Bayes' rule. 
An assessment $(\beta^*,\mu^*)$ is a {\bf sequential equilibrium} of a finite extensive-form game with perfect recall if it is sequentially rational and consistent.}
\end{definition} 

Throughout the remainder of this paper, we employ the term ``global sequential rationality" to refer to Kreps and Wilson's notion of sequential rationality. This distinction is necessary as we will introduce a corresponding concept termed ``local sequential rationality" in relation to global sequential rationality.
\newline
\textit{ {\bf Global Rationality at Information Set $I^j_i$}: An assessment $(\beta^*,\mu^*)$ possesses global rationality at information set $I^j_i$ if
 \(u^i(\beta^*,\mu^*|I^j_i)\ge u^i(\beta^i,\beta^{*-i},\mu^*|I^j_i)\)
 for every $\beta^i$ of player $i$. \newline
 {\bf Local Rationality at Information Set $I^j_i$}: An assessment $(\beta^*,\mu^*)$ possesses local rationality at information set $I^j_i$ if \(u^i(\beta^*,\mu^*|I^j_i)\ge u^i(\beta^i_{I^j_i},\beta^{*-I^j_i},\mu^*|I^j_i)\)
 for every $\beta^i_{I^j_i}$ of player $i$.\newline
 {\bf Global Sequential Rationality}: An assessment $(\beta^*,\mu^*)$ possesses global sequential rationality if it meets the global rationality at every information set. \newline
 {\bf Local Sequential Rationality}: An assessment $(\beta^*,\mu^*)$ possesses local sequential rationality if it meets the local rationality at every information set.}

To apply Definition~\ref{sed1} in the determination of all the sequential equilibria for an extensive-form game, there is a three-step procedure. Step 1: Compute all Nash equilibria of the agent normal-form or associated normal-form representation. Step 2: For each Nash equilibrium, find a belief system and determine whether there exists a convergent sequence of totally mixed behavioral strategy profiles that meet the consistency. Step 3: For each assessment, verify the global sequential rationality.
It is evident that Definition~\ref{sed1} lacks explicit guidance on how to construct a convergent sequence of totally mixed behavioral strategy profiles that satisfy the consistency condition for a Nash equilibrium in the agent normal-form or associated normal-form representation of an extensive-form game. This limitation serves as the primary motivation for the work presented in this paper.

\section{\large A Characterization of Sequential Equilibrium through $\varepsilon$-Perfect $\gamma$-Sequential Equilibrium}

This section presents a characterization of sequential equilibrium through $\varepsilon$-perfect $\gamma$-sequential equilibrium with local sequential rationality. To accomplish this task, this section is organized as follows. Since global sequential rationality is defined by nonconvex payoff functions $u^i(\beta^i,\beta^{*-i},\mu^*|I^j_i)$, one needs to solve a dynamic programming problem at each information set. To address this challenge, we first establish a necessary and sufficient condition for determining whether a given consistent assessment is a sequential equilibrium through local sequential rationality.
Next, to provide an effective framework for constructing a consistent assessment, we present an equivalent definition of sequential equilibrium by introducing the concept of $\varepsilon$-perfect $\gamma$-sequential equilibrium with local sequential rationality. This characterization serves as a practical tool for analytically identifying all sequential equilibria in small games. 
Finally, to establish a polynomial system as a necessary and sufficient condition for determining whether a totally mixed assessment qualifies as an $\varepsilon$-perfect $\gamma$-sequential equilibrium, we separate the perturbation from strategies and secure an equivalent definition of $\varepsilon$-perfect $\gamma$-sequential equilibrium. 


Let \(\widetilde{\mathscr{B}}=\{(\beta,\mu(\beta))|\beta>0,\;\mu(\beta)=(\mu^i_{I^j_i}(h|\beta):i\in N,j\in M_i,h\in I^j_i),\;\mu^i_{I^j_i}(h|\beta)=\frac{\omega(h|\beta)}{\omega(I^j_i|\beta)}\}\) and $\mathscr{B}$ be the closure of $\widetilde{\mathscr{B}}$. Clearly, $(\beta^*,\mu^*)\in \mathscr{B}$ is a consistent assessment. Using this notation, we arrive at a necessary and sufficient condition for determining whether a consistent assessment is a sequential equilibrium through local sequential rationality.
\begin{theorem}\label{edsethm0}{\em $(\beta^*,\mu^*)\in\mathscr{B}$ is a sequential equilibrium if and only if it meets the property that $\beta^{*i}_{I^j_i}(a')=0$ for any $i\in N$, $j\in M_i$ and $a',a''\in A(I^j_i)$ with $ u^i(a'',\beta^{*-I^j_i},\mu^*|I^j_i)-u^i(a',\beta^{*-I^j_i},\mu^*|I^j_i)>0$.
}
\end{theorem}
\begin{proof} In the proof of this theorem, the following notations will be required. For $i\in N$, $j\in M_i$, and $a\in A(I^j_i)$, let 
 \[\setlength{\abovedisplayskip}{1.2pt}
\setlength{\belowdisplayskip}{1.2pt}
\begin{array}{l}
M(a,I^j_i)=\left\{q\in M_i\left|\begin{array}{l}\text{for any
 $h=\langle a_1,\ldots,a_L\rangle\in I^q_i$, there exists $1\le \ell\le L$ such}\\
 \text{that $a_\ell=a$ and $\{a_{\ell+1},\ldots,a_L\}\cap A(I^p_i)=\emptyset$ for all $p\in M_i$}\end{array}\right.\right\},\\
 
 Z^0(a, I^j_i)=\left\{h=\langle a_1,\ldots,a_K\rangle\in Z\left|\begin{array}{l}
 \text{$a_\ell=a$ for some $1\le \ell\le K$ and}\\
 \text{$\{a_{\ell+1},\ldots,a_K\}\cap A(I^q_i)=\emptyset$ for all $q\in M_i$}\end{array}\right.\right\}. 
\end{array}\]
Let $M(I^j_i)=\mathop{\cup}\limits_{a\in A(I^j_i)}M(a,I^j_i)$. $\{I^q_i|q\in M(a, I^j_i)\}$ consists of all the information sets of player $i$ that closely follow $a\in A(I^j_i)$.
$Z^0(a, I^j_i)$ denotes the set of terminal histories that intersect with the action $a \in A(I^j_i)$, where $a$ is the last action taken by player $i$ along these terminal histories.
To distinguish the behavioral strategies of player $i$ on the information sets following $I^j_i$ from other strategies, for $i\in N$ and $j\in M_i$, let
$\varrho^i_{I^j_i}(\hat\beta,\tilde\beta)=(\varrho^i_{I^j_i}(\hat\beta^p_{I^q_p},\tilde\beta):p\in N,q\in M_p)$,  where  \begin{equation}\label{sethmeqBb}\setlength{\abovedisplayskip}{1.2pt}
\setlength{\belowdisplayskip}{1.2pt}
\varrho^i_{I^j_i}(\hat\beta^p_{I^q_p},\tilde\beta)=\left\{\begin{array}{ll}
\tilde\beta^i_{I^q_i} & \text{if $p=i$ and $h\cap A(I^j_i)\ne\emptyset$ for some $h\in I^q_i$,}\\
\hat\beta^p_{I^q_p} & \text{otherwise.}
\end{array}\right.\end{equation}

\noindent
$(\Leftarrow)$. Suppose that $(\beta^*,\mu^*)\in\mathscr{B}$ meets the property in Theorem~\ref{edsethm0}.  Then there exists a convergent sequence of $\{(\beta^\ell,\mu^\ell),\ell=1,2,\ldots\}$ such that $(\beta^*,\mu^*)=\lim\limits_{\ell\to\infty}(\beta^\ell,\mu^\ell)$, where $\beta^\ell=(\beta^{\ell i}_{I^j_i}(a):i\in N,j\in M_i,a\in A(I^j_i))>0$ and $\mu^\ell=(\mu^{\ell i}_{I^j_i}(h):i\in N, j\in M_i, h\in I^j_i)=\mu(\beta^\ell)=(\mu^i_{I^j_i}(h|\beta^\ell):i\in N,j\in M_i,h\in I^j_i)$ with  $\mu^{\ell i}_{I^j_i}(h)=\mu^i_{I^j_i}(h|\beta^\ell)=\frac{\omega(h|\beta^\ell)}{\omega(I^j_i|\beta^\ell)}$.   We next prove that,  for any $i\in N$ and $j\in M_i$,
\begin{equation}\label{sethmeqAa} \setlength{\abovedisplayskip}{1.2pt}
\setlength{\belowdisplayskip}{1.2pt} u^i(\beta^*,\mu^*|I^j_i)=\max\limits_{\beta^i}u^i(\beta^i,\beta^{*-i},\mu^*|I^j_i).\end{equation}
For $i\in N$ and $j\in M_i$, 
we have \begin{equation}\setlength{\abovedisplayskip}{1.2pt}
\setlength{\belowdisplayskip}{1.2pt}\label{sethmeqA}\begin{array}{rl}
 \max\limits_{\beta^i} u^i(\beta^i,\beta^{*-i},\mu^*|I^j_i) = & \max\limits_{\beta^i} \sum\limits_{a\in A(I^j_i)}\beta^i_{I^j_i}(a)  u^i(a,\beta^{i,-I^j_i},\beta^{*-i},\mu^*|I^j_i)\\
 = & \max\limits_{\beta^i_{I^j_i}} \sum\limits_{a\in A(I^j_i)}\beta^i_{I^j_i}(a) \max\limits_{\beta^i} u^i(a,\beta^{i,-I^j_i},\beta^{*-i},\mu^*|I^j_i),\end{array}\end{equation}
where the last equality comes from perfect recall.
In the following,
we derive through the backward induction the conclusion that,  for any $i\in N$ and $j\in M_i$,   \begin{equation}\setlength{\abovedisplayskip}{1.2pt}
\setlength{\belowdisplayskip}{1.2pt}\label{sethmeq3}\max\limits_{\beta^i_{I^j_i}}\sum\limits_{a\in  A(I^j_i)}\beta^{i}_{I^j_i}(a)\max\limits_{\beta^i}u^i(a,\beta^{i,-I^j_i},\beta^{*-i}, \mu^* |I^j_i)=\sum\limits_{a\in  A(I^j_i)}\beta^{*i}_{I^j_i}(a)u^i(a,\beta^{*-I^j_i}, \mu^* |I^j_i).\end{equation}
{\bf Case (1)}. Consider $i\in N$ and $j\in M_i$ with $M(I^j_i)=\emptyset$. We have
\begin{equation}\setlength{\abovedisplayskip}{1.2pt}
\setlength{\belowdisplayskip}{1.2pt}\label{sethmeq4} \begin{array}{rl}
 & \max\limits_{\beta^i_{I^j_i}}\sum\limits_{a\in A(I^j_i)}\beta^i_{I^j_i}(a)\max\limits_{\beta^i}u^i(a,\beta^{i,-I^j_i},\beta^{*-i},\mu^*|I^j_i)\\
 = & \max\limits_{\beta^i_{I^j_i}}\sum\limits_{a\in A(I^j_i)}\beta^i_{I^j_i}(a)u^i(a,\beta^{*-I^j_i},\mu^*|I^j_i)
 =  \sum\limits_{a\in A(I^j_i)}\beta^{*i}_{I^j_i}(a)u^i(a,\beta^{*-I^j_i},\mu^*|I^j_i),
\end{array}
\end{equation}
where the first equality comes from the fact that $u^i(a,\beta^{i,-I^j_i},\beta^{*-i},\mu^*|I^j_i)$ is independent of $\beta^i$ and the second equality comes from the property in Theorem~\ref{edsethm0}.\newline
{\bf Case (2)}. 
Consider $i\in N$ and $j\in M_i$ with $M(I^j_i)\ne\emptyset$ such that, for any $q\in M(I^j_i)$,  {\small
 \begin{equation}\setlength{\abovedisplayskip}{1.2pt}
\setlength{\belowdisplayskip}{1.2pt}\label{sethmeq5}\max\limits_{\beta^i_{I^q_i}}\sum\limits_{a'\in  A(I^q_i)}\beta^{i}_{I^q_i}(a')\max\limits_{\beta^i}u^i(a',\beta^{i,-I^q_i},\beta^{*-i}, \mu^* |I^q_i)=\sum\limits_{a'\in  A(I^q_i)}\beta^{*i}_{I^q_i}(a')u^i(a',\beta^{*-I^q_i}, \mu^* |I^q_i).\end{equation}}

\noindent
 For $a\in A(I^j_i)$ with $M(a, I^j_i)\ne\emptyset$, we have {\footnotesize
\begin{equation}\setlength{\abovedisplayskip}{1.2pt}
\setlength{\belowdisplayskip}{1.2pt}\label{sethmeq6}\begin{array}{rl}
 & u^i(a,\beta^{i,-I^j_i}, \beta^{\ell,-i},\mu^\ell|I^j_i)=\sum\limits_{a\in h\in Z}u^i(h)\nu^i_{I^j_i}(h|a, \beta^{i,-I^j_i}, \beta^{\ell,-i},\mu^\ell)\\
 
 = &  \sum\limits_{q\in M(a, I^j_i)}\sum\limits_{h\in Z}^{h\cap A(I^q_i)\ne\emptyset}u^i(h)\nu^i_{I^j_i}(h|a, \beta^{i,-I^j_i}, \beta^{\ell,-i},\mu^\ell) + \sum\limits_{h\in Z^0(a, I^j_i)}u^i(h)\nu^i_{I^j_i}(h|a,\beta^{i,-I^j_i}, \beta^{\ell,-i},\mu^\ell)\\
 
  = &  \sum\limits_{q\in M(a, I^j_i)}\frac{1}{\omega(I^j_i|\beta^{\ell})}\sum\limits_{h\in Z}^{h\cap A(I^q_i)\ne\emptyset}u^i(h)\omega(h|a, \varrho^i_{I^j_i}(\beta^{\ell,-I^j_i}, \beta)) + \sum\limits_{h\in Z^0(a, I^j_i)}u^i(h)\nu^i_{I^j_i}(h|a,\beta^{\ell,-I^j_i},\mu^\ell)\\
  
  = & \sum\limits_{q\in M(a, I^j_i)} \frac{\omega(I^q_i|\beta^{\ell,-I^j_i})}{\omega(I^j_i|\beta^\ell)}\sum\limits_{h\in Z}^{h\cap A(I^q_i)\ne\emptyset}u^i(h)\frac{\omega(h|a, \varrho^i_{I^j_i}(\beta^{\ell,-I^j_i}, \beta))}{\omega(I^q_i|\beta^{\ell,-I^j_i})} + \sum\limits_{h\in Z^0(a, I^j_i)}u^i(h)\nu^i_{I^j_i}(h|a,\beta^{\ell,-I^j_i},\mu^\ell)\\

  = & \sum\limits_{q\in M(a, I^j_i)} \frac{\omega(I^q_i|\beta^{\ell,-I^j_i})}{\omega(I^j_i|\beta^\ell)}\sum\limits_{h\in Z}^{h\cap A(I^q_i)\ne\emptyset}u^i(h)\nu^i_{I^q_i}(h|\beta^i,\beta^{\ell,-i},\mu^\ell)+ \sum\limits_{h\in Z^0(a, I^j_i)}u^i(h)\nu^i_{I^j_i}(h|a,\beta^{\ell,-I^j_i},\mu^\ell)\\

  = & \sum\limits_{q\in M(a, I^j_i)} \frac{\omega(I^q_i|\beta^{\ell,-I^j_i})}{\omega(I^j_i|\beta^\ell)}\sum\limits_{a'\in  A(I^q_i)}\beta^{i}_{I^q_i}(a')u^i(a',\beta^{i,-I^q_i},\beta^{\ell,-i}, \mu^\ell |I^q_i)+ \sum\limits_{h\in Z^0(a, I^j_i)}u^i(h)\nu^i_{I^j_i}(h|a,\beta^{\ell,-I^j_i},\mu^\ell),
 \end{array}\end{equation}}where the third equality comes from Eq.~(\ref{seedeqB}), Eq.~(\ref{sethmeqBb}), and the fact that $\nu^i_{I^j_i}(h|a,\beta^{i,-I^j_i}, \beta^{\ell,-i},\mu^\ell)$ is independent of $\beta^i$ for $h\in Z^0(a, I^j_i)$.
 Furthermore, for $q\in M(a, I^j_i)$, \begin{equation}
\label{sethmeqC}\setlength{\abovedisplayskip}{1.2pt}
\setlength{\belowdisplayskip}{1.2pt}
\frac{\omega(I^q_i|\beta^{\ell,-I^j_i})}{ \omega(I^j_i|\beta^\ell)}=\sum\limits^{\hat h =\langle a_1,\ldots,a_k\rangle\in I^j_i}_{h=\langle\hat h, a_{k+1},\ldots,a_L\rangle
\in I^q_i,\; a_{k+1}=a}\mu^{\ell i}_{I^j_i}(\hat h)\prod\limits_{g=k+1}^{L-1}\beta^{\ell, P(\langle a_1,\ldots,a_{g}\rangle)}_{\langle a_1,\ldots,a_{g}\rangle}(a_{g+1}).
\end{equation} 
Therefore, as a result of substituting Eq.~(\ref{sethmeqC}) into Eq.~(\ref{sethmeq6}), it holds that
\begin{equation}\setlength{\abovedisplayskip}{1.2pt}
\setlength{\belowdisplayskip}{1.2pt}\label{sethmeq8}\begin{array}{rl} 
& \max\limits_{\beta^i}u^i(a,\beta^{i,-I^j_i},\beta^{*-i},\mu^*|I^j_i)
=  \max\limits_{\beta^i} \lim\limits_{\ell\to\infty}  u^i(a,\beta^{i,-I^j_i},\beta^{\ell,-i},\mu^\ell|I^j_i) \\

= & \max\limits_{\beta^i} \lim\limits_{\ell\to\infty}\sum\limits_{q\in M(a, I^j_i)} \sum\limits^{\hat h =\langle a_1,\ldots,a_k\rangle\in I^j_i}_{h=\langle\hat h, a_{k+1},\ldots,a_L\rangle
\in I^q_i,\;a_{k+1}=a}\mu^{\ell i}_{I^j_i}(\hat h)\prod\limits_{g=k+1}^{L-1}\beta^{\ell, P(\langle a_1,\ldots,a_{g}\rangle)}_{\langle a_1,\ldots,a_{g}\rangle}(a_{g+1})\\
& \sum\limits_{a'\in  A(I^q_i)}\beta^{i}_{I^q_i}(a')u^i(a',\beta^{i,-I^q_i},\beta^{\ell,-i}, \mu^\ell |I^q_i) + \sum\limits_{h\in Z^0(a, I^j_i)}u^i(h)\nu^i_{I^j_i}(h|a,\beta^{\ell,-I^j_i},\mu^\ell) \\

= & \sum\limits_{q\in M(a, I^j_i)}\sum\limits^{\hat h =\langle a_1,\ldots,a_k\rangle\in I^j_i}_{h=\langle\hat h, a_{k+1},\ldots,a_L\rangle
\in I^q_i,\;a_{k+1}=a}\mu^{* i}_{I^j_i}(\hat h)\prod\limits_{g=k+1}^{L-1}\beta^{*, P(\langle a_1,\ldots,a_{g}\rangle)}_{\langle a_1,\ldots,a_{g}\rangle}(a_{g+1})\\

&\max\limits_{\beta^i_{I^q_i}} \sum\limits_{a'\in  A(I^q_i)}\beta^{i}_{I^q_i}(a')\max\limits_{\beta^i}u^i(a',\beta^{i,-I^q_i},\beta^{*-i}, \mu^* |I^q_i)
 + \sum\limits_{h\in Z^0(a, I^j_i)}u^i(h)\nu^i_{I^q_i}(h|a,\beta^{*-I^j_i},\mu^*)\\

= & \sum\limits_{q\in M(a, I^j_i)} \sum\limits^{\hat h =\langle a_1,\ldots,a_k\rangle\in I^j_i}_{h=\langle\hat h, a_{k+1},\ldots,a_L\rangle
\in I^q_i,\;a_{k+1}=a}\mu^{* i}_{I^j_i}(\hat h)\prod\limits_{g=k+1}^{L-1}\beta^{*, P(\langle a_1,\ldots,a_{g}\rangle)}_{\langle a_1,\ldots,a_{g}\rangle}(a_{g+1})\\
& \sum\limits_{a'\in  A(I^q_i)}\beta^{*i}_{I^q_i}(a')u^i(a',\beta^{*-I^q_i}, \mu^* |I^q_i) + \sum\limits_{h\in Z^0(a, I^j_i)}u^i(h)\nu^i_{I^j_i}(h|a,\beta^{*,-I^j_i},\mu^*),
\end{array}\end{equation}
where the second equality comes from Eq.~(\ref{sethmeq6}) and Eq.~(\ref{sethmeqC}) and the last equality comes from Eq.~(\ref{sethmeq5}).
Let
\begin{equation}\setlength{\abovedisplayskip}{1.2pt}
\setlength{\belowdisplayskip}{1.2pt}\label{sethmeqD}\begin{array}{rl} 
w(\beta,\mu)= & \sum\limits_{q\in M(a, I^j_i)} \sum\limits^{\hat h =\langle a_1,\ldots,a_k\rangle\in I^j_i}_{h=\langle\hat h, a_{k+1},\ldots,a_L\rangle
\in I^q_i,\;a_{k+1}=a}\mu^i_{I^j_i}(\hat h)\prod\limits_{g=k+1}^{L-1}\beta^{P(\langle a_1,\ldots,a_{g}\rangle)}_{\langle a_1,\ldots,a_{g}\rangle}(a_{g+1})\\
& \sum\limits_{a'\in  A(I^q_i)}\beta^i_{I^q_i}(a')u^i(a',\beta^{-I^q_i}, \mu|I^q_i) + \sum\limits_{h\in Z^0(a, I^j_i)}u^i(h)\nu^i_{I^j_i}(h|a,\beta^{-I^j_i},\mu),
\end{array}\end{equation}
which is a continuous function of $(\beta,\mu)$.
Then, as a result of Eq.~(\ref{sethmeq8}), we have \begin{equation}\label{Sqrsethm2} \setlength{\abovedisplayskip}{1.2pt}
\setlength{\belowdisplayskip}{1.2pt}w(\beta^*,\mu^*)=\max\limits_{\beta^i}u^i(a,\beta^{i,-I^j_i},\beta^{*-i},\mu^*|I^j_i).\end{equation} Thus,
due to the continuity of $w(\beta, \mu)$ and $\lim\limits_{\ell\to\infty}(\beta^\ell,\mu^\ell)=(\beta^*,\mu^*)$, it follows that \begin{equation} \setlength{\abovedisplayskip}{1.2pt}
\setlength{\belowdisplayskip}{1.2pt}\label{Ssethm1}\lim\limits_{\ell\to\infty}w(\beta^\ell,\mu^\ell)=w(\beta^*,\mu^*)=\max\limits_{\beta^i}u^i(a,\beta^{i,-I^j_i},\beta^{*-i},\mu^*|I^j_i).\end{equation} Moreover,
\begin{equation}\setlength{\abovedisplayskip}{1.2pt}
\setlength{\belowdisplayskip}{1.2pt}\label{sethmeq9}\begin{array}{rl} 
w(\beta^\ell,\mu^\ell)= & \sum\limits_{q\in M(a, I^j_i)} \sum\limits^{\hat h =\langle a_1,\ldots,a_k\rangle\in I^j_i}_{h=\langle\hat h, a_{k+1},\ldots,a_L\rangle
\in I^q_i,\;a_{k+1}=a}\mu^{\ell i}_{I^j_i}(\hat h)\prod\limits_{g=k+1}^{L-1}\beta^{\ell, P(\langle a_1,\ldots,a_{g}\rangle)}_{\langle a_1,\ldots,a_{g}\rangle}(a_{g+1})\\
& \sum\limits_{a'\in  A(I^q_i)}\beta^{\ell i}_{I^q_i}(a')u^i(a',\beta^{\ell,-I^q_i}, \mu^\ell |I^q_i)+ \sum\limits_{h\in Z^0(a, I^j_i)}u^i(h)\nu^i_{I^j_i}(h|a,\beta^{\ell,-I^j_i},\mu^\ell)\\
= &  \sum\limits_{q\in M(a, I^j_i)} \frac{\omega(I^q_i|\beta^{\ell,-I^j_i})}{\omega(I^j_i|\beta^\ell)}\sum\limits_{a'\in  A(I^q_i)}\beta^{\ell i}_{I^q_i}(a')u^i(a',\beta^{\ell,-I^q_i}, \mu^\ell |I^q_i)\\
& + \sum\limits_{h\in Z^0(a, I^j_i)}u^i(h)\nu^i_{I^j_i}(h|a,\beta^{\ell,-I^j_i},\mu^\ell)\\
 = & u^i(a,\beta^{\ell,-I^j_i},\mu^\ell|I^j_i).
\end{array} \end{equation}
 Eq.~(\ref{sethmeq9}) and Eq.~(\ref{Ssethm1}) together bring us the conclusion that
\[\setlength{\abovedisplayskip}{1.2pt}
\setlength{\belowdisplayskip}{1.2pt}\begin{array}{rl} & \lim\limits_{\ell\to\infty}w(\beta^\ell,\mu^\ell)=\lim\limits_{\ell\to\infty}u^i(a,\beta^{\ell,-I^j_i},\mu^\ell|I^j_i)\\
= & u^i(a,\beta^{*,-I^j_i},\mu^*|I^j_i)
= w(\beta^*,\mu^*)=\max\limits_{\beta^i}u^i(a,\beta^{i,-I^j_i},\beta^{*-i},\mu^*|I^j_i).\end{array}\]
This result together with the property in Theorem~\ref{edsethm0} implies that
\begin{equation}\setlength{\abovedisplayskip}{1.2pt}
\setlength{\belowdisplayskip}{1.2pt}\label{sethmeq10}
\begin{array}{rl}
 & \max\limits_{\beta^i_{I^j_i}}\sum\limits_{a\in  A(I^j_i)}\beta^{i}_{I^j_i}(a)\max\limits_{\beta^i}u^i(a,\beta^{i,-I^j_i},\beta^{*-i}, \mu^* |I^j_i)\\
 = & \max\limits_{\beta^i_{I^j_i}}
\sum\limits_{a\in  A(I^j_i)}\beta^i_{I^j_i}(a)u^i(a,\beta^{*-I^j_i}, \mu^* |I^j_i)
= \sum\limits_{a\in  A(I^j_i)}\beta^{*i}_{I^j_i}(a)u^i(a,\beta^{*-I^j_i}, \mu^* |I^j_i),\end{array}\end{equation}
which is exactly the desired conclusion in Eq.~(\ref{sethmeq3}). Therefore it follows from Eq.~(\ref{sethmeqA}) that 
\begin{equation}\setlength{\abovedisplayskip}{1.2pt}
\setlength{\belowdisplayskip}{1.2pt}\label{sethmeqB}\max\limits_{\beta^i} u^i(\beta^i,\beta^{*-i},\mu^*|I^j_i)= \sum\limits_{a\in  A(I^j_i)}\beta^{*i}_{I^j_i}(a)u^i(a,\beta^{*-I^j_i}, \mu^* |I^j_i)=u^i(\beta^*,\mu^*|I^j_i).\end{equation}
Hence, $(\beta^*,\mu^*)$ is a sequential equilibrium according to Definition~\ref{sed1}.

\noindent $(\Rightarrow)$. Let $(\beta^*,\mu^*)\in\mathscr{B}$ be a sequential equilibrium according to Definition~\ref{sed1}. Then, $u^i(\beta^*,\mu^*|I^j_i)=\max\limits_{\beta^i}u^i(\beta^i,\beta^{*-i},\mu^*|I^j_i)$ for any $i\in N$ and $j\in M_i$. Thus, $u^i(\beta^*,\mu^*|I^j_i)=u^i(\beta^{*i}_{I^j_i},\beta^{*-I^j_i},\mu^*|I^j_i)
\le \max\limits_{\beta^i_{I^j_i}}u^i(\beta^i_{I^j_i},\beta^{*i,-I^j_i},\beta^{*-i},\mu^*|I^j_i)\le\max\limits_{\beta^i}u^i(\beta^i,\beta^{*-i},\mu^*|I^j_i)
=u^i(\beta^*,\mu^*|I^j_i)$. Therefore, $u^i(\beta^*,\mu^*|I^j_i)=\max\limits_{\beta^i_{I^j_i}}u^i(\beta^i_{I^j_i},\beta^{*-I^j_i},\mu^*|I^j_i)$. The proof of Theorem~\ref{edsethm0} is completed.
\end{proof}

Given any convergent sequence of $\{(\beta^k,\mu^k),k=1,2,\ldots\}$ with $(\beta^*,\mu^*)=\lim\limits_{k\to\infty}(\beta^k,\mu^k)$, where $(\beta^k,\mu^k)$ is totally mixed for all $k$, we next apply Theorem~\ref{edsethm0} to establish a polynomial system as a necessary and sufficient condition for determining whether $(\beta^*,\mu^*)$ is a sequential equilibrium. 
 
\begin{theorem} \label{nscsethm1} {\em $(\beta^*,\mu^*)\in\mathscr{B}$ is a sequential equilibrium if and only if there exists a pair of $(\lambda^*,\zeta^*)$ together with $(\beta^*,\mu^*)$ satisfying the system,
{\small \begin{equation}\label{sensc1}\setlength{\abovedisplayskip}{1.2pt}
\setlength{\belowdisplayskip}{1.2pt}
 \begin{array}{l}
 u^i(a, \beta^{-I^j_i},\mu|I^j_i)+\lambda^i_{I^j_i}(a)-\zeta^i_{I^j_i}=0,\;i\in N,j\in M_i,a\in A(I^j_i),\\
\omega(I^j_i|\beta)\mu^i_{I^j_i}(h)=\omega(h|\beta),\;i\in N,j\in M_i,h\in I^j_i,\\
 \sum\limits_{a\in A(I^j_i)}\beta^i_{I^j_i}(a)=1,\;i\in N,j\in M_i,\\
 \beta^i_{I^j_i}(a)\lambda^i_{I^j_i}(a)=0,\;0\le\beta^i_{I^j_i}(a),\;0\le\lambda^i_{I^j_i}(a),\;i\in N,j\in M_i,a\in A(I^j_i).
 \end{array}
 \end{equation}}
 }
 \end{theorem}
 \begin{proof} $(\Rightarrow)$. Suppose that $(\beta^*,\mu^*)\in\mathscr{B}$ is a sequential equilibrium.  Let $\zeta^*=(\zeta^{*i}_{I^j_i}:i\in N,j\in M_i)$ with $\zeta^{*i}_{I^j_i}=\max\limits_{a\in A(I^j_i)}u^i(a,\beta^{*-I^j_i},\mu^*|I^j_i)$ and $\lambda^{*}=(\lambda^{*i}_{I^j_i}(a): i\in N,j\in M_i,a\in A(I^j_i))$ with $\lambda^{*i}_{I^j_i}(a)=\zeta^{*i}_{I^j_i}-u^i(a,\beta^{*-I^j_i},\mu^*|I^j_i)$. 
 Clearly, $\lambda^{*i}_{I^j_i}(a)\ge 0$. We denote by $a^*\in A(I^j_i)$ an action with $u^i(a^*,\beta^{*-I^j_i},\mu^*|I^j_i)=\max\limits_{a\in A(I^j_i)}u^i(a,\beta^{*-I^j_i},\mu^*|I^j_i)$.
 Let $a^0\in A(I^j_i)$ be an action with $\lambda^{*i}_{I^j_i}(a^0)>0$. Then, $u^i(a^*,\beta^{*-I^j_i},\mu^*|I^j_i)-u^i(a^0,\beta^{*-I^j_i},\mu^*|I^j_i)=\lambda^{*i}_{I^j_i}(a^0)>0$. Thus it follows from  Theorem~\ref{edsethm0} that $\beta^{*i}_{I^j_i}(a^0)=0$. 
 Therefore, $0\le\lambda^{*i}_{I^j_i}(a)$ and $\beta^{*i}_{I^j_i}(a)\lambda^{*i}_{I^j_i}(a)=0$ for all $i\in N$, $j\in M_i$, and $a\in A(I^j_i)$, and
 $(\beta^*,\mu^*,\lambda^*,\zeta^*)$ satisfies the system~(\ref{sensc1}).
 
$(\Leftarrow)$. Suppose that  $(\beta^*,\mu^*)\in\mathscr{B}$ and $(\beta^*,\mu^*,\lambda^*,\zeta^*)$ satisfies the system~(\ref{sensc1}). {\bf (a)}. Assume that $\lambda^{*i}_{I^j_i}(a)=0$ for all $i\in N$, $j\in M_i$, and $a\in A(I^j_i)$.  Then, $u^i(a', \beta^{*-I^j_i},\mu^*|I^j_i)=u^i(a'', \beta^{*-I^j_i},\mu^*|I^j_i)$ for any $a',a''\in A(I^j_i)$.  Thus, $(\beta^*,\mu^*)$ is a sequential equilibrium in accordance with Theorem~\ref{edsethm0}.
{\bf (b)}. Assume that there exists some $a^0\in A(I^j_i)$ such that 
 $\lambda^{*i}_{I^{j}_{i}}(a^0)>0$. Then there exists $a^*\in A(I^j_i)$ such that $u^i(a^*,\beta^{*-I^j_i},\mu^*|I^j_i)-u^i(a^0,\beta^{*-I^j_i},\mu^*|I^j_i)=\lambda^{*i}_{I^j_i}(a^0)>0$. Furthermore,  $\beta^{*i}_{I^{j}_{i}}(a^0)=0$ because of $\beta^{*i}_{I^{j}_{i}}(a^0)\lambda^{*i}_{I^{j}_{i}}(a^0)=0$ according to the system~(\ref{sensc1}).  
 Thus it follows from
 Theorem~\ref{edsethm0} that $(\beta^*,\mu^*)$ is a sequential equilibrium. This completes the proof.
 \end{proof}

As a corollary of Theorem~\ref{nscsethm1}, we come to the following conclusion.
\begin{corollary}\label{nscseco0} {\em $(\beta^*,\mu^*)\in\mathscr{B}$ is a sequential equilibrium if and only if $(\beta^*,\mu^*)$ satisfies the system,
{\small \begin{equation}\label{sensc2}\setlength{\abovedisplayskip}{1.2pt}
\setlength{\belowdisplayskip}{1.2pt}
 \begin{array}{l}
 \beta^i_{I^j_i}(a)(\sum\limits_{a'\in A(I^j_i)}\beta^i_{I^j_i}(a')u^i(a', \beta^{-I^j_i},\mu|I^j_i)-u^i(a, \beta^{-I^j_i},\mu|I^j_i))=0,\;i\in N,j\in M_i,a\in A(I^j_i),\\
  \omega(I^j_i|\beta)\mu^i_{I^j_i}(h)=\omega(h|\beta),\;i\in N,j\in M_i,h\in I^j_i,\\
 \sum\limits_{a\in A(I^j_i)}\beta^i_{I^j_i}(a)=1,\;i\in N,j\in M_i,\\
 0\le\sum\limits_{a'\in A(I^j_i)}\beta^i_{I^j_i}(a')u^i(a', \beta^{-I^j_i},\mu|I^j_i)-u^i(a, \beta^{-I^j_i},\mu|I^j_i),\;0\le\beta^i_{I^j_i}(a),\;i\in N,j\in M_i,a\in A(I^j_i).
 \end{array}
 \end{equation}}}
 \end{corollary}
 \begin{proof} Multiplying $\beta^i_{I^j_i}(a)$ to equations in the first group of the system~(\ref{sensc1}) and taking the sum over $a\in A(I^j_i)$, we obtain  
 \begin{equation}\label{nsces0}\setlength{\abovedisplayskip}{1.2pt}
 \setlength{\belowdisplayskip}{1.2pt}
 \zeta^i_{I^j_i}=\sum\limits_{a'\in A(I^j_i)}\beta^i_j(a')u^i(a', \beta^{-I^j_i},\mu|I^j_i),\;i\in N,j\in M_i.\end{equation}
 Replacing  $\zeta^i_{I^j_i}$ in the system~(\ref{sensc1}) by Eq.~(\ref{nsces0}), we come to an equivalent system,
 \begin{equation}\label{nsces00}\setlength{\abovedisplayskip}{1.2pt}
 \setlength{\belowdisplayskip}{1.2pt}
  \begin{array}{l}
  \lambda^i_{I^j_i}(a)=\sum\limits_{a'\in A(I^j_i)}\beta^i_j(a')u^i(a', \beta^{-I^j_i},\mu|I^j_i)-u^i(a, \beta^{-I^j_i},\mu|I^j_i),\;i\in N,j\in M_i,a\in A(I^j_i),\\
  \sum\limits_{a\in A(I^j_i)}\beta^i_{I^j_i}(a)=1,\;i\in N,j\in M_i,\\
  \beta^i_{I^j_i}(a)\lambda^i_{I^j_i}(a)=0,\;0\le\beta^i_{I^j_i}(a),\;0\le\lambda^i_{I^j_i}(a),\;i\in N,j\in M_i,a\in A(I^j_i).
  \end{array}
  \end{equation}
  Substituting $\lambda^i_{I^j_i}(a)$ in the first group of equations into the third group of equations in the system~(\ref{nsces00}), we arrive at the system~(\ref{sensc2}). The proof is completed.
  \end{proof}

Theorem~\ref{edsethm0} explains how to determine whether a consistent assessment constitutes a sequential equilibrium but does not provide guidance on constructing such a consistent assessment. The construction of a consistent assessment for a sequential equilibrium has remained a challenge. To fully resolve this problem, we introduce the notion of $\varepsilon$-perfect $\gamma$-sequential equilibrium, which is defined as follows.
\begin{definition}[\bf An Equivalent Definition of Sequential Equilibrium through $\varepsilon$-Perfect $\gamma$-Sequential Equilibrium with Local Sequential Rationality] \label{esedA} {\em For any given  $0<\gamma$ and $0<\varepsilon$, a totally mixed assessment  $(\beta(\gamma,\varepsilon),\mu(\gamma,\varepsilon))$ is  an $\varepsilon$-perfect $\gamma$-sequential equilibrium if $\beta^i_{I^j_i}(\gamma,\varepsilon; a')\le\varepsilon$ for any $i\in N$, $j\in M_i$, and $a',a''\in A(I^j_i)$ with
\begin{equation}\label{esedAeqAA}\setlength{\abovedisplayskip}{1.2pt}
\setlength{\belowdisplayskip}{1.2pt} u^i(a'',\beta^{-I^j_i}(\gamma,\varepsilon), \mu(\gamma,\varepsilon)|I^j_i)-u^i(a',\beta^{-I^j_i}(\gamma,\varepsilon), \mu(\gamma,\varepsilon)|I^j_i)>\gamma,\end{equation}
where $\mu(\gamma, \varepsilon)=(\mu^i_{I^j_i}(\gamma, \varepsilon;h):i\in N,j\in M_i,h\in I^j_i)$ with $\mu^i_{I^j_i}(\gamma, \varepsilon;h)=\frac{\omega(h|\beta(\gamma, \varepsilon))}{\omega(I^j_i|\beta(\gamma, \varepsilon))}$ for $h\in I^j_i$. 
For any given $\gamma>0$, $(\beta(\gamma),\mu(\gamma))$  is a perfect $\gamma$-sequential equilibrium if $(\beta(\gamma),\mu(\gamma))$ is a limit point of a sequence $\{(\beta(\gamma,\varepsilon_k), \mu(\gamma,\varepsilon_k)), k=1,2,\ldots\}$ with $\varepsilon_k>0$ and $\lim\limits_{k\to\infty}\varepsilon_k=0$, where $(\beta(\gamma,\varepsilon_k), \mu(\gamma,\varepsilon_k))$ is an $\varepsilon_k$-perfect $\gamma$-sequential equilibrium for every $k$. 
$(\beta^*,\mu^*)$ is a sequential equilibrium if it is a limit point of a sequence $\{(\beta(\gamma_q),\mu(\gamma_q)),q=1,2,\ldots\}$ with $\gamma_q>0$ and $\lim\limits_{q\to\infty}\gamma_q=0$, where $(\beta(\gamma_q),\mu(\gamma_q))$ is a perfect $\gamma_q$-sequential equilibrium for every $q$. 
}
\end{definition}
\begin{theorem}{\em Definition~\ref{esedA} and Definition~\ref{sed1} of sequential equilibrium are equivalent.
}
\end{theorem}
\begin{proof} $(\Rightarrow)$. We denote by $(\beta^*,\mu^*)$ a sequential equilibrium according to Definition~\ref{esedA}. Then there exists a convergent sequence $\{(\beta^*(\gamma_q),\mu^*(\gamma_q)),q=1,2,\ldots\}$ with $\gamma_q>0$ and $\lim\limits_{q\to\infty}\gamma_q=0$ such that  $(\beta^*,\mu^*)=\lim\limits_{q\to\infty}(\beta^*(\gamma_q),\mu^*(\gamma_q))$, where $(\beta^*(\gamma_q),\mu^*(\gamma_q))$ is the limit point of a convergent sequence $\{(\beta(\gamma_q,\varepsilon_k), \mu(\gamma_q,\varepsilon_k)), k=1,2,\ldots\}$ with $\varepsilon_k>0$ and $\lim\limits_{k\to\infty}\varepsilon_k=0$ satisfying that $(\beta(\gamma_q,\varepsilon_k), \mu(\gamma_q,\varepsilon_k))$ is an $\varepsilon_k$-perfect $\gamma_q$-sequential equilibrium for every $k$. 
Let $\{\delta_q>0,q=1,2,\ldots\}$ be a convergent sequence with $\lim\limits_{q\to\infty}\delta_q=0$. For each $q$, we choose $(\beta^*(\gamma_q, \varepsilon_{k_q}),\mu^*(\gamma_q, \varepsilon_{k_q}))$ such that $\|(\beta^*(\gamma_q, \varepsilon_{k_q}),\mu^*(\gamma_q, \varepsilon_{k_q}))-(\beta^*(\gamma_q),\mu^*(\gamma_q))\|<\delta_q$. Then, $\{(\beta^*(\gamma_q, \varepsilon_{k_q}),\mu^*(\gamma_q, \varepsilon_{k_q})),q=1,2,\ldots\}$ is a convergent sequence with $\lim\limits_{q\to\infty}(\beta^*(\gamma_q, \varepsilon_{k_q}),\mu^*(\gamma_q, \varepsilon_{k_q}))=(\beta^*,\mu^*)$.  For some $i\in N$ and $j\in M_i$,
suppose that there exist $a',a''\in A(I^j_i)$ such that  $\beta^{*i}_{I^j_i}(a')>0$ and $r_0=u^i(a'', \beta^{*-I^j_i},\mu^*|I^j_i)-u^i(a', \beta^{*-I^j_i},\mu^*|I^j_i)>0$. Then, as a result of the continuity of $u^i(a, \beta^{-I^j_i},\mu|I^j_i)$, there exists a sufficiently large $Q_0$ such that 
 $u^i(a'', \beta^{*-I^j_i}(\gamma_q, \varepsilon_{k_q}),\mu^*(\gamma_q$, $\varepsilon_{k_q})|I^j_i)-u^i(a', \beta^{*-I^j_i}(\gamma_q, \varepsilon_{k_q}),\mu^*(\gamma_q, \varepsilon_{k_q})|I^j_i)\ge r_0/2$ for all $q\ge Q_0$. Since $\lim\limits_{q\to\infty}\gamma_q=0$, there exists a sufficiently large $Q_1\ge Q_0$ such that $\gamma_q<r_0/2$ for all $q\ge Q_1$. This implies $\beta^i_{I^j_i}(\gamma_q, \varepsilon_{k_q};a')\le\varepsilon_{k_q}$ for all $q\ge Q_1$. Thus, $0<\beta^{*i}_{I^j_i}(a')=\lim\limits_{q\to\infty}\beta^i_{I^j_i}(\gamma_q, \varepsilon_{k_q};a')=0$. A contradiction occurs. 
Therefore, $(\beta^*,\mu^*)$ is a sequential equilibrium according to Theorem~\ref{edsethm0}.

$(\Rightarrow)$. We denote by $(\beta^*,\mu^*)$ a sequential equilibrium according to Definition~\ref{sed1}. Then, as a result of the consistency, there exists a convergent sequence, $\{(\beta^\ell,\mu^\ell),\ell=1,2,\ldots\}$, such that $\beta^*=\lim\limits_{\ell\to\infty}\beta^\ell$ and $\mu^*=\lim\limits_{\ell\to\infty}\mu^\ell$, where $\beta^\ell=(\beta^{\ell i}_{I^j_i}(a):i\in N,j\in M_i,a\in A(I^j_i))>0$ and $\mu^\ell=(\mu^{\ell i}_{I^j_i}(h):i\in N, j\in M_i, h\in I^j_i)=\mu(\beta^\ell)=(\mu^i_{I^j_i}(h|\beta^\ell):i\in N,j\in M_i,h\in I^j_i)$ with  $\mu^{\ell i}_{I^j_i}(h)=\mu^i_{I^j_i}(h|\beta^\ell)=\frac{\omega(h|\beta^\ell)}{\omega(I^j_i|\beta^\ell)}$.  For $i\in N$ and $j\in M_i$, let $A^0(I^j_i)=\{a\in A(I^j_i)|\beta^{*i}_{I^j_i}(a)=0\}$ and $A^+(I^j_i)=\{a\in A(I^j_i)|\max\limits_{\tilde a\in A(I^j_i)}u^i(\tilde a,\beta^{*-I^j_i},\mu^*|I^j_i)-u^i(a, \beta^{*-I^j_i},\mu^*|I^j_i)>0\}$. When $\mathop{\cup}\limits_{i\in N,\;j\in M_i}A^+(I^j_i)=\emptyset$, we have $\beta^*>0$ and consequently, $(\beta^*,\mu^*)$ is a sequential equilibrium according to Definition~\ref{esedA}. We assume that $\mathop{\cup}\limits_{i\in N,\;j\in M_i}A^+(I^j_i)\ne\emptyset$.
Let $
\gamma_0=
 \frac{1}{2}\min\limits_{i\in N,\;j\in M_i}\min\limits_{a\in A^+(I^j_i)}\max\limits_{\tilde a\in A(I^j_i)}u^i(\tilde a,\beta^{*-I^j_i},\mu^*|I^j_i)-u^i(a, \beta^{*-I^j_i},\mu^*|I^j_i)$.
We take $\{\varepsilon_\ell,\ell=1,2,\ldots\}$ to be a convergent sequence with $\varepsilon_\ell\ge\max\limits_{i\in N,\;j\in M_i}\max\limits_{a\in A^0(I^j_i)}\beta^{\ell i}_{I^j_i}(a)$ and $\lim\limits_{\ell\to\infty}\varepsilon_\ell=0$. 
Let $\beta(\gamma_0,\varepsilon_\ell)=(\beta^i_{I^j_i}(\gamma_0,\varepsilon_\ell; a): i\in N,j\in M_i,a\in A(I^j_i))$ with $\beta^i_{I^j_i}(\gamma_0,\varepsilon_\ell)=\beta^{\ell i}_{I^j_i}$ and $\mu(\gamma_0,\varepsilon_\ell)=(\mu^i_{I^j_i}(\gamma_0,\varepsilon_\ell; h): i\in N,j\in M_i,h\in I^j_i)$ with $\mu^i_{I^j_i}(\gamma_0,\varepsilon_\ell; h)=\mu^{\ell i}_{I^j_i}(h)$. We know from Theorem~\ref{edsethm0} that $\beta^{*i}_{I^j_i}(a')=0$ for any $i\in N$, $j\in M_i$ and $a',a''\in A(I^j_i)$ with $u^i(a'',\beta^{*-I^j_i},\mu^*|I^j_i)-u^i(a',\beta^{*-I^j_i},\mu^*|I^j_i)>0$. Thus it follows from the continuity of $u^i(a,\beta^{-I^j_i},\mu|I^j_i)$   that there exists a sufficiently large $L_0$ such that, for any $\ell\ge L_0$, $(\beta(\gamma_0,\varepsilon_\ell),\mu(\gamma_0,\varepsilon_\ell))$ is an $\varepsilon_\ell$-perfect $\gamma_0$-sequential equilibrium  according to Definition~\ref{esedA}. Therefore, $(\beta^*,\mu^*)$ is a sequential equilibrium according to Definition~\ref{esedA}. The proof is completed.
\end{proof}

To prove the existence of a sequential equilibrium, Kreps and Wilson~\cite{Kreps and Wilson (1982)} had to borrow the existence of perfect equilibrium from Selten~\cite{Selten (1975)}. As a direct application of Definition~\ref{esedA}, we secure a proof of the existence of a sequential equilibrium in the following theorem. 

\begin{theorem} {\em Every finite extensive-form game with perfect recall always has a sequential equilibrium.}
\end{theorem}
\begin{proof}  Let $\{\gamma_q>0, q=1,2,\ldots\}$ be a convergent sequence with  $\lim\limits_{q\to\infty}\gamma_q=0$ and $\{\varepsilon_k>0,k=1,2,\ldots\}$ a convergent sequence with $\lim\limits_{k\to\infty}\varepsilon_k=0$. We denote $m_0=\sum\limits_{i\in N}\sum\limits_{j\in M_i}|A(I^j_i)|$.
Let $\delta(\varepsilon_k)=\frac{\varepsilon^2_k}{m_0}$ and $\triangle(\varepsilon_k)=\mathop{\times}\limits_{i\in N,\;j\in M_i}\triangle^i_{I^j_i}(\varepsilon_k)$, where $\triangle^i_{I^j_i}(\varepsilon_k)=\{\beta^i_{I^j_i}(\varepsilon_k)\in\triangle^i_{I^j_i}|\beta^i_{I^j_i}(\varepsilon_k;a)\ge\delta(\varepsilon_k),a\in A(I^j_i)\}$. For $i\in N$ and $j\in M_i$, we define a point-to-set mapping $F^i_{I^j_i}(\cdot | \gamma_q):\triangle(\varepsilon_k)\to\triangle^i_{I^j_i}(\varepsilon_k)$ by {\small
\[\setlength{\abovedisplayskip}{1.2pt}
\setlength{\belowdisplayskip}{1.2pt}
F^i_{I^j_i}(\beta(\varepsilon_k)|\gamma_q)=\left\{\beta^{*i}_{I^j_i}(\varepsilon_k)\in\triangle^i_{I^j_i}(\varepsilon_k)\left|\begin{array}{ll}
\text{$\beta^{*i}_{I^j_i}(\varepsilon_k; a')\le\varepsilon_k$ for any $a',a''\in A(I^j_i)$ with}\\

\text{$u^i(a'',\beta^{-I^j_i}(\varepsilon_k),\mu(\varepsilon_k)|I^j_i)-u^i(a',\beta^{-I^j_i}(\varepsilon_k),\mu(\varepsilon_k)|I^j_i)> \gamma_q$}
\end{array}\right.
\right\},\]}where $\mu(\varepsilon_k)=(\mu^i_{I^j_i}(\varepsilon_k; h):i\in N,j\in M_i,h\in I^j_i)$ with $\mu^i_{I^j_i}(\varepsilon_k; h)=\mu^i_{I^j_i}(h|\beta(\varepsilon_k))=\frac{\omega(h|\beta(\varepsilon_k))}{\omega(I^j_i|\beta(\varepsilon_k))}$.
Clearly, for any $\beta(\varepsilon_k)\in\triangle(\varepsilon_k)$, the points in $F^i_{I^j_i}(\beta(\varepsilon_k)|\gamma_q)$ satisfy a finite number of linear inequalities. Thus, $F^i_{I^j_i}(\beta(\varepsilon_k)|\gamma_q)$ is a nonempty convex and compact set. The continuity of $u^i(a,\beta^{-I^j_i}(\varepsilon_k),\mu(\varepsilon_k)|I^j_i)$ on $\triangle(\varepsilon_k)$ implies that $F^i_{I^j_i}(\beta(\varepsilon_k)|\gamma_q)$ is upper-semicontinuous on $\triangle(\varepsilon_k)$. Let $F(\beta(\varepsilon_k)|\gamma_q)=\mathop{\times}\limits_{i\in N,\;j\in M_i}F^i_{I^j_i}(\beta(\varepsilon_k)|\gamma_q)$. Then, $F(\cdot|\gamma_q):\triangle(\varepsilon_k)\to\triangle(\varepsilon_k)$ meets all the requirements of Kakutani's fixed point theorem. Thus there exists some $\beta^*(\gamma_q, \varepsilon_k)\in \triangle(\varepsilon_k)$ such that $\beta^*(\gamma_q, \varepsilon_k)\in F(\beta^*(\gamma_q, \varepsilon_k)|\gamma_q)$. Since $\{(\beta^*(\gamma_q, \varepsilon_k),\mu^*(\gamma_q, \varepsilon_k)),k=1,2,\ldots\}$ is a bounded sequence, it has a convergent subsequence. For convenience, we still denote such a convergent subsequence by $\{(\beta^*(\gamma_q, \varepsilon_k),\mu^*(\gamma_q, \varepsilon_k)),k=1,2,\ldots\}$.  Let $(\beta^*(\gamma_q),\mu^*(\gamma_q))=\lim\limits_{k\to\infty}(\beta^*(\gamma_q, \varepsilon_k),\mu^*(\gamma_q, \varepsilon_k))$. Since $\{(\beta^*(\gamma_q),\mu^*(\gamma_q)),q=1,2,\ldots\}$ is a bounded sequence, it has a convergent subsequence. For convenience, we still denote such a convergent subsequence by $\{(\beta^*(\gamma_q),\mu^*(\gamma_q)),q=1,2,\ldots\}$. Let $(\beta^*,\mu^*)=\lim\limits_{q\to\infty}(\beta^*(\gamma_q),\mu^*(\gamma_q))$.  Therefore it follows from Definition~\ref{esedA} that
$(\beta^*,\mu^*)$ is a sequential equilibrium.  
The proof is completed.
\end{proof}

We illustrate with two examples in Section~\ref{aesed} how Definition~\ref{esedA} can be used to analytically identify all sequential equilibria for small finite extensive-form games with perfect recall.  

Although Definition~\ref{esedA} provides an effective scheme for constructing a convergent sequence that satisfies the consistency condition, obtaining a polynomial system as a necessary and sufficient condition for determining whether an assessment constitutes an $\varepsilon$-perfect $\gamma$-sequential equilibrium remains challenging. To address this difficulty, we introduce the following equivalent definition of $\varepsilon$-perfect $\gamma$-sequential equilibrium by separating perturbation from strategies. For any sufficiently small $\varepsilon>0$, let $\eta(\varepsilon)=(\eta^i_{I^j_i}(\varepsilon):i\in N,j\in M_i)$  be a vector with $\eta^i_{I^j_i}(\varepsilon)=(\eta^{i}_{I^j_i}(\varepsilon; a):a\in A(I^j_i))^\top$ such that $0<\eta^{i}_{I^j_i}(\varepsilon;a)\le \varepsilon$ and $\tau^i_{I^j_i}(\varepsilon)=\sum\limits_{a\in A(I^j_i)}\eta^{i}_{I^j_i}(\varepsilon; a)< 1$. 
We specify $\varpi(\beta,\eta(\varepsilon))=(\varpi(\beta^p_{I^\ell_p},\eta(\varepsilon)):p\in N,\ell\in M_p)$, where $\varpi(\beta^p_{I^\ell_p},\eta(\varepsilon))=(\varpi(\beta^p_{I^\ell_p}(a),\eta(\varepsilon)):a\in A(I^\ell_p))^\top$ with \[
\setlength{\abovedisplayskip}{1.2pt} 
\setlength{\belowdisplayskip}{1.2pt}
\varpi(\beta^p_{I^\ell_p}(a),\eta(\varepsilon))=(1-\tau^p_{I^\ell_p}(\varepsilon))\beta^p_{I^\ell_p}(a)+\eta^{p}_{I^\ell_p}(\varepsilon;a).\]
With $\varpi(\beta,\eta(\varepsilon))$, Definition~\ref{esedA} can be equivalently rewritten as below.
\begin{definition}[\bf An Equivalent Definition of Sequential Equilibrium through $\varepsilon$-Perfect $\gamma$-Sequential Equilibrium with Separation of Perturbation from Strategies] \label{esedB} {\em For any given $0<\gamma$ and $0<\varepsilon$, a totally mixed assessment  $(\varpi(\beta(\gamma),\eta(\varepsilon)),\mu(\gamma,\varepsilon))$ is  an $\varepsilon$-perfect $\gamma$-sequential equilibrium if $\beta^i_{I^j_i}(\gamma; a')=0$ for any $i\in N$, $j\in M_i$, and $a',a''\in A(I^j_i)$ with
\begin{equation}\label{esedBeqAA}\setlength{\abovedisplayskip}{1.2pt}
\setlength{\belowdisplayskip}{1.2pt} u^i(a'',\varpi(\beta^{-I^j_i}(\gamma),\eta(\varepsilon)), \mu(\gamma,\varepsilon)|I^j_i)-u^i(a',\varpi(\beta^{-I^j_i}(\gamma),\eta(\varepsilon)), \mu(\gamma,\varepsilon)|I^j_i)>\gamma,\end{equation}
where $\mu(\gamma, \varepsilon)=(\mu^i_{I^j_i}(\gamma, \varepsilon;h):i\in N,j\in M_i,h\in I^j_i)$ with $\mu^i_{I^j_i}(\gamma, \varepsilon;h)=\frac{\omega(h|\varpi(\beta(\gamma),\eta( \varepsilon)))}{\omega(I^j_i|\varpi(\beta(\gamma),\eta( \varepsilon)))}$ for $h\in I^j_i$. 
For any given $\gamma>0$, $(\beta(\gamma),\mu(\gamma))$ is a perfect $\gamma$-sequential equilibrium if $\mu(\gamma)$ is a limit point of a sequence $\{\mu(\gamma,\varepsilon_k), k=1,2,\ldots\}$ with $\varepsilon_k>0$ and $\lim\limits_{k\to\infty}\varepsilon_k=0$, where $(\varpi(\beta(\gamma),\eta(\varepsilon_k)), \mu(\gamma,\varepsilon_k))$ is an $\varepsilon_k$-perfect $\gamma$-sequential equilibrium for every $k$.
$(\beta^*,\mu^*)$ is a sequential equilibrium if it is a limit point of a sequence $\{(\beta(\gamma_q),\mu(\gamma_q)),q=1,2,\ldots\}$ with $\gamma_q>0$ and $\lim\limits_{q\to\infty}\gamma_q=0$, where $(\beta(\gamma_q),\mu(\gamma_q))$ is a perfect $\gamma_q$-sequential equilibrium for every $q$.}
\end{definition}
\begin{theorem}\label{esedBthm1}{\em Definition~\ref{esedB} and Definition~\ref{esedA} of sequential equilibrium are equivalent.
}
\end{theorem}
\begin{proof} $(\Rightarrow)$. We denote by $(\beta^*,\mu^*)$ a sequential equilibrium according to Definition~\ref{esedB}. Let $\tilde\beta(\gamma,\varepsilon)=\varpi(\beta(\gamma),\eta(\varepsilon))$. Then, $(\tilde\beta(\gamma,\varepsilon),\mu(\gamma,\varepsilon))$ constitutes an $\varepsilon$-perfect $\gamma$-sequential equilibrium according to Definition~\ref{esedA}. Thus, $(\beta^*,\mu^*)$ is a sequential equilibrium according to Definition~\ref{esedA}. 

\noindent  $(\Leftarrow)$. We denote by $(\beta^*,\mu^*)$ a sequential equilibrium according to Definition~\ref{esedA}. Then there exists a convergent sequence $\{(\beta(\gamma_q),\mu(\gamma_q)),q=1,2,\ldots\}$ with $\gamma_q>0$ and $\lim\limits_{q\to\infty}\gamma_q=0$ such that $(\beta^*,\mu^*)=\lim\limits_{q\to\infty} (\beta(\gamma_q),\mu(\gamma_q))$,
 where $(\beta(\gamma_q),\mu(\gamma_q))$ is the limit point of a convergent sequence $\{(\beta(\gamma_q,\varepsilon_k), \mu(\gamma_q,\varepsilon_k)), k=1,2,\ldots\}$ satisfying that $\varepsilon_k>0$, $\lim\limits_{k\to\infty}\varepsilon_k=0$, and $(\beta(\gamma_q,\varepsilon_k), \mu(\gamma_q,\varepsilon_k))$ is an $\varepsilon_k$-perfect $\gamma_q$-sequential equilibrium for every $k$ according to Definition~\ref{esedA}.

For $i\in N$ and $j\in M_i$,  let $A^0(I^j_i|\gamma_q)=\{a\in A(I^j_i)|\beta^i_{I^j_i}(\gamma_q; a)=0\}$. We have $A^0(\gamma_q)=\mathop{\cup}\limits_{i\in N,\;j\in M_i}A^0(I^j_i|\gamma_q)$. Consider the case of $A^0(\gamma_q)=\emptyset$.  Let $\{\varepsilon_k>0,k=1,2,\ldots\}$ be a convergent sequence with $\lim\limits_{k\to\infty}\varepsilon_k=0$. We take $\eta(\varepsilon_k)=\varepsilon_k\beta(\gamma_q)$. Then, $\varpi(\beta(\gamma_q),\eta(\varepsilon_k))=\beta(\gamma_q)$ and $\{\mu(\varpi(\beta(\gamma_q),\eta(\varepsilon_k)))>0,k=1,2,\ldots\}$ is a convergent sequence with $\lim\limits_{k\to\infty}\mu(\varpi(\beta(\gamma_q),\eta(\varepsilon_k)))=\mu(\gamma_q)$.  Thus, $(\beta^*,\mu^*)$ is a sequential equilibrium according to Definition~\ref{esedB}.  

Consider the case of $A^0(\gamma_q)\ne\emptyset$. Let \begin{equation}\label{esedBeqA}\setlength{\abovedisplayskip}{1.2pt} 
\setlength{\belowdisplayskip}{1.2pt}
\varepsilon_k=\max\limits_{i\in N,\;j\in M_i}\max\limits_{a\in A^0(I^j_i|\gamma_q)\ne\emptyset}\beta^{i}_{I^j_i}(\gamma_q,\varepsilon_k; a).\end{equation} Then, $\{\varepsilon_k,k=1,2,\ldots\}$ is a convergent sequence with $\varepsilon_k>0$ and $\lim\limits_{k\to\infty}\varepsilon_k=0$. We take $\eta(\varepsilon_k)=(\eta^i_{I^j_i}(\varepsilon_k; a):i\in N, j\in M_i, a\in A(I^j_i))$ to be a vector with
\begin{equation}\label{esedBeqB}\setlength{\abovedisplayskip}{1.2pt} 
\setlength{\belowdisplayskip}{1.2pt}
\eta^i_{I^j_i}(\varepsilon_k; a)=\left\{\begin{array}{ll}
\beta^{i}_{I^j_i}(\gamma_q,\varepsilon_k; a) & \text{if $a\in A^0(I^j_i|\gamma_q)$,}\\
\varepsilon_k\beta^{i}_{I^j_i}(\gamma_q; a) & \text{otherwise.}
\end{array}\right.
\end{equation}
Then, $\varpi(\beta^{i}_{I^j_i}(\gamma_q; a),\eta(\varepsilon_k))=\beta^{i}_{I^j_i}(\gamma_q,\varepsilon_k; a)$ for any $i\in N$, $j\in M_i$, and $a\in A^0(I^j_i)$. 
It follows from Eq.~(\ref{esedBeqA}) that $\eta^i_{I^j_i}(\varepsilon_k; a)\le\varepsilon_k$. As a result of Eq.~(\ref{esedBeqB}), for any $h=\langle a_1,a_2,\ldots,a_L\rangle\in H$, we have {\footnotesize
\begin{equation}\label{esedBeqC} 
\setlength{\abovedisplayskip}{1.2pt} 
\setlength{\belowdisplayskip}{1.2pt}
\frac{\omega(h|\varpi(\beta(\gamma_q),\eta(\varepsilon_k)))}{\omega(h|\beta(\gamma_q,\varepsilon_k))}=\prod\limits_{k=0}^{L-1}\frac{\varpi(\beta^{P(\langle a_1,\ldots,a_k\rangle)}_{\langle a_1,\ldots,a_k\rangle}(\gamma_q; a_{k+1}),\eta(\varepsilon_k))}{\beta^{P(\langle a_1,\ldots,a_k\rangle)}_{\langle a_1,\ldots,a_k\rangle}(\gamma_q,\varepsilon_k; a_{k+1})}=\mathop{\prod\limits_{k=0}^{L-1}}\limits_{a_{k+1}\notin A^0}\frac{\varpi(\beta^{P(\langle a_1,\ldots,a_k\rangle)}_{\langle a_1,\ldots,a_k\rangle}(\gamma_q; a_{k+1}),\eta(\varepsilon_k))}{\beta^{P(\langle a_1,\ldots,a_k\rangle)}_{\langle a_1,\ldots,a_k\rangle}(\gamma_q,\varepsilon_k; a_{k+1})}.
\end{equation}}Thus, $\lim\limits_{k\to\infty}\frac{\omega(h|\varpi(\beta(\gamma_q),\eta(\varepsilon_k)))}{\omega(h|\beta(\gamma_q,\varepsilon_k))}=1$ or $\omega(h|\varpi(\beta(\gamma_q),\eta(\varepsilon_k)))= \omega(h|\beta(\gamma_q,\varepsilon_k))+o(\omega(h|\beta(\gamma_q,\varepsilon_k)))$. Therefore, for any $i\in N$, $j\in M_i$, and $h\in I^j_i$, 
\begin{equation}\label{esedBeqD}\setlength{\abovedisplayskip}{1.2pt} 
\setlength{\belowdisplayskip}{1.2pt}
\begin{array}{rl} & \lim\limits_{k\to\infty}\frac{\omega(h|\varpi(\beta(\gamma_q),\eta(\varepsilon_k)))}{\omega(I^j_i|\varpi(\beta(\gamma_q),\eta(\varepsilon_k)))}=\lim\limits_{k\to\infty}\frac{\omega(h|\beta(\gamma_q,\varepsilon_k))+o(\omega(h|\beta(\gamma_q,\varepsilon_k)))}{\sum\limits_{h'\in I^j_i}(\omega(h'|\beta(\gamma_q,\varepsilon_k))+o(\omega(h'|\beta(\gamma_q,\varepsilon_k))))}\\
= & \lim\limits_{k\to\infty}\frac{\omega(h|\beta(\gamma_q,\varepsilon_k))}{\sum\limits_{h'\in I^j_i}\omega(h'|\beta(\gamma_q,\varepsilon_k))}=\mu^{*i}_{I^j_i}(h).\end{array}
\end{equation}
Hence, $(\beta^*,\mu^*)$ is a sequential equilibrium according to Definition~\ref{esedB}. The proof is completed.\end{proof}

Building upon Definition~\ref{esedB}, we next establish a necessary and sufficient condition for determining whether an assessment constitutes an $\varepsilon$-perfect $\gamma$-sequential equilibrium. One can attain the following conclusion in a similar way to the proof of Theorem~\ref{nscsethm1}.
\begin{theorem}\label{nscthmB} {\em $(\varpi(\beta(\gamma), \eta(\varepsilon)),\mu(\gamma,\varepsilon))>0$ is an $\varepsilon$-perfect $\gamma$-sequential equilibrium if and only if there exists a pair of $(\lambda(\gamma,\varepsilon),\zeta(\gamma,\varepsilon))$ together with $(\beta(\gamma),\eta(\varepsilon),\mu(\gamma,\varepsilon))$ satisfying the polynomial system,
\begin{equation}\label{esedBnsc}\setlength{\abovedisplayskip}{1.2pt} 
\setlength{\belowdisplayskip}{1.2pt}
\begin{array}{l}
-\gamma\le u^i(a,\varpi(\beta^{-I^j_i},\eta), \mu|I^j_i)+\lambda^i_{I^j_i}(a)-\zeta^i_{I^j_i}\le\gamma,\;i\in N,j\in M_i,a\in A(I^j_i),\\

\sum\limits_{a\in A(I^j_i)}\beta^i_{I^j_i}(a)=1,\;i\in N, j\in M_i,\\

\omega(I^j_i|\varpi(\beta^{-I^j_i},\eta))\mu^i_{I^j_i}(h)=\omega(h|\varpi(\beta^{-I^j_i},\eta)),\;i \in N, j\in M_i, h\in I^j_i,\\

\beta^i_{I^j_i}(a)\lambda^i_{I^j_i}(a)=0,\;0\le\beta^i_{I^j_i}(a),\;0\le\lambda^i_{I^j_i}(a),\;i\in N,j\in M_i,a\in A(I^j_i).
\end{array}
\end{equation}
}
\end{theorem}

We demonstrate with two examples in the Supplemental Appendix how Definition~\ref{esedB} can be used to identify all sequential equilibria for small finite extensive-form games with perfect recall.

Definition~\ref{esedA} explicitly reveals the significant differences between the concept of sequential equilibrium and the concept of extensive-form trembling hand perfect equilibrium proposed by Selten~\cite{Selten (1975)}, which takes the possibility of off-the-equilibrium actions into account by assuming that players may occasionally experience trembles with negligible probability.  Kreps and Wilson~\cite{Kreps and Wilson (1982)} stated that extensive-form trembling hand perfect equilibrium serves as a refinement of sequential equilibrium.  To show the difference between two concepts, we reproduce the definition of extensive-form trembling hand  perfect equilibrium here.
  \begin{definition}[{\bf Extensive-Form Trembling Hand Perfect Equilibrium}, Selten~\cite{Selten (1975)}] \label{defpe1} {\em  For any given $\varepsilon>0$, a totally mixed assessment $(\beta(\varepsilon),\mu(\varepsilon))$
   constitutes an $\varepsilon$-perfect equilibrium if $\beta^i_{I^j_i}(\varepsilon; a')\le\varepsilon$ for any $i\in N$, $j\in M_i$ and $a',a''\in A(I^j_i)$ with 
   \begin{equation}\label{defpe1eqAA}\setlength{\abovedisplayskip}{1.2pt}
\setlength{\belowdisplayskip}{1.2pt} u^i(a'',\beta^{-I^j_i}(\varepsilon),\mu(\varepsilon)|I^j_i)-u^i(a',\beta^{-I^j_i}(\varepsilon),\mu(\varepsilon)|I^j_i)>0,\end{equation}
where $\mu(\varepsilon)=(\mu^i_{I^j_i}(\varepsilon;h):i\in N,j\in M_i,h\in I^j_i)$ with $\mu^i_{I^j_i}(\varepsilon;h)=\frac{\omega(h|\beta(\varepsilon))}{\omega(I^j_i|\beta(\varepsilon))}$ for $h\in I^j_i$. 
 $(\beta^*, \mu^*)$ is an extnesive-form trembling hand perfect equilibrium if it is a limit point of a sequence $\{(\beta(\varepsilon_k),\mu(\varepsilon_k))>0$, $k=1,2,\ldots\}$ with $\varepsilon_k>0$ and $\lim\limits_{k\to\infty}\varepsilon_k=0$, where $(\beta(\varepsilon_k),\mu(\varepsilon_k))$ is an $\varepsilon_k$-perfect equilibrium for every $k$. }
 \end{definition}

 By utilizing our characterization of sequential equilibrium, it becomes easier to demonstrate how extensive-form trembling hand perfect equilibrium refines the concept of sequential equilibrium. 
 It follows from Definition~\ref{defpe1} of $\varepsilon$-perfect equilibrium that beliefs are generated  in the same way as that of $\varepsilon$-perfect $\gamma$-sequential equilibrium in Definition~\ref{esedA}. Given beliefs, denoted as $\mu(\varepsilon)$, that are determined by the formula $\mu^i_{I^j_i}(\varepsilon; h)=\frac{\omega(h|\beta(\varepsilon))}{\omega(I^j_i|\beta(\varepsilon))}$ at every information set, an $\varepsilon$-perfect equilibrium requires that $\beta^i_{I^j_i}(\varepsilon; a')\le\varepsilon$ if \(u^i(a'',\beta^{-I^j_i}(\varepsilon),\mu(\varepsilon)|I^j_i)-u^i(a',\beta^{-I^j_i}(\varepsilon),\mu(\varepsilon)|I^j_i)>0\). In a simpler term, this means that players' perturbed actions, $\beta(\varepsilon)$, must be optimal with respect to their beliefs.
 However, according to Definition~\ref{esedA} of $\varepsilon$-perfect $\gamma$-sequential equilibrium, given such beliefs $\mu(\gamma,\varepsilon)$, it only requires that $\beta^i_{I^j_i}(\gamma,\varepsilon; a')\le\varepsilon$ if \(u^i(a'',\beta^{-I^j_i}(\gamma,\varepsilon),\mu(\gamma,\varepsilon)|I^j_i)-u^i(a',\beta^{-I^j_i}(\gamma,\varepsilon),\mu(\gamma,\varepsilon)|I^j_i)>\gamma\). In other words, $\varepsilon$-perfect $\gamma$-sequential equilibrium demands optimal actions only when there is a constant difference in payoffs. This recognition highlights that the concept of sequential equilibrium is weaker than that of extensive-form trembling hand perfect equilibrium, although they are generically considered equivalent, as demonstrated by Blume and Zame~\cite{Blume and Zame (1994)}.

To further facilitate the applications of sequential equilibrium, one can capitalize on Theorem~\ref{nscsethm1},  Corollary~\ref{nscseco0}, Definition~\ref{esedA}, Defnition~\ref{esedB},  or Theorem~\ref{nscthmB} to devise numerical methods for computing sequential equilibria. The limiting behavior of sequential equilibrium naturally leads us to the development of a differentiable path-following method. We will exploit   Theorem~\ref{nscsethm1} in Section~\ref{efpm} to develop an entropy-barrier differentiable path-following method for computing such an equilibrium.

\section{\label{aesed}Illustrative Examples}

This section illustrates with two examples how one can employ Definition~\ref{esedA} to analytically find all the sequential equilibria for small extensive-form games. 
\begin{figure}[H]
    \centering
    \begin{minipage}{0.49\textwidth}
        \centering
        \includegraphics[width=0.8\textwidth, height=0.15\textheight]{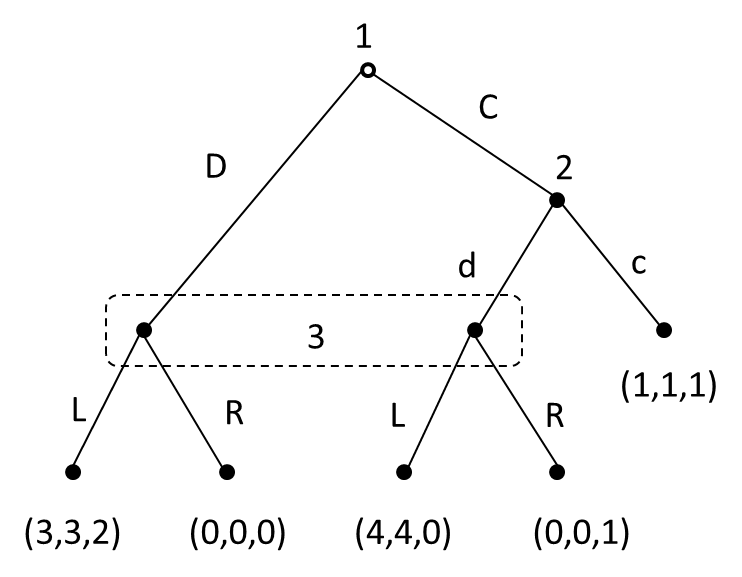}
                \caption{\label{TFigure3}\scriptsize An Extensive-Form Game (Selten's Horse) from Osborne and Rubinstein~\cite{Osborne and Rubinstein (1994)}}
\end{minipage}\hfill
    \begin{minipage}{0.49\textwidth}
        \centering
        \includegraphics[width=0.8\textwidth, height=0.15\textheight]{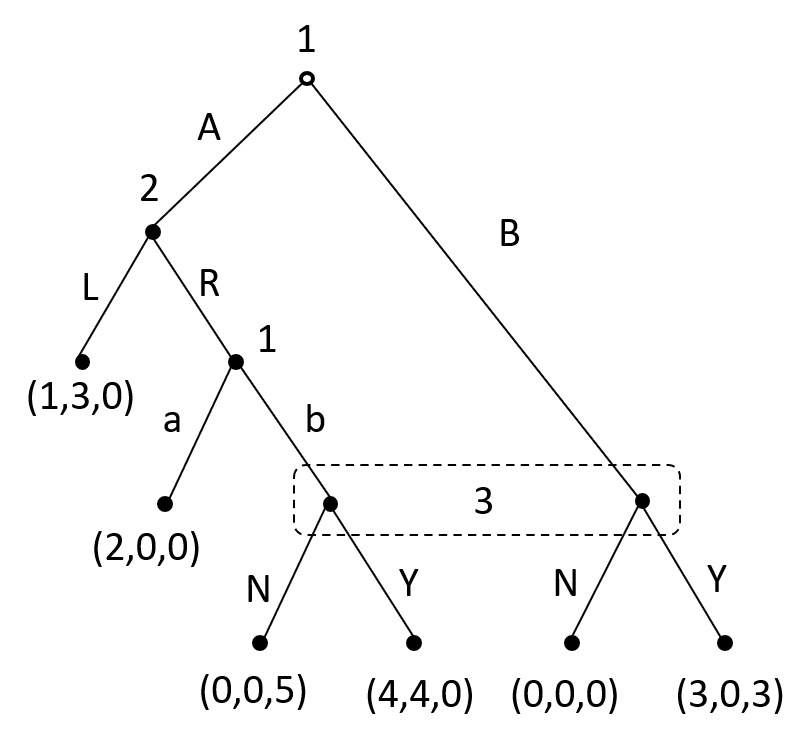}
\caption{\label{TFigure4}\scriptsize An Extensive-Form Game from Selten~\cite{Selten (1975)}}\end{minipage}
 \end{figure}
           
\begin{example} \label{esedAexm1}{\em
 Consider the game in Fig.~\ref{TFigure3}. The information sets consist of $I^1_1=\{\emptyset\}$, $I^1_2=\{\langle C\rangle\}$, and  $I^1_3=\{\langle D\rangle, \langle C, d\rangle\}$. 
 We denote by $(\beta^*,\mu^*)$ a sequential equilibrium, which is presented in the form of $\beta^*=((\beta^{*1}_{I^1_1}(D),\beta^{*1}_{I^1_1}(C)),(\beta^{*2}_{I^1_2}(d),\beta^{*2}_{I^1_2}(c)),(\beta^{*3}_{I^1_3}(L), \beta^{*3}_{I^1_3}(R)))$. Let  $(\beta(\gamma, \varepsilon),\mu(\gamma, \varepsilon))$ be an $\varepsilon$-perfect $\gamma$-sequential equilibrium. For simplicity, we will omit $(\gamma,\varepsilon)$ from $(\beta(\gamma, \varepsilon),\mu(\gamma, \varepsilon))$ in the rest of this example.
The conditional expected payoffs at $(\beta,\mu)$ on $I^j_i$ are given by 
{\footnotesize \[\setlength{\abovedisplayskip}{1.2pt}
\setlength{\belowdisplayskip}{1.2pt}
\begin{array}{l}
u^1(D,\beta^{-I^1_1},\mu|I^1_1)=3\beta^3_{I^1_3}(L),\;
 u^1(C,\beta^{-I^1_1},\mu|I^1_1)=\beta^2_{I^1_2}(c) +4\beta^2_{I^1_2}(d)\beta^3_{I^1_3}(L),\\

u^2(d,\beta^{-I^1_2},\mu|I^1_2)= 4\beta^3_{I^1_3}(L), \;\; u^2(c,\beta^{-I^1_2},\mu|I^1_2)=1,\\

u^3(L,\beta^{-I^1_3},\mu|I^1_3)=2\mu^3_{I^1_3}(\langle D\rangle), \;\; u^3(R,\beta^{-I^1_3},\mu|I^1_3)=\mu^3_{I^1_3}(\langle C, d\rangle),
\end{array}\]
where $\mu^3_{I^1_3}(\langle D\rangle)=\frac{\beta^1_{I^1_1}(D)}{\beta^1_{I^1_1}(D)+\beta^1_{I^1_1}(C)\beta^2_{I^1_2}(d)}$ and $\mu^3_{I^1_3}(\langle C, d\rangle)=\frac{\beta^1_{I^1_1}(C)\beta^2_{I^1_2}(d)}{\beta^1_{I^1_1}(D)+\beta^1_{I^1_1}(C)\beta^2_{I^1_2}(d)}$.}\newline
{\bf Case (1):} Suppose that $u^2(d,\beta^{-I^1_2},\mu|I^1_2)-u^2(c,\beta^{-I^1_2},\mu|I^1_2)>\gamma$. Then, $\beta^3_{I^1_3}(L)>\frac{1}{4}+\frac{1}{4}\gamma$ and $\beta^2_{I^1_2}(c)\le
\varepsilon$. Thus, $u^1(C,\beta^{-I^1_1},\mu|I^1_1)-u^1(D,\beta^{-I^1_1}$, $\mu|I^1_1)>\gamma$ and 
consequently, $\beta^1_{I^1_1}(D)\le\varepsilon$. Therefore, $u^3(R,\beta^{-I^1_3},\mu|I^1_3)-u^3(L,\beta^{-I^1_3},\mu|I^1_3)>\gamma$ and accordingly, $\beta^3_{I^1_3}(L)\le\varepsilon$. A contradiction occurs and the case is excluded. 
\newline
{\bf Case (2):} Suppose that $u^2(c,\beta^{-I^1_2},\mu|I^1_2)-u^2(d,\beta^{-I^1_2},\mu|I^1_2)>\gamma$.  Then, $\frac{1}{4}-\frac{1}{4}\gamma>\beta^3_{I^1_3}(L)$ and  $\beta^2_{I^1_2}(d)\le
\varepsilon$. Thus, $u^1(C,\beta^{-I^1_1},\mu|I^1_1)-u^1(D,\beta^{-I^1_1}$, $\mu|I^1_1)>\gamma$ and either $u^3(R,\beta^{-I^1_3},\mu|I^1_3)-u^3(L,\beta^{-I^1_3},\mu|I^1_3)>\gamma$ or $|u^3(R,\beta^{-I^1_3},\mu|I^1_3)-u^3(L,\beta^{-I^1_3},\mu|I^1_3)|\le\gamma$. Consequently, $\beta^1_{I^1_1}(D)\le\varepsilon$.

\noindent {\bf (a)}. Assume that $u^3(R,\beta^{-I^1_3},\mu|I^1_3)-u^3(L,\beta^{-I^1_3},\mu|I^1_3)>\gamma$. 
 Then,  $\mu^3_{I^1_3}(\langle C, d\rangle)>\frac{2}{3}+\frac{1}{3}\gamma$ and $\beta^3_{I^1_3}(L)\le\varepsilon$. Thus,  
 $\beta^1_{I^1_1}(C)\beta^2_{I^1_2}(d)> \frac{2+\gamma}{1-\gamma}\beta^1_{I^1_1}(D)$. The game has a class of sequential equilibria given by $(C,c,R)$ with $\mu^{*3}_{I^1_3}(\langle D\rangle)<\frac{1}{3}$.
 
\noindent {\bf (b)}. Assume that $|u^3(R,\beta^{-I^1_3},\mu|I^1_3)-u^3(L,\beta^{-I^1_3},\mu|I^1_3)|\le\gamma$. Then, 
$|\mu^3_{I^1_3}(\langle C, d\rangle)-\frac{2}{3}|\le\frac{1}{3}\gamma$, which leads to $|\beta^2_{I^1_2}(d)-(\beta^2_{I^1_2}(d)+2)\beta^1_{I^1_1}(D)|\le\gamma(\beta^2_{I^1_2}(d)+(1-\beta^2_{I^1_2}(d))\beta^1_{I^1_1}(D))$.
The game has a class of sequential equilibria given by  $(C,c, (\beta^{*3}_{I^1_3}(L),1-\beta^{*3}_{I^1_3}(L)))$ with $0\le\beta^{*3}_{I^1_3}(L)<\frac{1}{4}$ and 
 $\mu^{*3}_{I^1_3}(\langle D\rangle)=\frac{1}{3}$.
\newline
{\bf Case (3):} Suppose that $|u^2(c,\beta^{-I^1_2},\mu|I^1_2)-u^2(d,\beta^{-I^1_2},\mu|I^1_2)|\le\gamma$.  Then, $|\frac{1}{4}-\beta^3_{I^1_3}(L)|\le \frac{1}{4}\gamma$. Thus, $|u^3(R,\beta^{-I^1_3},\mu|I^1_3)-u^3(L,\beta^{-I^1_3},\mu|I^1_3)|\le \gamma$ and $u^1(C,\beta^{-I^1_1}$, $\mu|I^1_1)-u^1(D,\beta^{-I^1_1},\mu|I^1_1)>\gamma$. Consequently, $\beta^1_{I^1_1}(D)\le\varepsilon$ and
$|\mu^3_{I^1_3}(\langle C, d\rangle)-\frac{2}{3}|\le\frac{1}{3}\gamma$. 
Thus, $|\beta^2_{I^1_2}(d)-(\beta^2_{I^1_2}(d)+2)\beta^1_{I^1_1}(D)|\le\gamma(\beta^2_{I^1_2}(d)+(1-\beta^2_{I^1_2}(d))\beta^1_{I^1_1}(D))$.  
The game has a sequential equilibrium given by $(C,c,(\frac{1}{4},\frac{3}{4}))$ with $\mu^{*3}_{I^1_3}(\langle D\rangle)=\frac{1}{3}$.

The cases (1)-(3) together ensure that the game has two types of sequential equilibria given by: (1). $(C,c,R)$ with $\mu^{*3}_{I^1_3}(\langle D\rangle)<\frac{1}{3}$; (2). $(C,c, (\beta^{*3}_{I^1_3}(L),1-\beta^{*3}_{I^1_3}(L)))$ with $0\le\beta^{*3}_{I^1_3}(L)\le\frac{1}{4}$ and $\mu^{*3}_{I^1_3}(\langle D\rangle)=\frac{1}{3}$. }
\end{example}

\begin{example}\label{esedAexm2}
 {\em Consider the game in Fig.~\ref{TFigure4}. 
The information sets consist of $I^1_1=\{\emptyset\}$, $I^2_1=\{\langle A, R \rangle\}$, $I^1_2=\{\langle A\rangle\}$, and $I^1_3=\{\langle A, R,b
\rangle, \langle B
\rangle\}$. We denote by $(\beta^*,\mu^*)$ a sequential equilibrium, which is presented in the form of $\beta^*=((\beta^{*1}_{I^1_1}(A),\beta^{*1}_{I^1_1}(B)), (\beta^{*1}_{I^2_1}(a),\beta^{*1}_{I^2_1}(b))$,
$(\beta^{*2}_{I^1_2}(L),\beta^{*2}_{I^1_2}(R)),(\beta^{*3}_{I^1_3}(N), \beta^{*3}_{I^1_3}(Y)))$.
 Let  $(\beta(\gamma, \varepsilon),\mu(\gamma, \varepsilon))$ be an $\varepsilon$-perfect $\gamma$-sequential equilibrium. For simplicity, we will omit $(\gamma,\varepsilon)$ from $(\beta(\gamma, \varepsilon),\mu(\gamma, \varepsilon))$ in the rest of this example.
The conditional expected payoffs at $(\beta,\mu)$ on $I^j_i$ are given by {\footnotesize
\[\setlength{\abovedisplayskip}{1.2pt}
\setlength{\belowdisplayskip}{1.2pt}
\begin{array}{l}
u^1(A,\beta^{-I^1_1},\mu|I^1_1)=\beta^2_{I^1_2}(L)+2\beta^2_{I^1_2}(R)\beta^1_{I^2_1}(a)+4\beta^2_{I^1_2}(R)\beta^1_{I^2_1}(b)\beta^3_{I^1_3}(Y),\\

u^1(B,\beta^{-I^1_1},\mu| I^1_1)=3\beta^3_{I^1_3}(Y),\\

u^1(a,\beta^{-I^2_1},\mu| I^2_1) = 2,\;

u^1(b,\beta^{-I^2_1},\mu| I^2_1) = 4\beta^3_{I^1_3}(Y),\\

u^2(L,\beta^{-I^1_2},\mu| I^1_2)=3,\;

u^2(R,\beta^{-I^1_2},\mu| I^1_2)=4\beta^1_{I^2_1}(b)\beta^3_{I^1_3}(Y),\\

u^3(N,\beta^{-I^1_3},\mu| I^1_3)=5\mu^3_{I^1_3}(\langle A,R,b\rangle),\;

u^3(Y,\beta^{-I^1_3},\mu| I^1_3)=3\mu^3_{I^1_3}(\langle B\rangle),
\end{array}\]

\noindent where \[\setlength{\abovedisplayskip}{1.2pt}
\setlength{\belowdisplayskip}{1.2pt}\mu^3_{I^1_3}(\langle A,R,b\rangle)=\frac{\beta^1_{I^1_1}(A)\beta^2_{I^1_2}(R)\beta^1_{I^2_1}(b)}{\beta^1_{I^1_1}(A)\beta^2_{I^1_2}(R)\beta^1_{I^2_1}(b)+\beta^1_{I^1_1}(B)}, \text{ and } \mu^3_{I^1_3}(\langle B\rangle)=\frac{\beta^1_{I^1_1}(B)}{\beta^1_{I^1_1}(A)\beta^2_{I^1_2}(R)\beta^1_{I^2_1}(b)+\beta^1_{I^1_1}(B)}.\] }

\noindent
{\bf Case (1)}. Suppose that $u^3(N,\beta^{-I^1_3},\mu| I^1_3)-u^3((Y,\beta^{-I^1_3},\mu| I^1_3)>\gamma$. Then, $\mu^3_{I^1_3}(\langle A,R,b\rangle)>\frac{3}{8}+\frac{1}{8}\gamma$ and
 $\beta^3_{I^1_3}(Y)\le \varepsilon$. Thus,  $\beta^1_{I^1_1}(A)\beta^2_{I^1_2}(R)\beta^1_{I^2_1}(b)>\frac{3+\gamma}{5-\gamma}\beta^1_{I^1_1}(B)$, $u^1(a,\beta^{-I^2_1},\mu| I^2_1) - u^1(b,\beta^{-I^2_1},\mu| I^2_1) >\gamma$, and $u^2(L,\beta^{-I^1_2},\mu| I^1_2)-u^2(R,\beta^{-I^1_2},\mu| I^1_2)>\gamma$. Consequently, $\beta^1_{I^2_1}(b)\le \varepsilon$ and $\beta^2_{I^1_2}(R)\le  \varepsilon$. Therefore, $u^1(A,\beta^{-I^1_1},\mu|I^1_1)-u^1(B,\beta^{-I^1_1},\mu|I^1_1)>\gamma$ and accordingly, $\beta^1_{I^1_1}(B)\le \varepsilon$.
  The game has a class of sequential equilibria given by $(A, a, L, N)$ with $\mu^3_{I^1_3}(\langle A,R,b\rangle)>\frac{3}{8}$.

\noindent
{\bf Case (2)}. Suppose that $u^3(Y,\beta^{-I^1_3},\mu| I^1_3)-u^3((N,\beta^{-I^1_3},\mu| I^1_3)>\gamma$. Then, $\frac{3}{8}-\frac{1}{8}\gamma>\mu^3_{I^1_3}(\langle A,R,b\rangle)$ and
 $\beta^3_{I^1_3}(N)\le\varepsilon$. Thus, 
$\beta^1_{I^1_1}(A)\beta^2_{I^1_2}(R)\beta^1_{I^2_1}(b)<\frac{3-\gamma}{5+\gamma}\beta^1_{I^1_1}(B)$ and $u^1(b,\beta^{-I^2_1},\mu| I^2_1) - u^1(a,\beta^{-I^2_1},\mu| I^2_1) >\gamma$. Consequently, $\beta^1_{I^2_1}(a)\le\varepsilon$. Therefore,
 $u^2(R,\beta^{-I^1_2},\mu| I^1_2)-u^2(L,\beta^{-I^1_2},\mu| I^1_2)>\gamma$ and accordingly, $\beta^2_{I^1_2}(L)\le \varepsilon$. Hence, 
  $u^1(A,\beta^{-I^1_1}$, $\mu|I^1_1)-u^1(B,\beta^{-I^1_1}(,\mu|I^1_1)>\gamma$ and subsequently, $\beta^1_{I^1_1}(B)\le \varepsilon$. 
  These results together imply that $(1-\varepsilon)(1-\varepsilon)(1-\varepsilon)\le\beta^1_{I^1_1}(A)\beta^2_{I^1_2}(R)\beta^1_{I^2_1}(b)<\frac{3-\gamma}{5+\gamma}\beta^1_{I^1_1}(B)\le \frac{3-\gamma}{5+\gamma}\varepsilon$. 
   A contradiction occurs. The case is excluded.

 \noindent
{\bf Case (3)}. Suppose that $|u^3(Y,\beta^{-I^1_3},\mu| I^1_3)-u^3(N,\beta^{-I^1_3},\mu| I^1_3)|\le\gamma$. Then, $|\frac{3}{8}-\mu^3_{I^1_3}(\langle A,R,b\rangle)|\le\frac{1}{8}\gamma$, which leads to \begin{equation}\label{esedAeq1}\setlength{\abovedisplayskip}{1.2pt} 
\setlength{\belowdisplayskip}{1.2pt}\frac{3-\gamma}{5+\gamma}\beta^1_{I^1_1}(B)\le\beta^1_{I^1_1}(A)\beta^2_{I^1_2}(R)\beta^1_{I^2_1}(b)\le\frac{3+\gamma}{5-\gamma}\beta^1_{I^1_1}(B).\end{equation}
Thus either $u^1(A,\beta^{-I^1_1},\mu|I^1_1)-u^1(B,\beta^{-I^1_1},\mu|I^1_1)>\gamma$ or 
$|u^1(A,\beta^{-I^1_1},\mu|I^1_1)-u^1(B,\beta^{-I^1_1},\mu|I^1_1)|$ $\le\gamma$.

\noindent
{\bf (a)}. Assume that  $u^2(L,\beta^{-I^1_2},\mu| I^1_2)-u^2(R,\beta^{-I^1_2},\mu| I^1_2)>\gamma$. Then, $\beta^2_{I^1_2}(R)\le \varepsilon$.

\noindent {\bf (i)}. Consider the scenario that $u^1(A,\beta^{-I^1_1},\mu|I^1_1)-u^1(B,\beta^{-I^1_1},\mu|I^1_1)>\gamma$. Then, $\beta^1_{I^1_1}(B)\le\varepsilon$ and $\beta^3_{I^1_3}(Y)< \frac{1}{3}+2\varepsilon-\frac{1}{3}\gamma$. Thus, $u^1(a,\beta^{-I^2_1},\mu| I^2_1) - u^1(b,\beta^{-I^2_1},\mu| I^2_1) >\gamma$ and consequently, $\beta^1_{I^2_1}(b)\le\varepsilon$. The game has a  class of sequential equilibria given by $(A, a, L,  (1-\beta^3_{I^1_3}(Y), \beta^3_{I^1_3}(Y)))$ with $\beta^3_{I^1_3}(Y)<\frac{1}{3}$ and $\mu^3_{I^1_3}(\langle A,R,b\rangle)=\frac{3}{8}$.

\noindent {\bf (ii)}. Consider the scenario that $|u^1(A,\beta^{-I^1_1},\mu|I^1_1)-u^1(B,\beta^{-I^1_1},\mu|I^1_1)|\le \gamma$.
Then, $|\beta^3_{I^1_3}(Y)- \frac{1}{3}(1-\varepsilon)|\le\frac{1}{3}\gamma$. Thus,  $u^1(a,\beta^{-I^2_1},\mu| I^2_1) - u^1(b,\beta^{-I^2_1},\mu| I^2_1) >\gamma$ and consequently, $\beta^1_{I^2_1}(b)\le\varepsilon$. 
 The game has a  sequential equilibrium given by $(A, a, L,  (\frac{2}{3}, \frac{1}{3}))$ with $\mu^3_{I^1_3}(\langle A,R,b\rangle)=\frac{3}{8}$.

\noindent
{\bf (b)}. Assume that  $u^2(R,\beta^{-I^1_2},\mu| I^1_2)-u^2(L,\beta^{-I^1_2},\mu| I^1_2)>\gamma$. Then, $\beta^2_{I^1_2}(L)\le \varepsilon$ and $\beta^1_{I^2_1}(b)\beta^3_{I^1_3}(Y)>\frac{3}{4}+\frac{1}{4}\gamma$. Thus, $u^1(b,\beta^{-I^2_1},\mu| I^2_1) - u^1(a,\beta^{-I^2_1},\mu| I^2_1) >\gamma$ and consequently, $\beta^1_{I^2_1}(a)\le\varepsilon$. Therefore, $u^1(A,\beta^{-I^1_1},\mu|I^1_1)-u^1(B,\beta^{-I^1_1},\mu|I^1_1)>\gamma$ and accordingly, $\beta^1_{I^1_1}(B)\le\varepsilon$.
Hence it follows from Eq.~(\ref{esedAeq1}) that $(1-\varepsilon)(1-\varepsilon)(1-\varepsilon)\le\beta^1_{I^1_1}(A)\beta^2_{I^1_2}(R)\beta^1_{I^2_1}(b)\le\frac{3+\gamma}{5-\gamma}\beta^1_{I^1_1}(B)\le \frac{3+\gamma}{5-\gamma}\varepsilon$.
A contradiction occurs. The assumption cannot be sustained.

\noindent
{\bf (c)}. Assume that  $|u^2(L,\beta^{-I^1_2},\mu| I^1_2)-u^2(R,\beta^{-I^1_2},\mu| I^1_2)|\le\gamma$. Then,  $|\beta^1_{I^2_1}(b)\beta^3_{I^1_3}(Y)-\frac{3}{4}|\le\frac{1}{4}\gamma$. Thus, $u^1(b,\beta^{-I^2_1},\mu| I^2_1) - u^1(a,\beta^{-I^2_1},\mu| I^2_1) >\gamma$ and consequently, $\beta^1_{I^2_1}(a)\le\varepsilon$. Therefore, $\frac{1}{4(1-\varepsilon)}(3-\gamma)\le\beta^3_{I^1_3}(Y)\le\frac{1}{4(1-\varepsilon)}(3+\gamma)$.

 \noindent {\bf (i)}. Consider the scenario that $u^1(A,\beta^{-I^1_1},\mu|I^1_1)-u^1(B,\beta^{-I^1_1},\mu|I^1_1)>\gamma$. Then,  $\beta^1_{I^1_1}(B)\le\varepsilon$, which together with Eq.~(\ref{esedAeq1}) implies that $(1-\varepsilon)(1-\varepsilon)\beta^2_{I^1_2}(R)\le \frac{3+\gamma}{5-\gamma}\varepsilon$. Thus,
 $u^1(B,\beta^{-I^1_1},\mu|I^1_1)-u^1(A,\beta^{-I^1_1},\mu|I^1_1)> \gamma$. A contradiction occurs and the scenario is excluded.
 
 \noindent {\bf (ii)}. Consider the scenario that $|u^1(A,\beta^{-I^1_1},\mu|I^1_1)-u^1(B,\beta^{-I^1_1},\mu|I^1_1)|\le\gamma$. Then,  $|\beta^2_{I^1_2}(R)-\frac{5+4\varepsilon+3\gamma}{4(1-\varepsilon)(2-\gamma)}|\le \frac{\gamma}{2-\gamma}$. 
 This result together with Eq.~(\ref{esedAeq1}) tells us that $\beta^*$ satisfies 
 $\frac{3}{5}\beta^1_{I^1_1}(B)=\frac{5}{8}\beta^1_{I^1_1}(A)$ and $\beta^2_{I^1_2}(R)=\frac{5}{8}$. The game has a sequential equilibrium given by  $((\frac{24}{49}, \frac{25}{49}), b, (\frac{3}{8},\frac{5}{8}), (\frac{1}{4}, \frac{3}{4}))$ with $\mu^3_{I^1_3}(\langle A,R,b\rangle)=\frac{3}{8}$.
 
The cases (1)-(3) together show that the game has three types of sequential equilibria given by\newline
Type 1: $(A, a, L, N)$ with $\mu^3_{I^1_3}(\langle A,R,b\rangle)>\frac{3}{8}$;
\newline
Type 2: $((\frac{24}{49}, \frac{25}{49}), b, (\frac{3}{8},\frac{5}{8}), (\frac{1}{4}, \frac{3}{4}))$ with $\mu^3_{I^1_3}(\langle A,R,b\rangle)=\frac{3}{8}$;
\newline
Type 3:  $(A, a, L,  (1-\beta^3_{I^1_3}(Y), \beta^3_{I^1_3}(Y)))$ with $\beta^3_{I^1_3}(Y)\le\frac{1}{3}$ and $\mu^3_{I^1_3}(\langle A,R,b\rangle)=\frac{3}{8}$.
}
\end{example}
One can see from these two examples that Definition~\ref{esedA} explicitly tells us how to effectively find all sequential equilibria for a small finite extensive-form game through $\varepsilon$-perfect $\gamma$-sequential equilibria.

\section{\large An Entropy-Barrier Differentiable Path-Following Method for Computing Sequential Equilibria\label{stpm}}

\subsection{Motivations behind the Method}

The selection of a sequential equilibrium as pointed out in Harsanyi and Selten~\cite{Harsanyi and Selten (1988)}  plays a significant role in its applications. The concept of the logistic agent quantal response equilibrium (Logistic AQRE) was formulated in Mckelvey and Palfey~\cite{Mckelvey and Palfey (1998)}. For $\gamma\in [0,\infty)$, $\beta(\gamma)$ is a logistic AQRE if it is a solution to the system,
\begin{equation}\label{laqre0} \setlength{\abovedisplayskip}{1.2pt}\setlength{\belowdisplayskip}{1.2pt}
\beta^i_{I^j_i}(a)=\frac{\exp(\gamma u^i(a,\beta^{-I^j_i}|I^j_i))}{\sum\limits_{\tilde a\in A(I^j_i)}\exp(\gamma u^i(\tilde a,\beta^{-I^j_i}|I^j_i))},\;i\in N, j\in M_i, a\in A(I^j_i).
\end{equation}
 We denote by $a^0_{I^j_i}\in A(I^j_i)$ a given reference action. A simple manipulation to the system~(\ref{laqre0}) brings us the system,
\begin{equation}\label{laqre1}\setlength{\abovedisplayskip}{1.2pt}\setlength{\belowdisplayskip}{1.2pt}
\begin{array}{l}
\ln\beta^i_{I^j_i}(a^0_{I^j_i})-\ln\beta^i_{I^j_i}(a)-\gamma (u^i(a^0_{I^j_i},\beta^{-I^j_i}|I^j_i)- u^i(a,\beta^{-I^j_i}|I^j_i))=0,\\
\hspace{9.5cm}i\in N, j\in M_i, a\in A(I^j_i)\backslash\{a^0_{I^j_i}\}\\
\sum\limits_{a\in A(I^j_i)}\beta^i_{I^j_i}(a)=1,\;i\in N,j\in M_i.
\end{array}
\end{equation}
Let $v=(v^i_{I^j_i}(a):i\in N,j\in M_i,a\in A(I^j_i))$. We define  \begin{equation}\label{eqBhs1}\setlength{\abovedisplayskip}{1.2pt}\setlength{\belowdisplayskip}{1.2pt}c(\nu)=\exp(\nu).
 \end{equation}
Let
$\beta(v)=(\beta^i_{I^j_i}(v):i\in N,j\in M_i)$ with $\beta^i_{I^j_i}(v)=(\beta^i_{I^j_i}(v;a):a\in A(I^j_i))^\top$, where $\beta^i_{I^j_i}(v;a)=c(v^i_{I^j_i}(a))$. 
 Replacing $\beta$ in the system~(\ref{laqre1}) by $\beta(v)$, we arrive at Turocy's~\cite{Turocy (2010)} system of equations, 
\begin{equation}\label{eqAhs1}\setlength{\abovedisplayskip}{1.2pt}\setlength{\belowdisplayskip}{1.2pt}
\begin{array}{l}
v^i_{I^j_i}(a^0_{I^j_i})-v^i_{I^j_i}(a)-\gamma(u^i(a^0_{I^j_i}, \beta^{-I^j_i}(v)|I^j_i)-u^i(a, \beta^{-I^j_i}(v)|I^j_i))=0,\\
\hspace{9cm}i\in N,j\in M_i,a\in A(I^j_i)\backslash\{a^0_{I^j_i}\},\\
\sum\limits_{a\in A(I^j_i)}\beta^i_{I^j_i}(v;a)=1,\;i\in N,j\in M_i.
\end{array}
\end{equation}

\noindent
Following the argument in Mckelvey and Palfey~\cite{Mckelvey and Palfey (1998)},  one can conclude that the smooth path specified by the system~(\ref{eqAhs1}) approaches a sequential equilibrium as $\gamma$ goes from $0$ to $\infty$. Due to the variable transformation in Eq.~(\ref{eqBhs1}), some variables of $v^i_{I^j_i}(a)$, $i\in N,j\in M_i, a\in A(I^j_i)$, in the smooth path specified by the system~(\ref{eqAhs1}) may go to $-\infty$. 
Therefore the smooth path specified by the system~(\ref{eqAhs1}) is not contained in a bounded set. 

\begin{figure}[H]
    \centering
    \begin{minipage}{0.49\textwidth}
        \centering
        \includegraphics[width=0.95\textwidth, height=0.15\textheight]{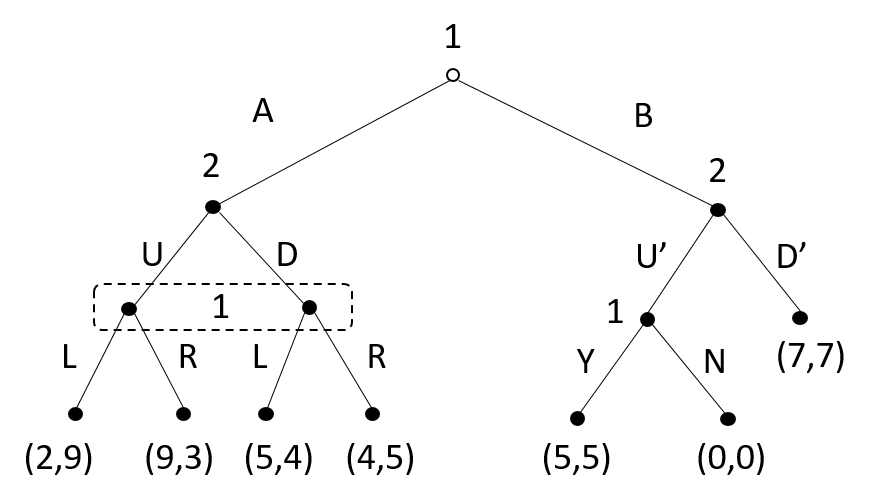}
                \caption{\label{TFigure1}\scriptsize An Extensive-Form Game}
\end{minipage}\hfill
    \begin{minipage}{0.49\textwidth}
       \centering
        \includegraphics[width=0.95\textwidth, height=0.15\textheight]{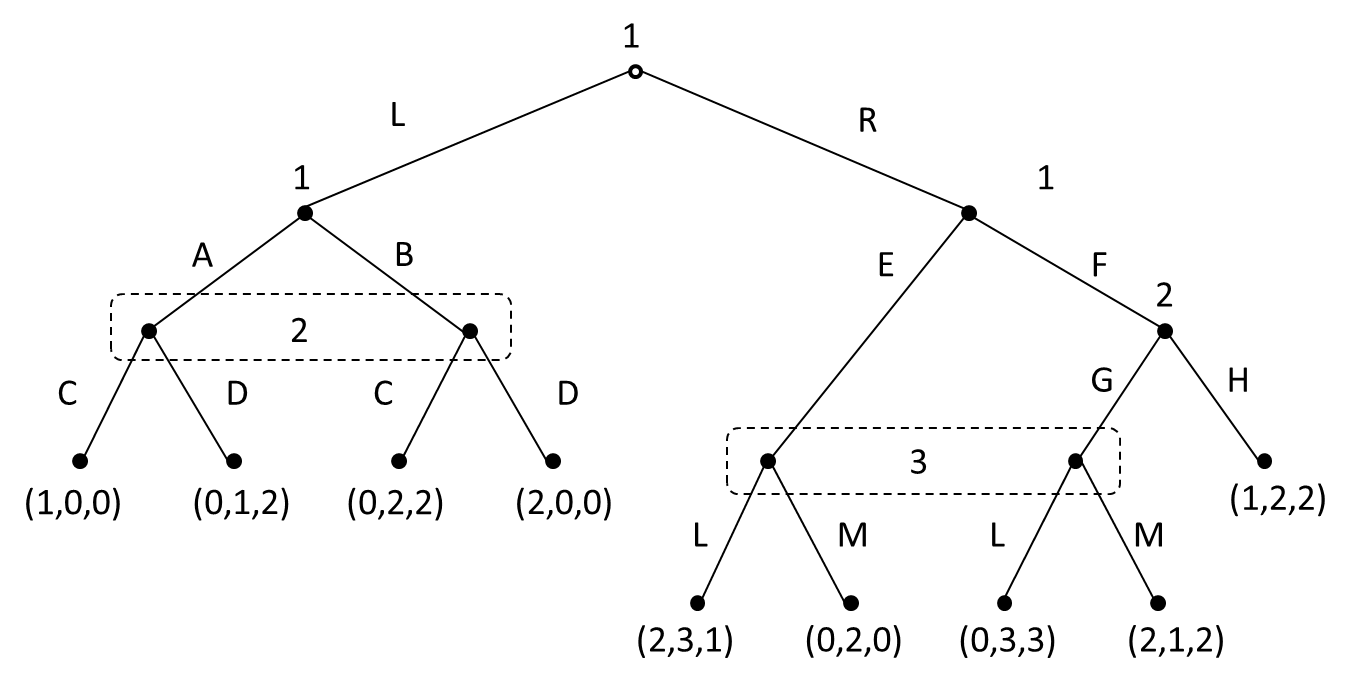}
              \caption{\label{TFigure2}\scriptsize An Extensive-Form Game from Bonanno~\cite{Bonanno (2018)}}
\end{minipage}
\end{figure}

\begin{table}[H]
{\footnotesize
$\begin{array}{l|c|c}\hline
& \text{Game in Fig.~\ref{TFigure1}} &   \text{Game in Fig.~\ref{TFigure2}} \\ \hline
\text{Exact Solution} & \beta^1_{I^2_1}(L)=\frac{2}{7},\;\beta^2_{I^1_2}(U)=\frac{1}{8} & \beta^1_{I^2_1}(A)=\frac{2}{3},\;\beta^2_{I^1_2}(C)=\frac{2}{3} \\ \hline
\text{Gambit 16.2.1 } & \beta^1_{I^2_1}(L)=0.284830,\;\beta^2_{I^1_2}(U)=0.125367 & \beta^1_{I^2_1}(A)=0.666252,\;\beta^2_{I^1_2}(C)=0.667079\\ 
\text{The System~(\ref{eqAhs1})} & \beta^1_{I^2_1}(L)=0.284789,\; \beta^2_{I^1_2}(U)=0.125384 & \beta^1_{I^2_1}(A)=0.665893,\;\beta^2_{I^1_2}(C)=0.667433 \\
  \hline
\end{array}$}
\caption{\label{nrlaqrese}\footnotesize Numerical Solutions by Gambit 16.2.1 and the System~(\ref{eqAhs1})  for the Games in Figs.~\ref{TFigure1}-\ref{TFigure2}}
\end{table}

We have adopted a standard predictor-corrector method as outlined in Eaves and Schmedders~\cite{Eaves and Schmedders (1999)}  for numerically tracing the smooth path specified by the system~(\ref{eqAhs1}) to a sequential equilibrium for the games in Figs.~\ref{TFigure1}-\ref{TFigure2}. Additionally, we utilized a tool for computing QRE based on Turocy's method~\cite{Turocy (2010)} within the Gambit software to solve these two games. 
Consider the game in Fig.~\ref{TFigure1}. The information sets consist of $I^1_1=\{\emptyset\}$, $I^2_1=\{\langle A, U\rangle, \langle A, D\rangle\}$, $I^3_1=\{\langle B, U'\rangle\}$, $I^1_2=\{\langle A\rangle\}$, and $I^2_2=\{\langle B\rangle\}$. This game has a unique sequential equilibrium denoted in the form of $((\beta^1_{I^1_1}(A),\beta^1_{I^1_1}(B)), (\beta^1_{I^2_1}(L),\beta^1_{I^2_1}(R)), (\beta^1_{I^3_1}(Y),\beta^1_{I^3_1}(N)), (\beta^2_{I^1_2}(U),\beta^2_{I^1_2}(D)), (\beta^2_{I^2_2}(U'),\beta^2_{I^2_2}(D')))= (B, (\frac{2}{7},\frac{5}{7}), Y, (\frac{1}{8},\frac{7}{8}), D')$ with $\mu^1_{I^2_1}(\langle A, U \rangle)=\frac{1}{8}$.
Consider the game in Fig.~\ref{TFigure2}. The information sets consist of $I^1_1=\{\emptyset\}$, $I^2_1=\{\langle L\rangle\}$, $I^3_1=\{\langle R\rangle\}$, $I^1_2=\{\langle L, A\rangle, \langle L, B\rangle\}$, $I^2_2=\{\langle R, F\rangle\}$, and $I^1_3=\{\langle R, E\rangle, \langle R, F, G\rangle\}$.  This game has a unique sequential equilibrium denoted in the form of $((\beta^1_{I^1_1}(L),\beta^1_{I^1_1}(R)), (\beta^1_{I^2_1}(A),\beta^1_{I^2_1}(B)),  (\beta^1_{I^3_1}(E),\beta^1_{I^3_1}(F))$, $(\beta^2_{I^1_2}(C),\beta^2_{I^1_2}(D)), (\beta^2_{I^2_2}(G),\beta^2_{I^2_2}(H)), (\beta^3_{I^1_3}(L),\beta^3_{I^1_3}(M)))=(R, (\frac{2}{3},\frac{1}{3}), E,(\frac{2}{3},\frac{1}{3}), G, L)$ with $\mu^2_{I^1_2}(\langle L, B\rangle)=\frac{1}{3}$ and $\mu^3_{I^1_3}(\langle R, F, G\rangle)=0$.  
The numerical results are reported in Table~\ref{nrlaqrese}. Table~\ref{nrlaqrese} highlights the differences between the exact solutions and the solutions generated by Gambit 16.2.1 and the system~(\ref{eqAhs1}). 
Based on the numerical results presented in Table~\ref{nrlaqrese}, it is evident that although the tool in Gambit and Turocy's~\cite{Turocy (2010)} method of the system~(\ref{eqAhs1}) have significantly advanced the development of the computation of a sequential equilibrium, they have difficulties to numerically converge with the desired accuracy to the unique sequential equilibrium for each of the games in Figs.~\ref{TFigure1}-\ref{TFigure2}. These comparisons inspire us to develop an effective and efficient differentiable path-following method to compute a sequential equilibrium.

\subsection{\label{efpm} An Entropy-Barrier Smooth Path to a Sequential Equilibrium}

The key component of a differentiable path-following method to compute a sequential equilibrium\footnote{\scriptsize A general framework for establishing a differentiable path-following method to compute a sequential equilibrium can be described as follows:
 \begin{description}
 \item[Step 1:] Constitute with an extra variable $t\in (0,1]$ an artificial extensive-form game $\Gamma(t)$ in which each player at each of his information sets solves against a given behavioral strategy profile a convex optimization problem. The game should continuously deform from a trivial game to the target game as $t$ descends from one to zero. 
 \item[Step 2:] Apply the optimality conditions to the convex optimization problems in the artificial game and acquire from the equilibrium condition an equilibrium system.
\item[Step 3:]  Check that the equilibrium system meets the following properties: The system with $t=1$ has a unique solution,  which can be easily computed, and the set of solutions to the system is bounded and the limit point of every convergent sequence of solutions to the system yields a sequential equilibrium.
\item[Step 4:] Verify that the closure of the set of solutions to the equilibrium system contains a connected component intersecting both the initial level of $t=1$ and the target level of $t=0$. 
\item[Step 5:] Ensure through the  Transversality Theorem in Eaves and Schmedders~\cite{Eaves and Schmedders (1999)} and the well-known implicit function theorem that the component is a smooth path that starts from the unique solution at $t=1$ and approaches a sequential equilibrium as $t\to 0$. 
 \item[Step 6:] Adopt a standard predictor-corrector method as outlined in Eaves and Schmedders~\cite{Eaves and Schmedders (1999)} for numerically tracing the smooth path to a sequential equilibrium. 
 \end{description}} is the existence of a smooth path that starts from an arbitrary totally mixed strategy profile and ends at a sequential equilibrium. 
This subsection will exploit Theorem~\ref{nscsethm1} to develop an entropy-barrier smooth path to a sequential equilibrium.  For $t\in (0,1]$, we constitute with  $\varpi(\beta,t)$ an entropy-barrier extensive-form game $\Gamma_E(t)$ in which player $i$ at his information set $I^j_i$ solves against a given behavioral strategy profile $\hat\beta$ the strictly convex optimization problem,
{\small
\begin{equation}\setlength{\abovedisplayskip}{1.2pt}\setlength{\belowdisplayskip}{1.2pt}\label{entbefgop1}
\begin{array}{rl}
\max\limits_{\beta^i_{I^j_i}} & (1-t)\sum\limits_{a\in A(I^j_i)}\beta^i_{I^j_i}(a)u^i((a, \varpi(\hat\beta^{-I^j_i},t))\land I^j_i)\\
& -
t\omega(I^j_i|\varpi(\hat\beta,t))\sum\limits_{a\in A(I^j_i)}\beta^i_{I^j_i}(a)(\ln\beta^i_{I^j_i}(a)-\ln\beta^{0i}_{I^j_i}(a)-1)\\

\text{s.t.} & \sum\limits_{a\in A(I^j_i)}\beta^i_{I^j_i}(a)=1.
\end{array}\end{equation}}Applying the optimality conditions to the problem~(\ref{entbefgop1}),  we acquire from the equilibrium condition of $\beta=\hat\beta$ the equilibrium  system of $\Gamma_E(t)$, 
{\small
\begin{equation}\setlength{\abovedisplayskip}{1.2pt}\setlength{\belowdisplayskip}{1.2pt}\label{entbefges1}
\begin{array}{l}
(1-t)u^i((a,\varpi(\beta^{-I^j_i},t))\land I^j_i)-t\omega(I^j_i|\varpi(\beta,t))(\ln\beta^i_{I^j_i}(a)-\ln\beta^{0i}_{I^j_i}(a))-\zeta^i_{I^j_i}=0,\\
\hspace{10cm}i\in N, j\in M_i, a\in A(I^j_i),\\

 \sum\limits_{a\in A(I^j_i)}\beta^i_{I^j_i}(a)=1,\;i\in N, j\in M_i.
 \end{array}
\end{equation}}To eliminate $\ln\beta^i_{I^j_i}(a)$ in the system~(\ref{entbefges1}), we make a variable transformation. Let  \[\setlength{\abovedisplayskip}{1.2pt}\setlength{\belowdisplayskip}{1.2pt}
\label{eqChs1}d(\nu)=\left\{\begin{array}{ll}
\exp\left(1-\frac{1}{\nu}\right) & \text{if $0<\nu$,}\\
0 & \text{otherwise.}
\end{array}\right.
\]  Let $w=(w^i_{I^j_i}(a):i\in N,j\in M_i,a\in A(I^j_i))$ and 
$\beta(w)=(\beta^i_{I^j_i}(w):i\in N,j\in M_i)$ with $\beta^i_{I^j_i}(w)=(\beta^i_{I^j_i}(w;a):a\in A(I^j_i))^\top$, where $\beta^i_{I^j_i}(w;a)=d(w^i_{I^j_i}(a))$. 
Substituting $\beta(w)$ into the system~(\ref{entbefges1}) for $\beta$, we reach the system, 
{\small
\begin{equation}\setlength{\abovedisplayskip}{1.2pt}\setlength{\belowdisplayskip}{1.2pt}\label{entbefges1a}
\begin{array}{l}
(1-t)u^i((a,\varpi(\beta^{-I^j_i}(w),t))\land I^j_i)+\omega(I^j_i|\varpi(\beta(w),t))(t/w^i_{I^j_i}(a)-t(1-\ln\beta^{0i}_{I^j_i}(a)))\\
\hspace{7.5cm}-\zeta^i_{I^j_i}=0,\;i\in N, j\in M_i, a\in A(I^j_i),\\

 \sum\limits_{a\in A(I^j_i)}\beta^i_{I^j_i}(w; a)=1,\;i\in N, j\in M_i,\\
 
0<w^i_{I^j_i}(a),\;i\in N, j\in M_i, a\in A(I^j_i).
 \end{array}
\end{equation}}
Replacing $t/w^i_{I^j_i}(a)$ by $\lambda^i_{I^j_i}(a)$ in the system~(\ref{entbefges1a}), we come to the system,
{\small
\begin{equation}\setlength{\abovedisplayskip}{1.2pt}\setlength{\belowdisplayskip}{1.2pt}\label{entbefges2}
\begin{array}{l}
(1-t)u^i((a,\varpi(\beta^{-I^j_i}(w),t))\land I^j_i)+\omega(I^j_i|\varpi(\beta(w),t))(\lambda^i_{I^j_i}(a)-t(1-\ln\beta^{0i}_{I^j_i}(a)))\\
\hspace{7.5cm}-\zeta^i_{I^j_i}=0,\;i\in N, j\in M_i, a\in A(I^j_i),\\

 \sum\limits_{a\in A(I^j_i)}\beta^i_{I^j_i}(w; a)=1,\;i\in N, j\in M_i,\\
 
 w^i_{I^j_i}(a) \lambda^i_{I^j_i}(a)=t,\;0<w^i_{I^j_i}(a),\;0< \lambda^i_{I^j_i}(a),\;i\in N, j\in M_i, a\in A(I^j_i).
 \end{array}
\end{equation}}To get rid of the inequalities and reduce the number of variables in the system~(\ref{entbefges2}), we make the transformations on variables as outlined in Cao et al.~\cite{Cao et al. (2022)}. For any given $\kappa_0>2$ and $\gamma_0>0$, let $\psi_1(\tau,\theta;\kappa_0,\gamma_0)=\left(\frac{\tau+\sqrt{\tau^2+4\gamma_0\theta}}{2}\right)^{\kappa_0}$ and 
$\psi_2(\tau,\theta;\kappa_0,\gamma_0)=\left(\frac{-\tau+\sqrt{\tau^2+4\gamma_0\theta}}{2}\right)^{\kappa_0}$. Then,
\(\psi_1(\tau,\theta;\kappa_0,\gamma_0)\psi_2(\tau,\theta;\kappa_0,\gamma_0)=(\gamma_0\theta)^{\kappa_0}\).
Substituting $w^i_{I^j_i}(z,t;a)=\psi_1(z^i_{I^j_i}(a),t^{1/\kappa_0};\kappa_0,1)$ and $\lambda^i_{I^j_i}(z,t;a)=\psi_2(z^i_{I^j_i}(a),t^{1/\kappa_0};\kappa_0, 1)$ into the system~(\ref{entbefges2}) for $w^i_{I^j_i}(a)$ and $\lambda^i_{I^j_i}(a)$, we attain an equivalent system with much fewer variables,
{\small
\begin{equation}\setlength{\abovedisplayskip}{1.2pt}\setlength{\belowdisplayskip}{1.2pt}\label{entbefges3}
\begin{array}{l}
(1-t)u^i((a,\varpi(\beta^{-I^j_i}(w(z,t)),t))\land I^j_i)+\omega(I^j_i|\varpi(\beta(w(z,t)),t))(\lambda^i_{I^j_i}(z,t; a)\\
\hspace{4cm}-t(1-\ln\beta^{0i}_{I^j_i}(a)))-\zeta^i_{I^j_i}=0,\;i\in N, j\in M_i, a\in A(I^j_i),\\

 \sum\limits_{a\in A(I^j_i)}\beta^i_{I^j_i}(w(z,t); a)=1,\;i\in N, j\in M_i.
 \end{array}
\end{equation}}Let $a^0_{I^j_i}\in A(I^j_i)$ be a given reference action. Subtracting the equation corresponding to action $a^0_{I^j_i}$ from the first group of equations in the system~(\ref{entbefges3}), we arrive at the system,
{\small
\begin{equation}\setlength{\abovedisplayskip}{1.2pt}\setlength{\belowdisplayskip}{1.2pt}\label{entbefges4}
\begin{array}{l}
(1-t)(u^i((a,\varpi(\beta^{-I^j_i}(w(z,t)),t))\land I^j_i)-u^i((a^0_{I^j_i},\varpi(\beta^{-I^j_i}(w(z,t)),t))\land I^j_i))\\
 \hspace{3cm}+\omega(I^j_i|\varpi(\beta(w(z,t)),t))(\lambda^i_{I^j_i}(z,t; a)-\lambda^i_{I^j_i}(z,t; a^0_{I^j_i})\\
\hspace{3cm}-t(\ln\beta^{0i}_{I^j_i}(a^0_{I^j_i})-\ln\beta^{0i}_{I^j_i}(a)))=0,\;i\in N, j\in M_i, a\in A(I^j_i)\backslash\{a^0_{I^j_i}\},\\

 \sum\limits_{a\in A(I^j_i)}\beta^i_{I^j_i}(w(z,t); a)=1,\;i\in N, j\in M_i.
 \end{array}
\end{equation}}

\noindent When $t=1$, the system~(\ref{entbefges4}) has a unique solution given by $z^*(1)$ with $z^{*i}_{I^j_i}(1;a)=(1-\ln\beta^{0i}_{I^j_i}(a))^{-1/\kappa_0}-(1-\ln\beta^{0i}_{I^j_i}(a))^{1/\kappa_0}$.

Let $\widetilde{\mathscr{S}}_E$ be the set of all $(z, t)$ satisfying the system~(\ref{entbefges4}) with $t>0$ and $\mathscr{S}_E$ the closure of $\widetilde{\mathscr{S}}_E$. One can get that $\mathscr{S}_E$ is a nonempty compact set. We denote by $\{(z^k,  t_k)\in\widetilde{\mathscr{S}}_E,k=1,2,\ldots\}$ a convergent sequence with $0<t_k\le 1$ and $(z^*,0)=\lim\limits_{k\to\infty}(z^k,t_k)$. Due to the equivalence between the system~(\ref{entbefges2}) and the system~(\ref{entbefges4}), it follows from Theorem~\ref{nscsethm1} that $(\beta(w(z^*,0)),\mu(\beta(w(z^*,0))))$ is a sequential equilibrium.
An application of a well-known fixed point theorem in Mas-Colell~\cite{Mas-Colell (1974)} (see also, Herings~\cite{Herings (2000)}) shows that $\mathscr{S}_E$ contains a unique connected component $\mathscr{S}_E^c$ such that $\mathscr{S}_E^c\cap(\triangle\times\{0\})\ne\emptyset$ and $\mathscr{S}_E^c\cap(\triangle\times\{1\})\ne\emptyset$.
Let $m_0=\sum\limits_{i\in N}\sum\limits_{j\in M_i}|A(I^j_i)|$. 
Let $g_0(z,t)$ denote the left-hand sides of equations in the system~(\ref{entbefges4}). Subtracting a perturbation term of $t(1-t)\alpha\in\mathbb{R}^{m_0}$ from $g_0(z,t)$, we arrive at the system,
\(g_0(z,t)-t(1-t)\alpha=0.\)
Let $g(z,t;\alpha)=g_0(z,t)-t(1-t)\alpha$. For any given $\alpha\in\mathbb{R}^{m_0}$, we denote $g_\alpha(z,t)=g(z,t;\alpha)$.
Let $\tilde {\cal G}_\alpha=\{(z,t)|g_\alpha(z,t)=0\text{ with }0< t\le 1\}$ and ${\cal G}_\alpha$ the closure of $\tilde {\cal G}_\alpha$.
One can obtain that ${\cal G}_\alpha$ is a compact set and $g(z,t;\alpha)$ is continuously differentiable on $\mathbb{R}^{m_0}\times (0,1)\times \mathbb{R}^{m_0}$ with $D_{\alpha}g(z,t;\alpha)=t(1-t)I_{m_0}$, where $I_{m_0}$ is an identity matrix. It is easy to see that, as $0<t<1$, $D_{\alpha}g(z,t;\alpha)$ is nonsingular. Furthermore, when $t=1$, $D_{z}g_0(z,1)$ is nonsingular. When $\|\alpha\|$ is sufficiently small, the continuity of $g(z,t;\alpha)$ ensures us that there is a unique connected component in ${\cal G}_\alpha$ intersecting both $\mathbb{R}^{m_0}\times\{1\}$ and $\mathbb{R}^{m_0}\times\{0\}$.
These results together with the Transversality Theorem in Eaves and Schmedders~\cite{Eaves and Schmedders (1999)} lead us to the following conclusion. For generic choice of $\alpha$ with sufficiently small $\|\alpha\|$, there exists a smooth path  $P_{\alpha}\subseteq {\cal G}_{\alpha}$ that starts from the unique solution $(z^*(1),1)$ on the level of $t=1$ and approaches a sequential equilibrium of $\Gamma$ on the target level of $t=0$. The rigorous proofs of these conclusions closely resemble those found in Cao et al.~\cite{Cao et al. (2024)}.
 
Next, we describe an alternative way to reduce the number of variables.  Multiplying $w^i_{I^j_i}(a)$ to the first group of equations in the system~(\ref{entbefges1a}), we come to the system, {\small
 \begin{equation}\label{aentbefges1}\setlength{\abovedisplayskip}{1.2pt}\setlength{\belowdisplayskip}{1.2pt}
\begin{array}{l}
(1-t)w^i_{I^j_i}(a)u^i((a,\varpi(\beta^{-I^j_i}(w),t))\land I^j_i)+t\omega(I^j_i|\varpi(\beta(w),t))(1-w^i_{I^j_i}(a)(1-\ln\beta^{0i}_{I^j_i}(a)))\\
\hspace{7cm}-w^i_{I^j_i}(a)\zeta^i_{I^j_i}=0,\;i\in N,j\in M_i, a\in A(I^j_i),\\
\sum\limits_{a\in A(I^j_i)}\beta^i_{I^j_i}(w; a) = 1,\;i\in N,j\in M_i.
\end{array}
\end{equation}}Taking the sum of the first group of equations in the system~(\ref{aentbefges1}) over $A(I^j_i)$, we get
\begin{equation}\label{aentbefges2}\setlength{\abovedisplayskip}{1.2pt}\setlength{\belowdisplayskip}{1.2pt}
\begin{array}{rl}
\zeta^i_{I^j_i}= & \frac{1}{\sum\limits_{a'\in A(I^j_i)}w^i_{I^j_i}(a')}((1-t)\sum\limits_{a'\in A(I^j_i)}w^i_{I^j_i}(a')u^i((a',\varpi(\beta^{-I^j_i}(w),t))\land I^j_i)\\
& +t\omega(I^j_i|\varpi(\beta(w),t))\sum\limits_{a'\in A(I^j_i)}(1-w^i_{I^j_i}(a')(1-\ln\beta^{0i}_{I^j_i}(a'))). \end{array} \end{equation}
Let $a^0_{I^j_i}\in A(I^j_i)$ be a given reference action. Substituting $\zeta^i_{I^j_i}$ of the system~(\ref{aentbefges2}) into the system~(\ref{aentbefges1}), we get the system with fewer variables,
 \begin{equation}\label{aentbefges3}\setlength{\abovedisplayskip}{1.2pt}\setlength{\belowdisplayskip}{1.2pt}
\begin{array}{l}
(1-t)w^i_{I^j_i}(a)(\sum\limits_{a'\in A(I^j_i)}w^i_{I^j_i}(a')(u^i((a,\varpi(\beta^{-I^j_i}(w),t))\land I^j_i)\\
\hspace{1.2cm}-u^i((a',\varpi(\beta^{-I^j_i}(w),t))\land I^j_i))+t\omega(I^j_i|\varpi(\beta(w),t))(w^i_{I^j_i}(a)\\
\hspace{1cm}\sum\limits_{a'\in A(I^j_i)}w^i_{I^j_i}(a')(\ln(\beta^{0i}_{I^j_i}(a)-\ln\beta^{0i}_{I^j_i}(a'))+\sum\limits_{a'\in A(I^j_i)}(w^i_{I^j_i}(a')-w^i_{I^j_i}(a)))=0,\\
 \hspace{7.6cm}i\in N,j\in M_i, a\in A(I^j_i)\backslash\{a^0_{I^j_i}\},\\
\sum\limits_{a\in A(I^j_i)}\beta^i_{I^j_i}(w; a) = 1,\;i\in N,j\in M_i.
\end{array}
\end{equation}
When $t=1$, the system~(\ref{aentbefges3}) has a unique solution given by $w^*(1)$ with $w^{*i}_{I^j_i}(1;a)=1/(1-\ln \beta^{0i}_{I^j_i}(a))$.

\begin{figure}[H]
    \centering
    \begin{minipage}{0.49\textwidth}
        \centering
        \includegraphics[width=1\textwidth, height=0.2\textheight]{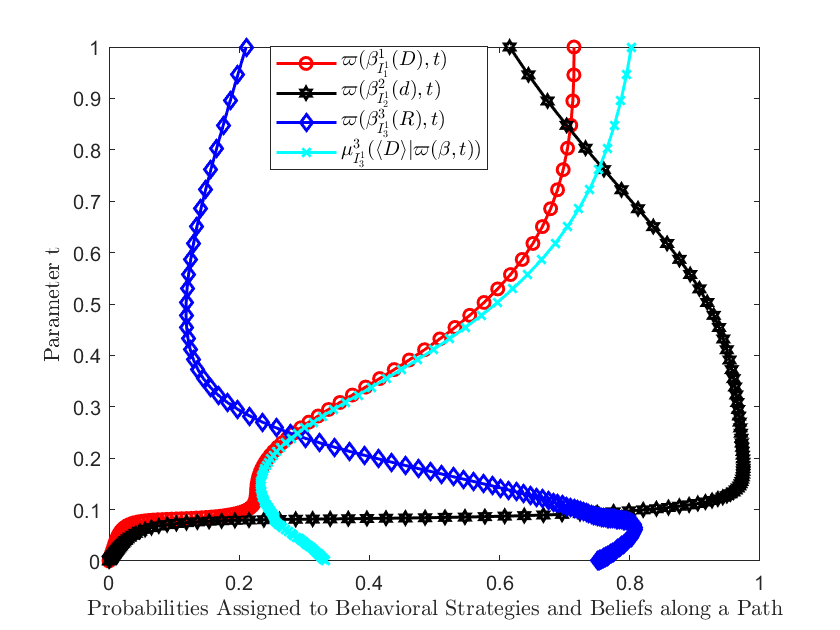}
                \caption{\label{entbTFigure3S}\scriptsize The smooth path specified by the system~(\ref{entbefges4}) for the game in Fig.~\ref{TFigure3}}
\end{minipage}\hfill
    \begin{minipage}{0.49\textwidth}
       \centering
        \includegraphics[width=1\textwidth, height=0.2\textheight]{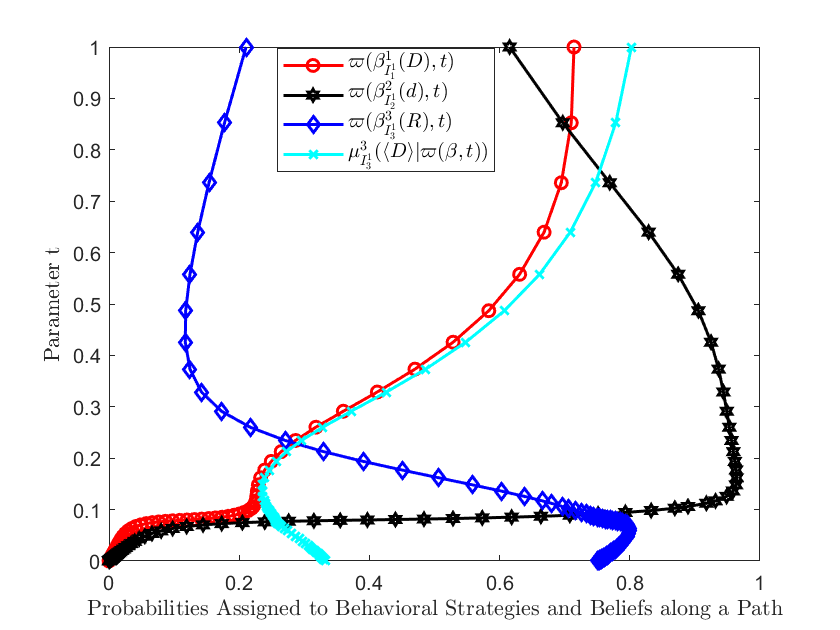}
              \caption{\label{entbTFigure3M}\scriptsize The smooth path specified by the system~(\ref{aentbefges3}) for the game in Fig.~\ref{TFigure3}}
\end{minipage}
\end{figure}

\begin{figure}[H]
    \centering
    \begin{minipage}{0.49\textwidth}
        \centering
        \includegraphics[width=1\textwidth, height=0.2\textheight]{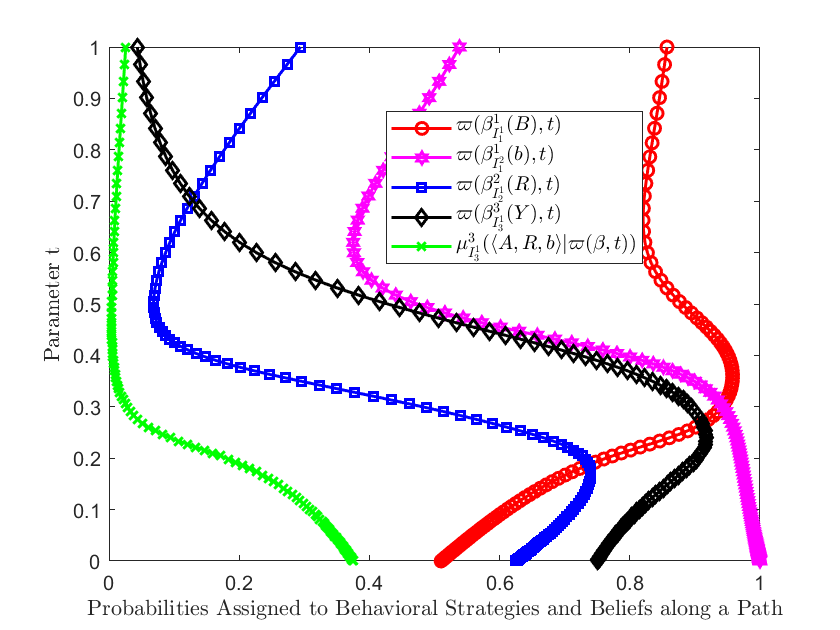}
                \caption{\label{entbTFigure4S}\scriptsize The smooth path specified by the system~(\ref{entbefges4}) for the game in Fig.~\ref{TFigure4}}
\end{minipage}\hfill
    \begin{minipage}{0.49\textwidth}
       \centering
        \includegraphics[width=1\textwidth, height=0.2\textheight]{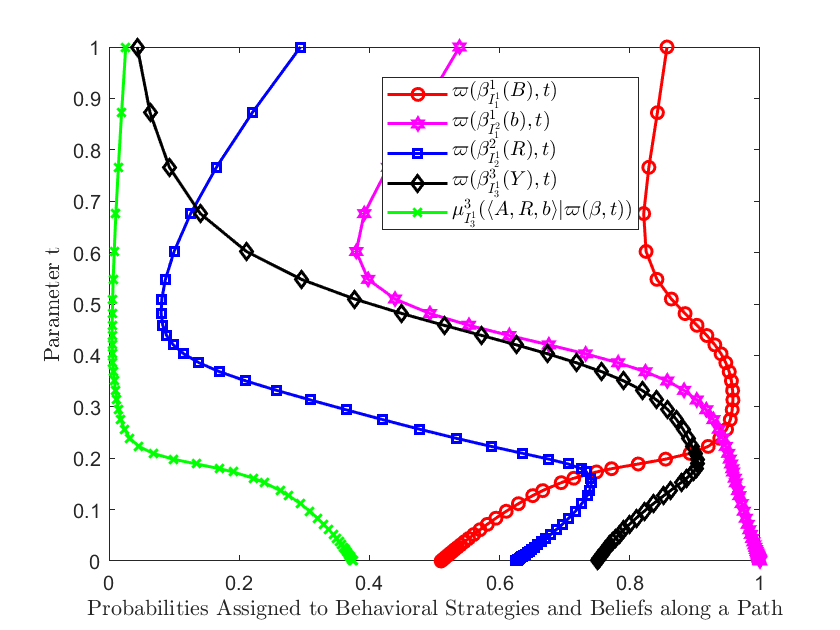}
              \caption{\label{entbTFigure4M}\scriptsize The smooth path specified by the system~(\ref{aentbefges3}) for the game in Fig.~\ref{TFigure4}}
\end{minipage}
\end{figure}

\begin{figure}[H]
    \centering
    \begin{minipage}{0.49\textwidth}
        \centering
        \includegraphics[width=1\textwidth, height=0.2\textheight]{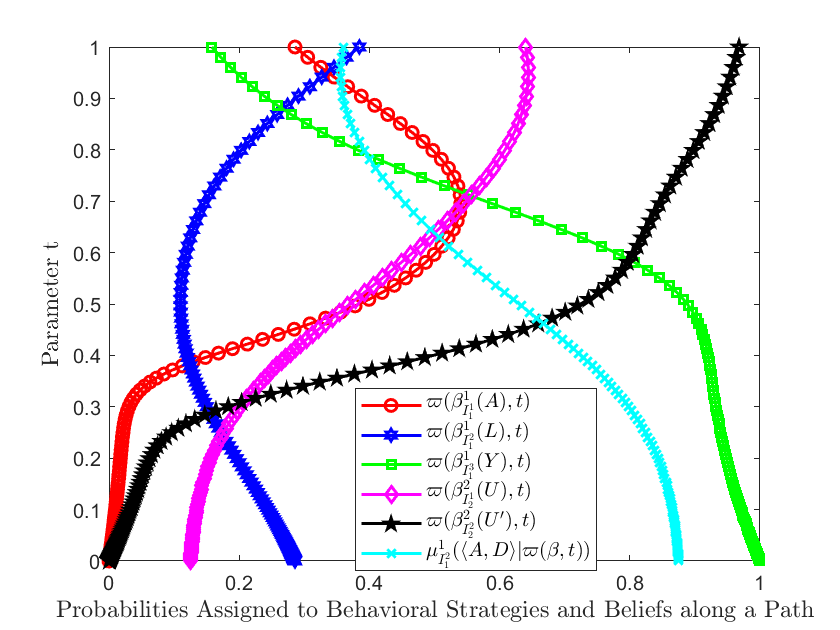}
                \caption{\label{entbTFigure1S}\scriptsize The smooth path specified by the system~(\ref{entbefges4}) for the game in Fig.~\ref{TFigure1}}
\end{minipage}\hfill
    \begin{minipage}{0.49\textwidth}
       \centering
        \includegraphics[width=1\textwidth, height=0.2\textheight]{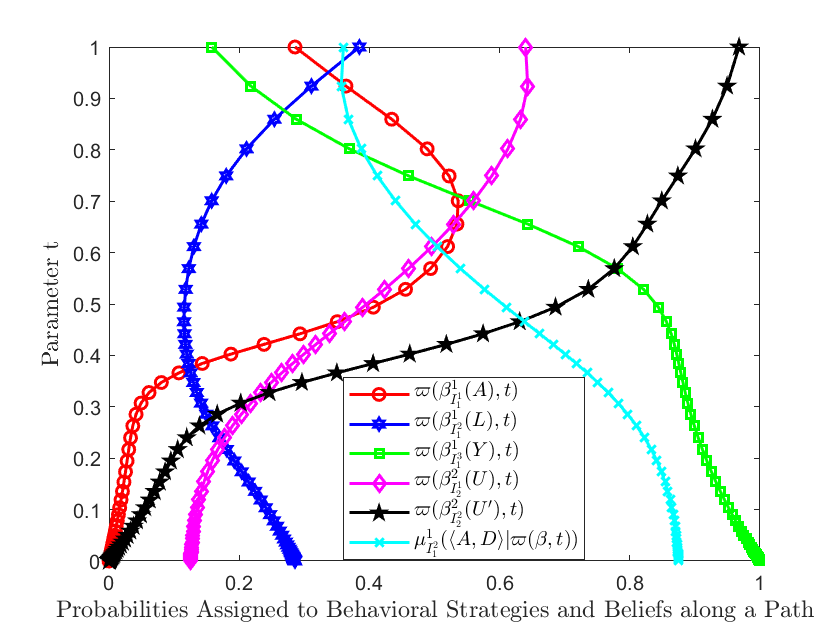}
              \caption{\label{entbTFigure1M}\scriptsize The smooth path specified by the system~(\ref{aentbefges3}) for the game in Fig.~\ref{TFigure1}}
\end{minipage}
\end{figure}

\begin{figure}[H]
    \centering
    \begin{minipage}{0.49\textwidth}
        \centering
        \includegraphics[width=1\textwidth, height=0.2\textheight]{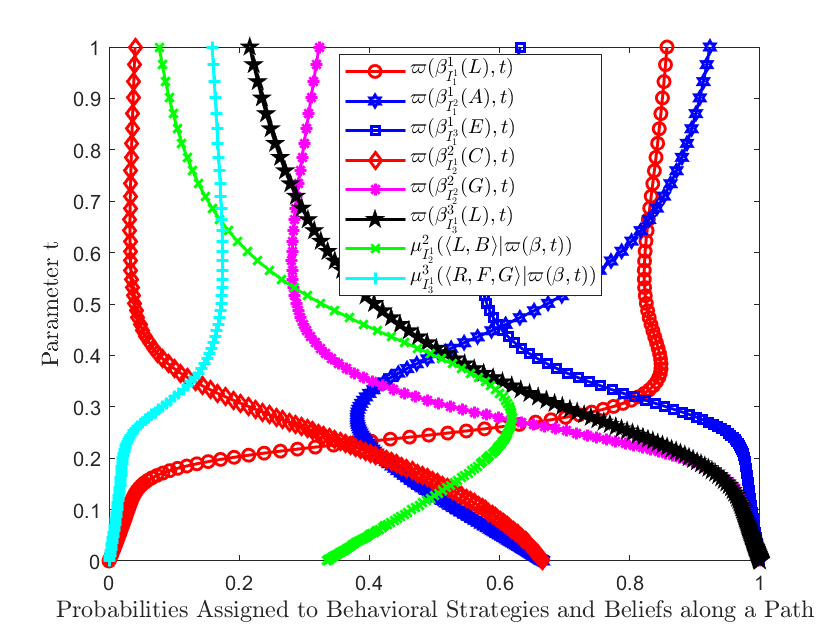}
                \caption{\label{entbTFigure2S}\scriptsize The smooth path specified by the system~(\ref{entbefges4}) for the game in Fig.~\ref{TFigure2}}
\end{minipage}\hfill
    \begin{minipage}{0.49\textwidth}
       \centering
        \includegraphics[width=1\textwidth, height=0.2\textheight]{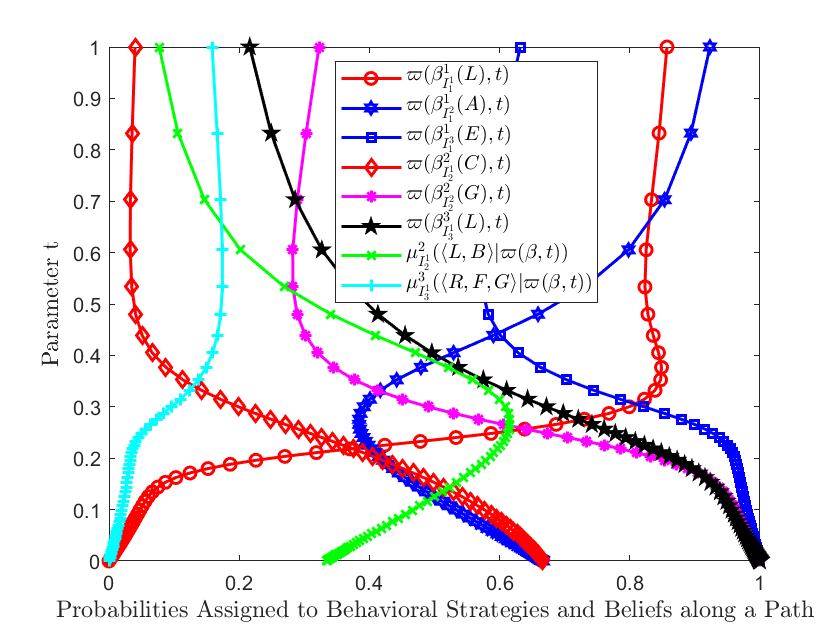}
              \caption{\label{entbTFigure2M}\scriptsize The smooth path specified by the system~(\ref{aentbefges3}) for the game in Fig.~\ref{TFigure2}}
\end{minipage}
\end{figure}

We have adopted a standard predictor-corrector method as outlined in Eaves and Schmedders~\cite{Eaves and Schmedders (1999)} for numerically tracing the smooth paths characterized by the systems~(\ref{entbefges4}) and~(\ref{aentbefges3}) to a sequential equilibrium. 
The smooth paths specified by the systems~(\ref{entbefges4}) and~(\ref{aentbefges3}) for the games in Figs.~\ref{TFigure3}-\ref{TFigure2}  are illustrated in Figs.~\ref{entbTFigure3S}-\ref{entbTFigure2M}. From these figures, it is apparent that each of these smooth paths leads to a sequential equilibrium in the respective game.  As $t<10^{-5}$, some of the numerical results for the systems~(\ref{entbefges4}) and~(\ref{aentbefges3}) for the games in Figs.~\ref{TFigure1}-\ref{TFigure2} are summarized in Table~\ref{nrlaqreseA}, which indicate the proposed methods have higher accuracy.

\begin{table}[H]
{\footnotesize
$\begin{array}{l|c|c}\hline
& \text{Game in Fig.~\ref{TFigure1}} &   \text{Game in Fig.~\ref{TFigure2}} \\ \hline
\text{Exact Solution} & \beta^1_{I^2_1}(L)=\frac{2}{7},\;\beta^2_{I^1_2}(U)=\frac{1}{8} & \beta^1_{I^2_1}(A)=\frac{2}{3},\;\beta^2_{I^1_2}(C)=\frac{2}{3} \\ \hline
\text{Gambit 16.2.1 } & \beta^1_{I^2_1}(L)=0.284830,\;\beta^2_{I^1_2}(U)=0.125367 & \beta^1_{I^2_1}(A)=0.666252,\;\beta^2_{I^1_2}(C)=0.667079\\ 
\text{Turocy's System~(\ref{eqAhs1})} & \beta^1_{I^2_1}(L)=0.284789,\; \beta^2_{I^1_2}(U)=0.125384 & \beta^1_{I^2_1}(A)=0.665893,\;\beta^2_{I^1_2}(C)=0.667433 \\
\text{The System~(\ref{entbefges4})} & \beta^1_{I^2_1}(L)=0.285710,\; \beta^2_{I^1_2}(U)=0.125000 & \beta^1_{I^2_1}(A)=0.666654,\;\beta^2_{I^1_2}(C)=0.666660 \\
\text{The System~(\ref{aentbefges3})} & \beta^1_{I^2_1}(L)=0.285710,\; \beta^2_{I^1_2}(U)=0.125000 & \beta^1_{I^2_1}(A)=0.666654,\;\beta^2_{I^1_2}(C)=0.666660 \\
\hline
\end{array}$}
\caption{\label{nrlaqreseA}\footnotesize Numerical Solutions by Gambit 16.2.1, Turocy's System~(\ref{eqAhs1}), the System~(\ref{entbefges4}), and the System~(\ref{aentbefges3})  for the Games in Figs.~\ref{TFigure1}-\ref{TFigure2}}
\end{table}

\section{Numerical Performance}

We have adapted a standard predictor-corrector method as outlined in Eaves and Schmedders~\cite{Eaves and Schmedders (1999)} for numerically tracing the smooth paths specified by the systems~(\ref{entbefges4}) and ~(\ref{aentbefges3})  in this paper. The predictor-corrector methods have been coded in MATLAB. The parameter values in the method are set as follows: the predictor step size $=$ $0.2\times10^{0.2\ln{t}}$ and accuracy of a starting point for a corrector step $=$  $0.1\times 10^{0.5\ln{t}}$. The methods terminate as soon as the criterion of $t<10^{-5}$ is met. 
As for the computational complexity, Goldberg et al.~\cite{Goldberg et al. (2013)} revealed that the widely utilized path-following method for finding the Brouwer fixpoint is PSPACE-complete to implement. This finding extends to the methods proposed in this paper and Turocy's~\cite{Turocy (2010)} work for identifying sequential equilibria. 
We have implemented the methods with the systems~(\ref{entbefges4}) and ~(\ref{aentbefges3}) to solve randomly generated extensive-form games, demonstrating their efficiency. The efficiency of the methods is measured based on the computational time and number of iterations. The computation was executed on a device with a 64-bit operating system, an x64-based Intel Xeon Gold 6426Y processor running at 2.50 GHz (with 2 processors), and 512 GB of installed RAM.

\begin{figure}[H]
    \centering
    \begin{minipage}{0.49\textwidth}
        \centering
        \includegraphics[width=0.7\textwidth, height=0.1\textheight]{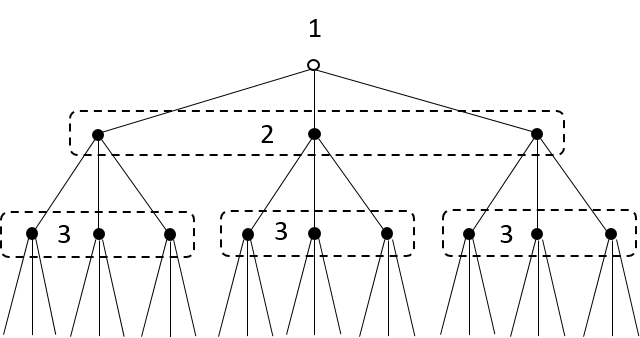}
                \caption{\label{RG1}\scriptsize Type A Extensive-Form Games with three players}
\end{minipage}\hfill
    \begin{minipage}{0.49\textwidth}
       \centering
        \includegraphics[width=0.7\textwidth, height=0.1\textheight]{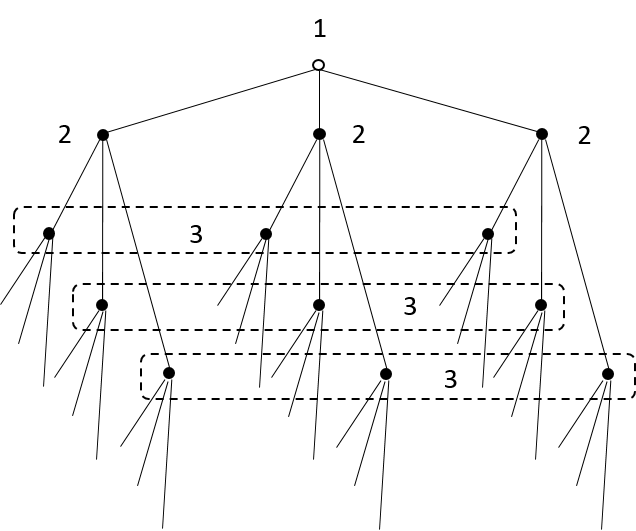}
              \caption{\label{RG2}\scriptsize Type B Extensive-Form Games with three players}
\end{minipage}
\end{figure}

We have employed our methods specified by the systems ~(\ref{entbefges4}) and ~(\ref{aentbefges3}) to solve three types of extensive-form games, namely Type A, Type B, and Type C.
When a game has three players and each player at an information set has three actions, the game trees of Type A and Type B are illustrated in Fig.~\ref{RG1} and  Fig.~\ref{RG2}, respectively.  In a game of Type A, both players $1$ and $2$ have one information set and the number of information sets for player $i$ with $i\ge 3$ equals $\prod_{k=1}^{i-2}|A(I_{k})|$. In a game of Type B, player $1$ has one information set and the number of information sets for player $i$ with $i\ge 2$ equals $|A(I_{i-1})|$.  A Type C game shares a similar structure with a Type A game, with the distinction that Type C games consist of multiple layers, denoted as $L$. Within each layer, actions are sequentially taken by players from player $1$ to player $n$.

In our computational experiments, each player has an identical number of actions in all of his information sets, and each payoff for a terminal history is an integer drawn uniformly from $-10$ to $10$ and assigned to zero with a random probability ranging from $0$ to $50\%$. For each given triple of $(n,m_i,|A(I_i)|)$ and  $(n,m_i,|A(I_i)|, L)$, 10 random games were generated and solved. We denote by EntBM1 the method with the system~(\ref{entbefges4}), and by EntBM2 the method with the system~(\ref{aentbefges3}).
The computational time (in seconds) and number of iterations are listed in Tables~\ref{Table1}-\ref{Table3}. 
The numerical results in Tables~\ref{Table1}-\ref{Table3} demonstrate that in terms of the average computation time and the average number of iterations, EntBM2 surpasses EntBM1 in efficiency when computing sequential equilibria. These findings suggest that EntBM2 is a promising approach for determining sequential equilibria, especially within extensive-form games of significant size.



\begin{table}[H]
\linespread{1} 
\scriptsize
\centering
\caption{{\footnotesize Numerical Performances of the Methods for Type A Extensive-Form Games}}\label{Table1}
\begin{tabular*}{\hsize}{@{}@{\extracolsep{\fill}}cc|cc|cc@{}}
\hline
\multicolumn{2}{c}{} & \multicolumn{2}{c}{Computational Time} & \multicolumn{2}{c}{Number of Iterations}\\
 $n,m_i,|A(I_i)|$ & & EntBM1 & EntBM2 &  EntBM1 & EntBM2 \\
\hline
  $3,(1,1,2),(2,10,10)$ & avg & 101.96 & 36.46  & 343.20 & 107  \\
 & min & 66.18 & 24.92 & 256 & 87 \\
 & max & 153.58 & 60.77 & 495 & 134 \\
 $3,(1,1,2),(2,15,15)$ & avg & 209.06 & 69.50 & 411.30 & 127\\
 & min & 139.74 & 44.90 & 317 & 98 \\
 & max & 399.77 & 123.04 & 664 & 179 \\
 $3,(1,1,2),(2,20,20)$ & avg & 311.95 & 122.89 & 419.67 & 150.56 \\
 & min & 178.39 & 91.99 & 266 & 102\\
 & max & 454.08 & 235.13 & 601 & 292 \\
 $3,(1,1,2),(2,25,25)$ & avg & 462.22 & 189.32 & 451.20 & 170.10 \\
 & min & 392.51 & 141.85 & 397 & 126 \\
 & max & 581.33 & 253.11 & 534 & 236 \\
\hline
\end{tabular*}
\end{table}

\begin{table}[H]
\linespread{1} 
\scriptsize
\centering
\caption{{\footnotesize Numerical Performances of the Methods for Type B Extensive-Form Games}}\label{Table2}
\begin{tabular*}{\hsize}{@{}@{\extracolsep{\fill}}cc|cc|cc@{}}
\hline
\multicolumn{2}{c}{} & \multicolumn{2}{c}{Computational Time} & \multicolumn{2}{c}{Number of Iterations}\\
 $n,m_i,|A(I_i)|$ & & EntBM1 & EntBM2 & EntBM1 & EntBM2 \\
\hline
 $4,(1,2,2,5),(2,2,5,3)$ & avg & 91.68 & 28.63 & 327 & 95.10 \\
 & min & 56.90 & 17.20 & 229 & 61 \\
 & max & 114.52 & 63.47 & 417 & 206 \\
 $5,(1,2,2,2,5),(2,2,2,5,3)$ & avg & 435.13 & 182.91 & 514.50 & 195.80 \\
 & min & 192.75 & 64.79 & 254 & 77 \\
 & max & 1372.36 & 607.63 & 1638 & 652 \\
 $6,(1,2,2,2,2,5),(2,2,2,2,5,3)$ & avg & 886.20 & 494.53 & 392.20 & 160.50\\
 & min & 438.73 & 184.50 & 219 & 90 \\
 & max & 1836.46 & 791.29 & 894 & 304 \\
 $7,(1,2,2,2,2,2,5),(2,2,2,2,2,5,3)$ & avg & 3784.51 & 2225.68 & 548.80 & 342.67 \\
 & min & 1341.35 & 580.65 & 262 & 97 \\
 & max & 6372.71 & 8207.21 & 989 & 1403 \\
\hline
\end{tabular*}
\end{table}

\begin{table}[H]
\linespread{1} 
\scriptsize
\centering
\caption{{\footnotesize Numerical Performances of the Methods for Type C Extensive-Form Games}}\label{Table3}
\begin{tabular*}{\hsize}{@{}@{\extracolsep{\fill}}cc|cc|cc@{}}
\hline
\multicolumn{2}{c}{} & \multicolumn{2}{c}{Computational Time} & \multicolumn{2}{c}{Number of Iterations}\\
 $n,m_i,|A(I_i)|,L$ & & EntBM1 & EntBM2 & EntBM1 & EntBM2\\
\hline
 $2,(11,21),(2,2),3$ & avg & 226.53 & 65.36  & 396.10 & 103.10 \\
 & min & 133.23 & 49.57 & 302 & 86 \\
 & max & 313.99 & 97.25 & 494 & 139\\
 $2,(43,85),(2,2),4$ & avg & 2092.80 & 680.53 & 584.60 & 173.40\\
 & min & 1431.84 & 386.33 & 474 & 124 \\
 & max & 3210.67 & 1282.56 & 869 & 287 \\
 $2,(4,10),(3,3),2$ & avg & 1052.67 & 303.99  & 503 & 137.30 \\
 & min & 540.89 & 137 & 356 & 87 \\
 & max & 2180.07 & 660.44 & 975 & 364 \\
 $3,(5,9,18),(2,2,2),2$ & avg & 830.72 & 227.02 & 440.50 & 114.80 \\
 & min & 528.31 & 135.94 & 342 & 81 \\
 & max & 1464.11 & 466.81 & 740 & 221 \\
\hline
\end{tabular*}
\end{table}

\section{Conclusion}

We have developed in this paper a characterization of sequential equilibrium in extensive-form games through the construction of an $\varepsilon$-perfect $\gamma$-sequential equilibrium with local sequential rationality. As a result of this characterization, we have attained polynomial systems that serve as a necessary and sufficient condition to ascertain whether a consistent assessment is a sequential equilibrium and whether an assessment is an $\varepsilon$-perfect $\gamma$-sequential equilibrium. As demonstrated in Examples~\ref{esedAexm1}-\ref{esedAexm2}, it is much easier to work with the characterization than with that of Kreps and Wilson~\cite{Kreps and Wilson (1982)} to analytically find all the sequential equilibria for a small extensive-form game. To boost the applications of sequential equilibrium, we have exploited the characterization to devise a differentiable path-following method to compute sequential equilibria. Numerical experiments corroborate the effectiveness and efficiency of the method.




\begin{thebibliography}{00}
\setlength{\itemsep}{-2mm}
\setstretch{1.0}
\small

\bibitem{Adao and Temzelides (1998)} Adao, B., Temzelides, T.: Sequential equilibrium and competition in a Diamond–Dybvig banking model. Review of Economic Dynamics 1(4), 859-877 (1998)

\bibitem{Battigalli (1996)} Battigalli, P.: Strategic independence and perfect Bayesian equilibria. Journal of Economic Theory 70(1), 201-234 (1996)

\bibitem{Battigalli and Dufwenberg (2009)} Battigalli, P., Dufwenberg, M.: Dynamic psychological games. Journal of Economic Theory 144(1), 1-35 (2009)

\bibitem{Besanko and Spulber (1990)} Besanko, D., Spulber, D.F.: Are treble damages neutral? Sequential equilibrium and private antitrust enforcement. The American Economic Review 870-887 (1990)

\bibitem{Blume and Zame (1994)} Blume, L.E., Zame, W.R.: The algebraic geometry of perfect and sequential equilibrium. Econometrica 62, 783-794 (1994)

\bibitem{Bonanno (2018)} Bonanno, G.:  Game Theory: Volume 1: Basic Concepts. CreateSpace Independent Publishing Platform (2018)

\bibitem{Browder (1960)} Browder, F.E.:  On continuity of fixed points under deformations of continuous mappings. Summa Brasiliensis Mathematicae 4, 183-191 (1960)

\bibitem{Camerer and Weigelt (1988)} Camerer, C.,  Weigelt, K.: Experimental tests of a sequential equilibrium reputation model. Econometrica 1-36 (1988)

\bibitem{Cao et al. (2024)} Cao, Y., Chen, Y., Dang, C.: A Differentiable Path-Following Method with a Compact Formulation to Compute Proper Equilibria. INFORMS Journal on Computing 36(2), 377-396 (2024)

\bibitem{Cao et al. (2022)} Cao, Y., Dang, C., Xiao, Z.: A differentiable path-following method to compute subgame perfect equilibria in stationary strategies in robust stochastic games and its applications. European Journal of Operational Research 298(3), 1032-1050 (2022)

\bibitem{Chen and Dang (2021)} Chen, Y., Dang, C.: A differentiable homotopy method to compute perfect equilibria. Mathematical Programming 185, 77-109 (2021)

\bibitem{Chakrabarti and Topolyan (2016)} Chakrabarti, S.K., Topolyan, I.: An extensive form-based proof of the existence of sequential equilibrium. Economic Theory Bulletin 4(2), 355-365 (2016)

\bibitem{Choi et al. (2008)} Choi, S., Gale, D., Kariv, S.: Sequential equilibrium in monotone games: A theory-based analysis of experimental data. Journal of Economic Theory 143(1), 302-330 (2008)

\bibitem{van Damme (1991)} van Damme, E.: Stability and Perfection of Nash Equilibria (Vol. 339). Berlin: Springer-Verlag (1991)

\bibitem{Dang et al. (2022)} Dang, C., Herings, P. J. J., Li, P.: An interior-point differentiable path-following method to compute stationary equilibria in stochastic games. INFORMS Journal on Computing 34(3), 1403-1418 (2022)

\bibitem{Eaves and Schmedders (1999)} Eaves, B.C., Schmedders, K.: General equilibrium models and homotopy methods. Journal of Economic Dynamics \& Control 23, 1249-1279 (1999)

\bibitem{Eibelshauser et al. (2023)} Eibelsh$\ddot{a}$user, S., Klockmann, V., Poensgen, D., von Schenk, A.: The logarithmic stochastic tracing procedure: A homotopy method to compute stationary equilibria of stochastic games. INFORMS Journal on Computing 35(6), 1511-1526 (2023)

\bibitem{Fiacco (1983)}  Fiacco, A.V.: Introduction to Sensitivity and Stability Analysis in Nonlinear Programming. Academic Press (1983)

\bibitem{Fudenberg and Kreps (1995)} Fudenberg, D., Kreps, D.M.: Learning in extensive-form games I. Self-confirming equilibria. Games and Economic Behavior 8(1), 20-55 (1995)

\bibitem{Fudenberg and Tirole (1991)} Fudenberg, D., Tirole, J.: Perfect Bayesian equilibrium and sequential equilibrium. Journal of Economic Theory 53(2), 236-260 (1991)

\bibitem{Garcia and Zangwill (1979)} Garcia, C. B., Zangwill, W. I.: Determining all solutions to certain systems of nonlinear equations. Mathematics of Operations Research 4(1), 1-14 (1979)

\bibitem{Garcia and Zangwill (1981)} Garcia, C.B.,  Zangwill, W.I.: Pathways to Solutions, Fixed Points, and Equilibria. Series in Comp. Math., Prentice-Hall, New Jersey (1981)

\bibitem{Geanakoplos et al. (1989)} Geanakoplos, J., Pearce, D., Stacchetti, E.: Psychological games and sequential rationality. Games and Economic Behavior 1(1), 60-79 (1989)

\bibitem{Goldberg et al. (2013)} Goldberg, P. W., Papadimitriou, C. H., Savani, R.: The complexity of the homotopy method, equilibrium selection, and Lemke-Howson solutions. ACM Transactions on Economics and Computation (TEAC) 1(2), 1-25 (2013)



\bibitem{Halpern (2009)} Halpern, J.Y.: A nonstandard characterization of sequential equilibrium, perfect equilibrium, and proper equilibrium. International Journal of Game Theory 38(1), 37-49 (2009)

\bibitem{Hansen et al. (2010)} Hansen, K. A., Miltersen, P. B., S{\o}rensen, T. B.: The computational complexity of trembling hand perfection and other equilibrium refinements. In Algorithmic Game Theory: Third International Symposium, SAGT 2010, Athens, Greece, October 18-20, 2010. Proceedings 3 (pp. 198-209). Springer Berlin Heidelberg.

\bibitem{Harsanyi (1975)} Harsanyi, J.C.: The tracing procedure: a Bayesian approach to defining a solution for n-person noncooperative games. International Journal of Game Theory 4(2), 61-94 (1975)

\bibitem{Harsanyi and Selten (1988)} Harsanyi, J.C., Selten, R.: A General Theory of Equilibrium Selection in Games. MIT Press, Cambridge (1988) 

\bibitem{Herings (2000)} Herings, P.J.J.: Two simple proofs of the feasibility of the linear tracing procedure. Economic Theory 15, 485-490 (2000)

\bibitem{Herings and Peeters (2001)} Herings, P.J.J., Peeters, R.J.A.P.: A differentiable homotopy method to compute Nash equilibria of $n$-person games. Economic Theory 18, 159-185 (2001)

\bibitem{Huang and Werner (2000)} Huang, K.X., Werner, J.: Asset price bubbles in Arrow-Debreu and sequential equilibrium. Economic Theory 15(2), 253-278 (2000)

\bibitem{Kohlberg and Mertens (1986)} Kohlberg, E., Mertens, J.F.: On the strategic stability of equilibria. Econometrica 54(5), 1003-1037 (1986)

\bibitem{Kohlberg and Reny (1997)} Kohlberg, E., Reny, P.J.: Independence on relative probability spaces and consistent assessments in game trees. Journal of Economic Theory 75(2), 280-313 (1997)

\bibitem{Kreps (1990)} Kreps, D.M.: A Course in Microeconomic Theory. Princeton University Press (1990)

\bibitem{Kreps and Wilson (1982)} Kreps, D.M., Wilson, R.: Sequential equilibria. Econometrica 50(4), 863-894 (1982)


\bibitem{Laan and Talman (1982)}  Laan, G. van der, Talman, A.J.J.:  On the computation of fixed points in the product space of unit simplices and an application to noncooperative N-person games, Mathematics of Operations Research 7, 1-13 (1982)

\bibitem{Manelli (1996)} Manelli, A.M.: Cheap talk and sequential equilibria in signaling games. Econometrica 917-942 (1996)

\bibitem{Maschler et al. (2020)} Maschler, M., Zamir, S., Solan, E.: Game Theory. Cambridge University Press (2020)

\bibitem{Mas-Colell (1974)} Mas-Colell, A.: A note on a theorem of F. Browder. Mathematical Programming 6, 229-233 (1974)

\bibitem{Mas-Colell et al. (1995)} Mas-Colell, A., Whinston, M.D., Green, J.R.: Microeconomic Theory (Vol. 1). New York: Oxford University Press (1995)

\bibitem{Mckelvey and Palfey (1998)} McKelvey, R.D., Palfrey, T.R.: Quantal response equilibria for extensive form games. Experimental Economics 1, 9-41 (1998)



\bibitem{Miltersen and Sorensen (2010)} Miltersen, P.B., Sorensen, T.B.: Computing a quasi-perfect equilibrium of a two-player game. Economic Theory 42, 175-192 (2010)

\bibitem{Myerson (1986)} Myerson, R.B.: Multistage games with communication. Econometrica 54(2), 323-358 (1986)


\bibitem{Nash (1951)} Nash, Jr.J.F.: Noncooperative games. Annals of Math 54, 289-295 (1951)

\bibitem{Osborne and Rubinstein (1994)} Osborne, M.J., Rubinstein, A.: A Course in Game Theory. MIT Press (1994)


\bibitem{Selten (1975)} Selten, R.: Reexamination of the perfectness concept for equilibrium points in extensive games. International Journal Game Theory 4, 25-55 (1975)

\bibitem{von Stengel (1996)} von Stengel, B.: Efficient computation of behavior strategies. Games and Economic Behavior 14(2), 220-246 (1996)

\bibitem{von Stengel et al. (2002)} von Stengel, B., van den Elzen, A., Talman, D.: Computing normal form perfect equilibria for extensive two‐person games. Econometrica 70(2), 693-715 (2002)

\bibitem{Turocy (2010)} Turocy, T.L.: Computing sequential equilibria using agent quantal response equilibria. Economic Theory 42, 255-269 (2010)

\bibitem{Watson (2023)} Watson, J.: Partially Constructed Sequential Equilibrium (Note). Available at SSRN 4656890 (2023)

\bibitem{Wilson (1972)} Wilson, R.: Computing equilibria of two-person games from the extensive form. Management Science 18(7), 448-460 (1972)

\end{thebibliography}

\end{document}



\title{\Large Supplemental Appendix: A Characterization of Sequential Equilibrium through $\varepsilon$-Perfect $\gamma$-Sequential Equilibrium with Local Sequential Rationality and Its Computation}

\author{Yiyin Cao\textsuperscript{\ref{fnote1}} and Chuangyin Dang\textsuperscript{\ref{fnote2}}\footnote{Corresponding Author. Email: mecdang@cityu.edu.hk}}

\date{}
\maketitle             

\footnotetext[1]{School of Management, Xi'an Jiaotong University, Xi'an, China, yiyincao2-c@my.cityu.edu.hk\label{fnote1}}
\footnotetext[2]{Department of Systems Engineering, City University of Hong Kong, Kowloon, Hong Kong  \label{fnote2}}

\maketitle

\begin{figure}[H]
    \centering
    \begin{minipage}{0.49\textwidth}
        \centering
        \includegraphics[width=0.8\textwidth, height=0.15\textheight]{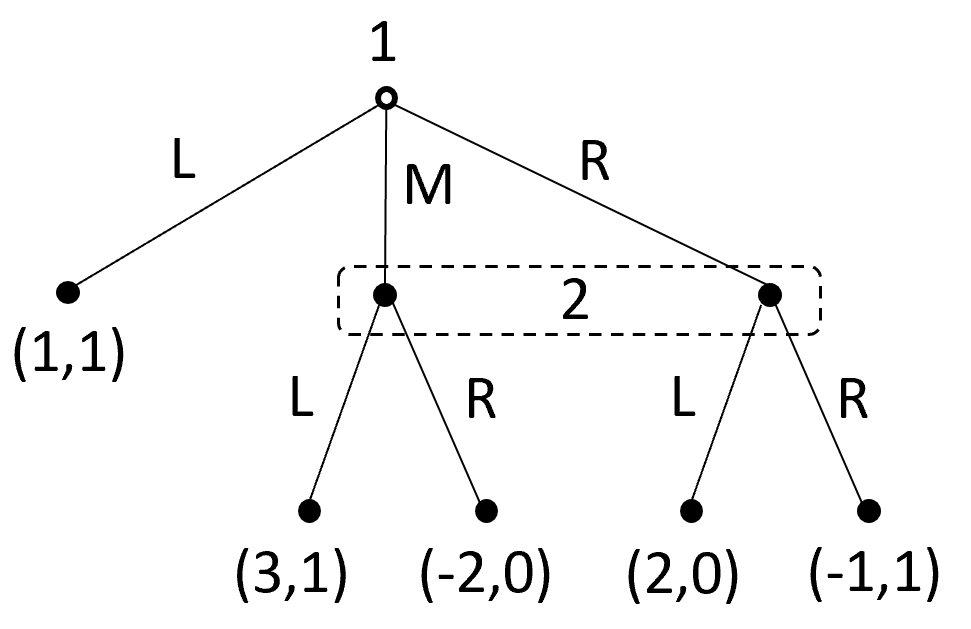}
                \caption{\label{TFigure5}\scriptsize An Extensive-Form Game from Osborne and Rubinstein~\cite{Osborne and Rubinstein (1994)}}
\end{minipage}\hfill
    \begin{minipage}{0.49\textwidth}
        \centering
        \includegraphics[width=0.8\textwidth, height=0.15\textheight]{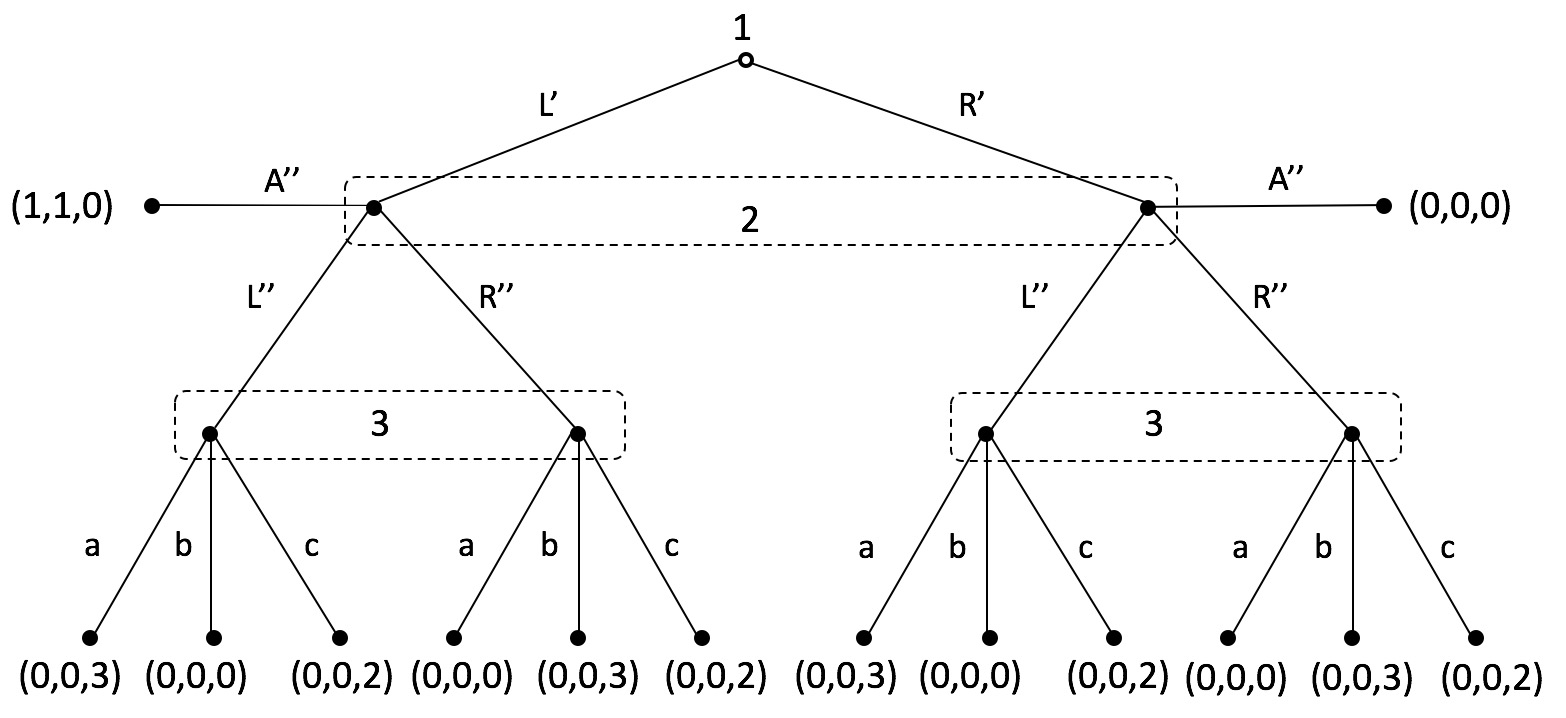}
\caption{\label{TFigure6}\scriptsize An Extensive-Form Game from Battigalli~\cite{Battigalli (1996)}}\end{minipage}
 \end{figure}

This appendix illustrates with two examples how one can employ Definition~3 to analytically find all the sequential equilibria for small extensive-form games. 
The more complicated game in Fig.~\ref{TFigure6} is deliberately chosen to demonstrate the effectiveness of Definition~3. 

\begin{example}\label{esedBexm1} {\em Consider the game in Fig.~\ref{TFigure5}. The information sets consist of $I^1_1=\{\emptyset\}$ and $I^1_2=\{\langle M\rangle, \langle R\rangle\}$. We denote by $(\beta^*,\mu^*)$ a sequential equilibrium in the form of $\beta^*=((\beta^{*1}_{I^1_1}(L),\beta^{*1}_{I^1_1}(M), \beta^{*1}_{I^1_1}(R)), (\beta^{*2}_{I^1_2}(L),\beta^{*2}_{I^1_2}(R)))$.
 Let  $(\varpi(\beta(\gamma), \eta(\varepsilon)),\mu(\gamma, \varepsilon))$ be an $\varepsilon$-perfect $\gamma$-sequential equilibrium. For simplicity, we drop $\gamma$ and $\varepsilon$ from $\beta(\gamma)$ and $\eta(\varepsilon)$ and $(\gamma, \varepsilon)$ from $\mu(\gamma, \varepsilon)$ in the rest of this example.
 The conditional expected payoffs at $(\varpi(\beta, \eta),\mu)$ on $I^j_i$ are given by
{\footnotesize 
\[\setlength{\abovedisplayskip}{1.2pt} 
\setlength{\belowdisplayskip}{1.2pt}\begin{array}{l}
u^1(L,\varpi(\beta^{-I^1_1},\eta), \mu|I^1_1)=1,\\
 u^1(M,\varpi(\beta^{-I^1_1},\eta), \mu|I^1_1)=3\varpi(\beta^2_{I^1_2}( L),\eta) -2\varpi(\beta^2_{I^1_2}( R),\eta),\\

u^1(R,\varpi(\beta^{-I^1_1},\eta), \mu|I^1_1)=2\varpi(\beta^2_{I^1_2}( L),\eta)-\varpi(\beta^2_{I^1_2}( R),\eta),\\

u^2(L,\varpi(\beta^{-I^1_2},\eta), \mu|I^1_2)=\mu^2_{I^1_2}(\langle M\rangle),\;

 u^2(R,\varpi(\beta^{-I^1_2},\eta), \mu|I^1_2)=\mu^2_{I^1_2}(\langle R\rangle),
\end{array}\]
where $\mu^2_{I^1_2}( \langle M\rangle)=\frac{\varpi(\beta^1_{I^1_1}( M),\eta)}{\varpi(\beta^1_{I^1_1}( M),\eta)+\varpi(\beta^1_{I^1_1}( R),\eta)}$ and $\mu^2_{I^1_2}(\langle R\rangle)=\frac{\varpi(\beta^1_{I^1_1}( R),\eta)}{\varpi(\beta^1_{I^1_1}( M),\eta)+\varpi(\beta^1_{I^1_1}( R),\eta)}$.}

\noindent {\bf Case (1)}. Suppose that $u^2(L,\varpi(\beta^{-I^1_2},\eta), \mu|I^1_2)-u^2(R,\varpi(\beta^{-I^1_2},\eta), \mu|I^1_2)>\gamma$. Then, $\beta^2_{I^1_2}( R)=0$ and $\mu^2_{I^1_2}(\langle M\rangle)>\frac{1}{2}+\frac{1}{2}\gamma$. Thus, {\footnotesize $u^1(M,\varpi(\beta^{-I^1_1},\eta), \mu|I^1_1)-u^1(L,\varpi(\beta^{-I^1_1},\eta), \mu|I^1_1)>\gamma$ and $u^1(M,\varpi(\beta^{-I^1_1},\eta), \mu|I^1_1)-u^1(R,\varpi(\beta^{-I^1_1},\eta), \mu|I^1_1)>\gamma$.} Consequently, $\beta^1_{I^1_1}( L)=0$ and $\beta^1_{I^1_1}( R)=0$. The game has a  sequential equilibrium given by $(M, L)$ with $\mu^2_{I^1_2}(\langle M\rangle)=1$.

\noindent {\bf Case (2)}. Suppose that $u^2(R,\varpi(\beta^{-I^1_2},\eta), \mu|I^1_2)-u^2(L,\varpi(\beta^{-I^1_2},\eta), \mu|I^1_2)>\gamma$. Then, $\beta^2_{I^1_2}( L)=0$ and $\mu^2_{I^1_2}(\gamma,\varepsilon; \langle R\rangle)>\frac{1}{2}+\frac{1}{2}\gamma$. Thus, {\footnotesize $u^1(L,\varpi(\beta^{-I^1_1},\eta), \mu|I^1_1)-u^1(M,\varpi(\beta^{-I^1_1},\eta), \mu|I^1_1)>\gamma$ and $u^1(L,\varpi(\beta^{-I^1_1},\eta), \mu|I^1_1)-u^1(R,\varpi(\beta^{-I^1_1},\eta), \mu|I^1_1)>\gamma$.} Consequently, $\beta^1_{I^1_1}( M)=0$ and $\beta^1_{I^1_1}( R)=0$. It follows from $\mu^2_{I^1_2}(\gamma,\varepsilon; \langle R\rangle)>\frac{1}{2}+\frac{1}{2}\gamma$ that $\eta^1_{I^1_1}(\varepsilon; R)>\frac{1+\gamma}{1-\gamma}\eta^1_{I^1_1}(\varepsilon; M)$. 
The game has a class of sequential equilibria given by $(L, R)$ with $\mu^2_{I^1_2}(\langle R\rangle)>\frac{1}{2}$.

\noindent {\bf Case (3)}. Suppose that $|u^2(R,\varpi(\beta^{-I^1_2},\eta), \mu|I^1_2)-u^2(L,\varpi(\beta^{-I^1_2},\eta), \mu|I^1_2)|\le \gamma$. Then, {\small $|\mu^2_{I^1_2}(\gamma,\varepsilon; \langle R\rangle)-\frac{1}{2}|\le\frac{1}{2}\gamma$ and $|\varpi(\beta^1_{I^1_1}( R),\eta)-\varpi(\beta^1_{I^1_1}( M),\eta)|\le \gamma(\varpi(\beta^1_{I^1_1}( M),\eta)+\varpi(\beta^1_{I^1_1}( R),\eta))$.} Thus, \begin{equation}\label{esedBeq1}\setlength{\abovedisplayskip}{1.2pt} 
\setlength{\belowdisplayskip}{1.2pt}\frac{1-\gamma}{1+\gamma}\varpi(\beta^1_{I^1_1}( M),\eta)\le \varpi(\beta^1_{I^1_1}( R),\eta)\le\frac{1+\gamma}{1-\gamma}\varpi(\beta^1_{I^1_1}( M),\eta).\end{equation}

\noindent {\bf (a)}. Assume that $u^1(L,\varpi(\beta^{-I^1_1},\eta), \mu|I^1_1)-u^1(R,\varpi(\beta^{-I^1_1},\eta), \mu|I^1_1)>\gamma$. Then, $\varpi(\beta^2_{I^1_2}( L),\eta)<\frac{2}{3}-\frac{1}{3}\gamma$ and $\beta^1_{I^1_1}( R)=0$, which together with Eq.~(\ref{esedBeq1}) implies $\beta^1_{I^1_1}( M)=0$. Thus either {\small 
$u^1(L,\varpi(\beta^{-I^1_1},\eta), \mu|I^1_1)-u^1(M,\varpi(\beta^{-I^1_1},\eta), \mu|I^1_1)>\gamma$ or $|u^1(L,\varpi(\beta^{-I^1_1},\eta), \mu|I^1_1)-u^1(M,\varpi(\beta^{-I^1_1},\eta), \mu|I^1_1)|\le \gamma$.}

\noindent {\bf (i)}. Consider the scenario that {\small $u^1(L,\varpi(\beta^{-I^1_1},\eta), \mu|I^1_1)-u^1(M,\varpi(\beta^{-I^1_1},\eta), \mu|I^1_1)>\gamma$.} Then,  $\varpi(\beta^2_{I^1_2}( L),\eta)<\frac{3}{5}-\frac{1}{5}\gamma$.  The game has a class of sequential equilibria given by $(L, (\beta^2_{I^1_2}(L), 1-\beta^2_{I^1_2}(L)))$ with $\beta^2_{I^1_2}(L)<\frac{3}{5}$ and $\mu^2_{I^1_2}( \langle R\rangle)=\frac{1}{2}$.
 
 \noindent {\bf (ii)}. Consider the scenario that {\footnotesize $|u^1(L,\varpi(\beta^{-I^1_1},\eta), \mu|I^1_1)-u^1(M,\varpi(\beta^{-I^1_1},\eta), \mu|I^1_1)|\le\gamma$.} Then,  $|\varpi(\beta^2_{I^1_2}( L),\eta)-\frac{3}{5}|\le\frac{1}{5}\gamma$. The game has a sequential equilibrium given by $(L, (\frac{3}{5}, \frac{2}{5}))$ with $\mu^2_{I^1_2}( \langle R\rangle)=\frac{1}{2}$.

\noindent {\bf (b)}. Assume that $u^1(R,\varpi(\beta^{-I^1_1},\eta), \mu|I^1_1)-u^1(L,\varpi(\beta^{-I^1_1},\eta), \mu|I^1_1)>\gamma$. Then, $\beta^1_{I^1_1}( L)=0$ and it follows from Eq.~(\ref{esedBeq1}) that  
$|u^1(M,\varpi(\beta^{-I^1_1},\eta), \mu|I^1_1)-u^1(R,\varpi(\beta^{-I^1_1},\eta), \mu|I^1_1)|\le \gamma$. Thus, $|\varpi(\beta^2_{I^1_2}( L),\eta)-\frac{1}{2}|\le\frac{1}{2}\gamma$. Therefore, $u^1(L,\varpi(\beta^{-I^1_1},\eta), \mu|I^1_1)-u^1(R,\varpi(\beta^{-I^1_1},\eta), \mu|I^1_1)>\gamma$. A contradiction occurs and the assumption is excluded.

\noindent {\bf (c)}. Assume that $|u^1(L,\varpi(\beta^{-I^1_1},\eta), \mu|I^1_1)-u^1(R,\varpi(\beta^{-I^1_1},\eta), \mu|I^1_1)|\le \gamma$. Then, $|\varpi(\beta^2_{I^1_2}( L),\eta)-\frac{2}{3}|\le\frac{1}{3}\gamma$. Thus, {\small
$u^1(M,\varpi(\beta^{-I^1_1},\eta), \mu|I^1_1)-u^1(R,\varpi(\beta^{-I^1_1},\\ \eta), \mu|I^1_1)>\gamma$ and 
$u^1(M,\varpi(\beta^{-I^1_1},\eta), \mu|I^1_1)-u^1(L,\varpi(\beta^{-I^1_1},\eta), \mu>\gamma$.}
 Consequently, $\beta^1_{I^1_1}( L)=0$ and
$\beta^1_{I^1_1}( R)=0$, which contradicts Eq.~(\ref{esedBeq1}).
The assumption is excluded.

The cases (1)-(3) together show that the game has three types of sequential equilibria given by \newline
(1).  $(M, L)$ with $\mu^2_{I^1_2}(\langle M\rangle)=1$.\newline
 (2).  $(L, R)$ with $\mu^2_{I^1_2}(\langle R\rangle)>\frac{1}{2}$. \newline
(3).  $(L, (\beta^2_{I^1_2}(L), 1-\beta^2_{I^1_2}(L)))$ with $\beta^2_{I^1_2}(L)\le\frac{3}{5}$ and $\mu^2_{I^1_2}( \langle R\rangle)=\frac{1}{2}$.
}
\end{example}
\begin{example} \label{esedBexm2} {\em Consider the game in Fig.~\ref{TFigure6}. The information sets consist of $I^1_1=\{\emptyset\}$, $I^1_2=\{\langle L'\rangle, \langle R'\rangle\}$, $I^1_3=\{\langle L', L''\rangle, \langle L', R''\rangle\}$, and $I^2_3=\{\langle R', L''\rangle, \langle R', R''\rangle\}$. We denote by $(\beta^*,\mu^*)$ a sequential equilibrium in the form of $\beta^*=((\beta^{*1}_{I^1_1}(L'),\beta^{*1}_{I^1_1}(R'))$, $(\beta^{*2}_{I^1_2}(A''), \beta^{*2}_{I^1_2}(L''),\beta^{*2}_{I^1_2}(R'')), (\beta^{*3}_{I^1_3}(a),\beta^{*3}_{I^1_3}(b), \beta^{*3}_{I^1_3}(c)),  (\beta^{*3}_{I^2_3}(a),\beta^{*3}_{I^2_3}(b), \beta^{*3}_{I^2_3}(c)))$.
 Let  $(\varpi(\beta(\gamma), \eta(\varepsilon))$, $\mu(\gamma, \varepsilon))$ be an $\varepsilon$-perfect $\gamma$-sequential equilibrium. The conditional expected payoffs at $(\varpi(\beta, \eta),\mu(\gamma, \varepsilon))$ on $I^j_i$ are given by
{\footnotesize
\[\setlength{\abovedisplayskip}{1.2pt} 
\setlength{\belowdisplayskip}{1.2pt}
\begin{array}{l}
u^1(L', \varpi(\beta^{-I^1_1},\eta),\mu|I^1_1)=\varpi(\beta^2_{I^1_2}( A''),\eta),\;

u^1(R', \varpi(\beta^{-I^1_1},\eta),\mu|I^1_1)=0,\\

u^2(A'', \varpi(\beta^{-I^1_2},\eta),\mu|I^1_2)=\mu^2_{I^1_2}(\gamma,\varepsilon;\langle L'\rangle),\;

u^2(L'', \varpi(\beta^{-I^1_2},\eta),\mu|I^1_2)=0,\\

u^2(R'', \varpi(\beta^{-I^1_2},\eta),\mu|I^1_2)=0,\\

u^3(a, \varpi(\beta^{-I^1_3},\eta),\mu|I^1_3)=3\mu^3_{I^1_3}(\gamma,\varepsilon;\langle L', L''\rangle),\;

u^3(b, \varpi(\beta^{-I^1_3},\eta),\mu|I^1_3)=3\mu^3_{I^1_3}(\gamma,\varepsilon;\langle L', R''\rangle),\\

u^3(c, \varpi(\beta^{-I^1_3},\eta),\mu|I^1_3)=2,\\

u^3(a, \varpi(\beta^{-I^2_3},\eta),\mu|I^2_3)=3\mu^3_{I^2_3}(\gamma,\varepsilon;\langle R', L''\rangle),\;

u^3(b, \varpi(\beta^{-I^2_3},\eta),\mu|I^2_3)=3\mu^3_{I^2_3}(\gamma,\varepsilon;\langle R', R''\rangle),\\

u^3(c, \varpi(\beta^{-I^2_3},\eta),\mu|I^2_3)=2,
\end{array}\]
where $\mu^2_{I^1_2}(\langle L'\rangle)=\varpi(\beta^1_{I^1_1}( L'),\eta)$,
$\mu^3_{I^1_3}(\langle L',L''\rangle)=\frac{\varpi(\beta^2_{I^1_2}( L''),\eta)}{\varpi(\beta^2_{I^1_2}( L''),\eta)+\varpi(\beta^2_{I^1_2}( R''),\eta)}$,
$\mu^3_{I^1_3}(\langle L',R''\rangle)=\frac{\varpi(\beta^2_{I^1_2}( R''),\eta)}{\varpi(\beta^2_{I^1_2}( L''),\eta)+\varpi(\beta^2_{I^1_2}( R''),\eta)}$, $\mu^3_{I^2_3}(\langle R',L''\rangle)=\mu^3_{I^1_3}(\langle L',L''\rangle)$, and
$\mu^3_{I^1_3}(\langle R',R''\rangle)=\mu^3_{I^1_3}(\langle L',R''\rangle)$.}

\noindent {\bf Case (1)}. Suppose that $u^2(A'', \varpi(\beta^{-I^1_2},\eta),\mu|I^1_2)-u^2(L'', \varpi(\beta^{-I^1_2},\eta),\mu|I^1_2)>\gamma$. Then, $\beta^2_{I^1_2}( L'')=\beta^2_{I^1_2}( R'')=0$. Thus, $u^1(L', \varpi(\beta^{-I^1_1},\eta),\mu|I^1_1)-u^1(R', \varpi(\beta^{-I^1_1},\eta),\mu|I^1_1)>\gamma$ and consequently, $\beta^1_{I^1_1}( R')=0$. 

\noindent {\bf (a)}. Assume that $u^3(a, \varpi(\beta^{-I^1_3},\eta),\mu|I^1_3)-u^3(b, \varpi(\beta^{-I^1_3},\eta),\mu|I^1_3)>\gamma$. Then, $\beta^3_{I^1_3}( b)=\beta^3_{I^2_3}( b)=0$ and $\frac{1}{2}-\frac{1}{6}\gamma>\mu^3_{I^1_3}(\gamma,\varepsilon;\langle L',R''\rangle)$. Thus, $\frac{3-\gamma}{3+\gamma}\eta^2_{I^1_2}(\varepsilon; L'')>\eta^2_{I^1_2}(\varepsilon; R'')$.

\noindent {\bf (i)}. Consider the scenario that $u^3(a, \varpi(\beta^{-I^1_3},\eta),\mu|I^1_3)-u^3(c, \varpi(\beta^{-I^1_3},\eta),\mu|I^1_3)>\gamma$. Then, $\beta^1_{I^1_3}( c)=\beta^1_{I^2_3}( c)=0$ and $\mu^3_{I^1_3}(\gamma,\varepsilon;\langle L',R''\rangle)<\frac{1}{3}-\frac{1}{3}\gamma$. Thus, $\frac{1-\gamma}{2+\gamma}\eta^2_{I^1_2}(\varepsilon; L'')>\eta^2_{I^1_2}(\varepsilon; R'')$. The game has a class of sequential equilibria given by $(L', A'', a, a)$ with $\mu^3_{I^2_3}(\langle R',R''\rangle)=\mu^3_{I^1_3}(\langle L',R''\rangle)<\frac{1}{3}$. 

\noindent {\bf (ii)}. Consider the scenario that $u^3(c, \varpi(\beta^{-I^1_3},\eta),\mu|I^1_3)-u^3(a, \varpi(\beta^{-I^1_3},\eta),\mu|I^1_3)>\gamma$. Then, $\beta^1_{I^1_3}( a)=\beta^1_{I^2_3}( a)=0$ and $\mu^3_{I^1_3}(\gamma,\varepsilon;\langle L',R''\rangle)>\frac{1}{3}-\frac{1}{3}\gamma$. Thus, $\frac{1-\gamma}{2+\gamma}\eta^2_{I^1_2}(\varepsilon; L'')<\eta^2_{I^1_2}(\varepsilon; R'')$. The game has a class of sequential equilibria given by $(L', A'', c, c)$ with $\frac{1}{3}<\mu^3_{I^2_3}(\langle R',R''\rangle)=\mu^3_{I^1_3}(\langle L',R''\rangle)<\frac{1}{2}$. 

\noindent {\bf (iii)}. Consider the scenario that $|u^3(a, \varpi(\beta^{-I^1_3},\eta),\mu|I^1_3)-u^3(c, \varpi(\beta^{-I^1_3},\eta),\mu|I^1_3)|\le\gamma$. Then, $|\mu^3_{I^1_3}(\gamma,\varepsilon;\langle L',R''\rangle)-\frac{1}{3}|\le \frac{1}{3}\gamma$. Thus, $\frac{1-\gamma}{2+\gamma}\eta^2_{I^1_2}(\varepsilon; L'')\le\eta^2_{I^1_2}(\varepsilon; R'')\le\frac{1+\gamma}{2-\gamma}\eta^2_{I^1_2}(\varepsilon; L'')$. The game has a class of sequential equilibria given by $(L', A'', (\beta^3_{I^1_3}(a), 0, 1-\beta^3_{I^1_3}(a)), (\beta^3_{I^2_3}(a), 0, 1-\beta^3_{I^2_3}(a)))$ with $\mu^3_{I^2_3}(\langle R',R''\rangle)=\mu^3_{I^1_3}(\langle L',R''\rangle)=\frac{1}{3}$. 

\noindent {\bf (b)}. Assume that $u^3(b, \varpi(\beta^{-I^1_3},\eta),\mu|I^1_3)-u^3(a, \varpi(\beta^{-I^1_3},\eta),\mu|I^1_3)>\gamma$. Then, $\beta^3_{I^1_3}( a)=\beta^3_{I^2_3}( a)=0$ and $\frac{1}{2}-\frac{1}{6}\gamma>\mu^3_{I^1_3}(\gamma,\varepsilon;\langle L',L''\rangle)$. Thus, $\frac{3-\gamma}{3+\gamma}\eta^2_{I^1_2}(\varepsilon; R'')>\eta^2_{I^1_2}(\varepsilon; L'')$.

\noindent {\bf (i)}. Consider the scenario that $u^3(b, \varpi(\beta^{-I^1_3},\eta),\mu|I^1_3)-u^3(c, \varpi(\beta^{-I^1_3},\eta),\mu|I^1_3)>\gamma$. Then, $\beta^1_{I^1_3}( c)=\beta^1_{I^2_3}( c)=0$ and $\mu^3_{I^1_3}(\gamma,\varepsilon;\langle L',L''\rangle)<\frac{1}{3}-\frac{1}{3}\gamma$. Thus, $\frac{1-\gamma}{2+\gamma}\eta^2_{I^1_2}(\varepsilon; R'')>\eta^2_{I^1_2}(\varepsilon; L'')$. The game has a class of sequential equilibria given by $(L', A'', b, b)$ with $\mu^3_{I^2_3}(\langle R',L''\rangle)=\mu^3_{I^1_3}(\langle L',L''\rangle)<\frac{1}{3}$. 

\noindent {\bf (ii)}. Consider the scenario that $u^3(c, \varpi(\beta^{-I^1_3},\eta),\mu|I^1_3)-u^3(b, \varpi(\beta^{-I^1_3},\eta),\mu|I^1_3)>\gamma$. Then, $\beta^1_{I^1_3}( b)=\beta^1_{I^2_3}( b)=0$ and $\mu^3_{I^1_3}(\gamma,\varepsilon;\langle L',L''\rangle)>\frac{1}{3}-\frac{1}{3}\gamma$. Thus, $\frac{1-\gamma}{2+\gamma}\eta^2_{I^1_2}(\varepsilon; R'')<\eta^2_{I^1_2}(\varepsilon; L'')$. The game has a class of sequential equilibria given by $(L', A'', c, c)$ with $\frac{1}{3}<\mu^3_{I^2_3}(\langle R',L''\rangle)=\mu^3_{I^1_3}(\langle L',L''\rangle)<\frac{1}{2}$. 

\noindent {\bf (iii)}. Consider the scenario that $|u^3(b, \varpi(\beta^{-I^1_3},\eta),\mu|I^1_3)-u^3(c, \varpi(\beta^{-I^1_3},\eta),\mu|I^1_3)|\le\gamma$. Then, $|\mu^3_{I^1_3}(\gamma,\varepsilon;\langle L',L''\rangle)-\frac{1}{3}|\le \frac{1}{3}\gamma$. Thus, $\frac{1-\gamma}{2+\gamma}\eta^2_{I^1_2}(\varepsilon; R'')\le\eta^2_{I^1_2}(\varepsilon; L'')\le\frac{1+\gamma}{2-\gamma}\eta^2_{I^1_2}(\varepsilon; R'')$. The game has a class of sequential equilibria given by $(L', A'', (0,\beta^3_{I^1_3}(b),  1-\beta^3_{I^1_3}(b)), (0,\beta^3_{I^2_3}(b), 1-\beta^3_{I^2_3}(b)))$ with $\mu^3_{I^2_3}(\langle R',L''\rangle)=\mu^3_{I^1_3}(\langle L',L''\rangle)=\frac{1}{3}$. 

\noindent {\bf (c)}. Assume that $|u^3(a, \varpi(\beta^{-I^1_3},\eta),\mu|I^1_3)-u^3(b, \varpi(\beta^{-I^1_3},\eta),\mu|I^1_3)|\le\gamma$. Then, $|\mu^3_{I^1_3}(\gamma,\varepsilon;\langle L',L''\rangle)-\frac{1}{2}|\le\frac{1}{6}\gamma$. Thus, $\frac{3-\gamma}{3+\gamma}\eta^2_{I^1_2}(\varepsilon; R'')\le \eta^2_{I^1_2}(\varepsilon; L'')\le \frac{3+\gamma}{3-\gamma}\eta^2_{I^1_2}(\varepsilon; R'')$.
Therefore, $u^3(c, \varpi(\beta^{-I^1_3},\eta),\mu|I^1_3)-u^3(a, \varpi(\beta^{-I^1_3},\eta),\mu|I^1_3)>\gamma$ and $u^3(c, \varpi(\beta^{-I^1_3},\eta),\mu|I^1_3)-u^3(b, \varpi(\beta^{-I^1_3},\eta),\mu|I^1_3)>\gamma$. Consequently, $\beta^1_{I^1_3}( a)=\beta^1_{I^2_3}( a)=0$ and $\beta^1_{I^1_3}( b)=\beta^1_{I^2_3}( b)=0$. The game has a sequential equilibrium given by $(L', A'', c, c)$ with $\mu^3_{I^2_3}(\langle R',L''\rangle)=\mu^3_{I^1_3}(\langle L',L''\rangle)=\frac{1}{2}$. 

\noindent {\bf Case (2)}. Suppose that $0\le u^2(A'', \varpi(\beta^{-I^1_2},\eta),\mu|I^1_2)-u^2(L'', \varpi(\beta^{-I^1_2},\eta),\mu|I^1_2)\le \gamma$. Then, $\varpi(\beta^1_{I^1_1}( L'),\eta)\le\gamma$. Thus, $0\le u^1(L', \varpi(\beta^{-I^1_1},\eta),\mu|I^1_1)-u^1(R', \varpi(\beta^{-I^1_1},\eta),\mu|I^1_1)\le\gamma$. Therefore, $\varpi(\beta^2_{I^1_2}( A''),\eta)\le\gamma$.

\noindent {\bf (a)}. Assume that $u^3(a, \varpi(\beta^{-I^1_3},\eta),\mu|I^1_3)-u^3(b, \varpi(\beta^{-I^1_3},\eta),\mu|I^1_3)>\gamma$. Then, $\beta^3_{I^1_3}( b)=\beta^3_{I^2_3}( b)=0$ and $\frac{1}{2}-\frac{1}{6}\gamma>\mu^3_{I^1_3}(\gamma,\varepsilon;\langle L',R''\rangle)$, which yields that $\frac{3-\gamma}{3+\gamma}\varpi(\beta^2_{I^1_2}( L''),\eta)>\varpi(\beta^2_{I^1_2}( R''),\eta)$.

\noindent {\bf (i)}. Consider the scenario that $u^3(a, \varpi(\beta^{-I^1_3},\eta),\mu|I^1_3)-u^3(c, \varpi(\beta^{-I^1_3},\eta),\mu|I^1_3)>\gamma$. Then, $\beta^1_{I^1_3}( c)=\beta^1_{I^2_3}( c)=0$ and $\mu^3_{I^1_3}(\gamma,\varepsilon;\langle L',R''\rangle)<\frac{1}{3}-\frac{1}{3}\gamma$. Thus, $\frac{1-\gamma}{2+\gamma}\varpi(\beta^2_{I^1_2}( L''),\eta)>\varpi(\beta^2_{I^1_2}( R''),\eta)$. The game has a class of sequential equilibria given by $(R', (0,1-\beta^2_{I^1_2}(R''), \beta^2_{I^1_2}(R'')), a, a)$ with $\beta^2_{I^1_2}(R'')<\frac{1}{3}$ and $\mu^3_{I^2_3}(\langle R',R''\rangle)=\mu^3_{I^1_3}(\langle L',R''\rangle)<\frac{1}{3}$. 

\noindent {\bf (ii)}. Consider the scenario that $u^3(c, \varpi(\beta^{-I^1_3},\eta),\mu|I^1_3)-u^3(a, \varpi(\beta^{-I^1_3},\eta),\mu|I^1_3)>\gamma$. Then, $\beta^1_{I^1_3}( a)=\beta^1_{I^2_3}( a)=0$ and $\mu^3_{I^1_3}(\gamma,\varepsilon;\langle L',R''\rangle)>\frac{1}{3}-\frac{1}{3}\gamma$. Thus, $\frac{1-\gamma}{2+\gamma}\varpi(\beta^2_{I^1_2}( L''),\eta)<\varpi(\beta^2_{I^1_2}( R''),\eta)$. The game has a class of sequential equilibria given by $(R', (0,1-\beta^2_{I^1_2}(R''), \beta^2_{I^1_2}(R'')), c, c)$ with $\frac{1}{2}>\beta^2_{I^1_2}(R'')>\frac{1}{3}$  and $\frac{1}{3}<\mu^3_{I^2_3}(\langle R',R''\rangle)=\mu^3_{I^1_3}(\langle L',R''\rangle)<\frac{1}{2}$. 

\noindent {\bf (iii)}. Consider the scenario that $|u^3(a, \varpi(\beta^{-I^1_3},\eta),\mu|I^1_3)-u^3(c, \varpi(\beta^{-I^1_3},\eta),\mu|I^1_3)|\le\gamma$. Then, $|\mu^3_{I^1_3}(\gamma,\varepsilon;\langle L',R''\rangle)-\frac{1}{3}|\le \frac{1}{3}\gamma$, which yields that $\frac{1-\gamma}{2+\gamma}\varpi(\beta^2_{I^1_2}( L''),\eta)\le\varpi(\beta^2_{I^1_2}( R''),\eta)\le\frac{1+\gamma}{2-\gamma}\varpi(\beta^2_{I^1_2}( L''),\eta)$. The game has a class of sequential equilibria given by $(R', (0,\frac{2}{3}, \frac{1}{3}), (\beta^3_{I^2_3}(a), 0, 1-\beta^3_{I^2_3}(a)), (\beta^3_{I^2_3}(a), 0, 1-\beta^3_{I^2_3}(a)))$ with $\mu^3_{I^2_3}(\langle R',R''\rangle)=\mu^3_{I^1_3}(\langle L',R''\rangle)=\frac{1}{3}$. 

\noindent {\bf (b)}. Assume that $u^3(b, \varpi(\beta^{-I^1_3},\eta),\mu|I^1_3)-u^3(a, \varpi(\beta^{-I^1_3},\eta),\mu|I^1_3)>\gamma$. Then, $\beta^3_{I^1_3}( a)=\beta^3_{I^2_3}( a)=0$ and $\frac{1}{2}-\frac{1}{6}\gamma>\mu^3_{I^1_3}(\gamma,\varepsilon;\langle L',L''\rangle)$. Thus, $\frac{3-\gamma}{3+\gamma}\varpi(\beta^2_{I^1_2}( R''),\eta)>\varpi(\beta^2_{I^1_2}( L''),\eta)$.

\noindent {\bf (i)}. Consider the scenario that $u^3(b, \varpi(\beta^{-I^1_3},\eta),\mu|I^1_3)-u^3(c, \varpi(\beta^{-I^1_3},\eta),\mu|I^1_3)>\gamma$. Then, $\beta^1_{I^1_3}( c)=\beta^1_{I^2_3}( c)=0$ and $\mu^3_{I^1_3}(\gamma,\varepsilon;\langle L',L''\rangle)<\frac{1}{3}-\frac{1}{3}\gamma$. Thus, $\frac{1-\gamma}{2+\gamma}\varpi(\beta^2_{I^1_2}( R''),\eta)>\varpi(\beta^2_{I^1_2}( L''),\eta)$. The game has a class of sequential equilibria given by $(R', (\beta^2_{I^1_2}(L''), 1-\beta^2_{I^1_2}(L'')), b, b)$ with $\beta^2_{I^1_2}(L'')<\frac{1}{3}$ and $\mu^3_{I^2_3}(\langle R',L''\rangle)=\mu^3_{I^1_3}(\langle L',L''\rangle)<\frac{1}{3}$. 

\noindent {\bf (ii)}. Consider the scenario that $u^3(c, \varpi(\beta^{-I^1_3},\eta),\mu|I^1_3)-u^3(b, \varpi(\beta^{-I^1_3},\eta),\mu|I^1_3)>\gamma$. Then, $\beta^1_{I^1_3}( b)=\beta^1_{I^2_3}( b)=0$ and $\mu^3_{I^1_3}(\gamma,\varepsilon;\langle L',L''\rangle)>\frac{1}{3}-\frac{1}{3}\gamma$. Thus, $\frac{1-\gamma}{2+\gamma}\varpi(\beta^2_{I^1_2}( R''),\eta)<\varpi(\beta^2_{I^1_2}( L''),\eta)$. The game has a class of sequential equilibria given by $(R', (\beta^2_{I^1_2}(L''), 1-\beta^2_{I^1_2}(L'')), c, c)$ with $\frac{1}{2}>\beta^2_{I^1_2}(L'')>\frac{1}{3}$ and $\frac{1}{3}<\mu^3_{I^2_3}(\langle R',L''\rangle)=\mu^3_{I^1_3}(\langle L',L''\rangle)<\frac{1}{2}$. 

\noindent {\bf (iii)}. Consider the scenario that $|u^3(b, \varpi(\beta^{-I^1_3},\eta),\mu|I^1_3)-u^3(c, \varpi(\beta^{-I^1_3},\eta),\mu|I^1_3)|\le\gamma$. Then, $|\mu^3_{I^1_3}(\gamma,\varepsilon;\langle L',L''\rangle)-\frac{1}{3}|\le \frac{1}{3}\gamma$. Thus, $\frac{1-\gamma}{2+\gamma}\varpi(\beta^2_{I^1_2}( R''),\eta)\le\varpi(\beta^2_{I^1_2}( L''),\eta)\le\frac{1+\gamma}{2-\gamma}\varpi(\beta^2_{I^1_2}( R''),\eta)$. The game has a class of sequential equilibria given by $(R', (0,\frac{1}{3},\frac{2}{3}), (0,\beta^3_{I^1_3}(b), 1-\beta^3_{I^1_3}(b)), (0, \beta^3_{I^2_3}(b), 1-\beta^3_{I^2_3}(b)))$ with $\mu^3_{I^2_3}(\langle R',L''\rangle)=\mu^3_{I^1_3}(\langle L',L''\rangle)=\frac{1}{3}$. 

\noindent {\bf (c)}. Assume that $|u^3(a, \varpi(\beta^{-I^1_3},\eta),\mu|I^1_3)-u^3(b, \varpi(\beta^{-I^1_3},\eta),\mu|I^1_3)|\le\gamma$. Then, $|\mu^3_{I^1_3}(\gamma,\varepsilon;\langle L',L''\rangle)-\frac{1}{2}|\le\frac{1}{6}\gamma$. Thus, \[\setlength{\abovedisplayskip}{1.2pt} 
\setlength{\belowdisplayskip}{1.2pt}
\frac{3-\gamma}{3+\gamma}\varpi(\beta^2_{I^1_2}( R''),\eta)\le \varpi(\beta^2_{I^1_2}( L''),\eta)\le \frac{3+\gamma}{3-\gamma}\varpi(\beta^2_{I^1_2}( R''),\eta).\]
Therefore, $u^3(c, \varpi(\beta^{-I^1_3},\eta),\mu|I^1_3)-u^3(a, \varpi(\beta^{-I^1_3},\eta),\mu|I^1_3)>\gamma$ and $u^3(c, \varpi(\beta^{-I^1_3},\eta),\mu|I^1_3)-u^3(b, \varpi(\beta^{-I^1_3},\eta),\mu|I^1_3)>\gamma$. Consequently, $\beta^1_{I^1_3}( a)=\beta^1_{I^2_3}( a)=0$ and $\beta^1_{I^1_3}( b)=\beta^1_{I^2_3}( b)=0$. The game has a sequential equilibrium given by $(R', (0,\frac{1}{2},\frac{1}{2}), c, c)$ with $\mu^3_{I^2_3}(\langle R',L''\rangle)=\mu^3_{I^1_3}(\langle L',L''\rangle)=\frac{1}{2}$. 

The cases (1) and (2) together ensure us that the game has fourteen types of sequential equilibria given by\\
{\scriptsize
(1). $(L', A'', a, a)$ with $\mu^3_{I^2_3}(\langle R',R''\rangle)=\mu^3_{I^1_3}(\langle L',R''\rangle)<\frac{1}{3}$. \\
(2). $(L', A'', c, c)$ with $\frac{1}{3}<\mu^3_{I^2_3}(\langle R',R''\rangle)=\mu^3_{I^1_3}(\langle L',R''\rangle)<\frac{1}{2}$. \\
(3). $(L', A'', (\beta^3_{I^1_3}(a), 0, 1-\beta^3_{I^1_3}(a)), (\beta^3_{I^2_3}(a), 0, 1-\beta^3_{I^2_3}(a)))$ with $\mu^3_{I^2_3}(\langle R',R''\rangle)=\mu^3_{I^1_3}(\langle L',R''\rangle)=\frac{1}{3}$. \\
(4).  $(L', A'', b, b)$ with $\mu^3_{I^2_3}(\langle R',L''\rangle)=\mu^3_{I^1_3}(\langle L',L''\rangle)<\frac{1}{3}$. \\
(5). $(L', A'', c, c)$ with $\frac{1}{3}<\mu^3_{I^2_3}(\langle R',L''\rangle)=\mu^3_{I^1_3}(\langle L',L''\rangle)<\frac{1}{2}$. \\
(6). $(L', A'', (0,\beta^3_{I^1_3}(b),  1-\beta^3_{I^1_3}(b)), (0,\beta^3_{I^2_3}(b), 1-\beta^3_{I^2_3}(b)))$ with $\mu^3_{I^2_3}(\langle R',L''\rangle)=\mu^3_{I^1_3}(\langle L',L''\rangle)=\frac{1}{3}$. \\
(7). $(L', A'', c, c)$ with $\mu^3_{I^2_3}(\langle R',L''\rangle)=\mu^3_{I^1_3}(\langle L',L''\rangle)=\frac{1}{2}$. \\
(8). $(R', (0,1-\beta^2_{I^1_2}(R''), \beta^2_{I^1_2}(R'')), a, a)$ with $\beta^2_{I^1_2}(R'')<\frac{1}{3}$ and $\mu^3_{I^2_3}(\langle R',R''\rangle)=\mu^3_{I^1_3}(\langle L',R''\rangle)<\frac{1}{3}$. \\
(9). $(R', (0,1-\beta^2_{I^1_2}(R''), \beta^2_{I^1_2}(R'')), c, c)$ with $\frac{1}{2}>\beta^2_{I^1_2}(R'')>\frac{1}{3}$  and $\frac{1}{3}<\mu^3_{I^2_3}(\langle R',R''\rangle)=\mu^3_{I^1_3}(\langle L',R''\rangle)<\frac{1}{2}$. \\
(10). $(R', (0,\frac{2}{3}, \frac{1}{3}), (\beta^3_{I^2_3}(a), 0, 1-\beta^3_{I^2_3}(a)), (\beta^3_{I^2_3}(a), 0, 1-\beta^3_{I^2_3}(a)))$ with $\mu^3_{I^2_3}(\langle R',R''\rangle)=\mu^3_{I^1_3}(\langle L',R''\rangle)=\frac{1}{3}$. \\
(11). $(R', (\beta^2_{I^1_2}(L''), 1-\beta^2_{I^1_2}(L'')), b, b)$ with $\beta^2_{I^1_2}(L'')<\frac{1}{3}$ and $\mu^3_{I^2_3}(\langle R',L''\rangle)=\mu^3_{I^1_3}(\langle L',L''\rangle)<\frac{1}{3}$. \\
(12). $(R', (\beta^2_{I^1_2}(L''), 1-\beta^2_{I^1_2}(L'')), c, c)$ with $\frac{1}{2}>\beta^2_{I^1_2}(L'')>\frac{1}{3}$ and $\frac{1}{3}<\mu^3_{I^2_3}(\langle R',L''\rangle)=\mu^3_{I^1_3}(\langle L',L''\rangle)<\frac{1}{2}$. \\
(13). $(R', (0,\frac{1}{3},\frac{2}{3}), (0,\beta^3_{I^1_3}(b), 1-\beta^3_{I^1_3}(b)), (0, \beta^3_{I^2_3}(b), 1-\beta^3_{I^2_3}(b)))$ with $\mu^3_{I^2_3}(\langle R',L''\rangle)=\mu^3_{I^1_3}(\langle L',L''\rangle)=\frac{1}{3}$. \\
(14). $(R', (0,\frac{1}{2},\frac{1}{2}), c, c)$ with $\mu^3_{I^2_3}(\langle R',L''\rangle)=\mu^3_{I^1_3}(\langle L',L''\rangle)=\frac{1}{2}$. }
}
\end{example}

